\documentclass{emulateapj}
\usepackage{graphicx,color}
\usepackage{epstopdf}

\usepackage{natbib}
\newcommand{\be}{\begin{eqnarray}}
\newcommand{\ee}{\end{eqnarray}}
\newcommand{\ds}{\displaystyle}
\usepackage{natbib}
\newcommand\gcm{g~cm$^{-3}$}
\newcommand\simgreater{\,\lower0.7ex\hbox{$\stackrel{>}{\sim}$}\,}
\newcommand\simless{\,\lower0.7ex\hbox{$\stackrel{<}{\sim}$}\,}
\newcommand\msol{$M_\odot$}

\newcommand{\ye}{$Y_{{\rm e}}$}
\newcommand{\yl}{$Y_{{\ell}}$}

\newcommand{\nue}{\ensuremath{\nu_{{\rm e}}}}
\newcommand{\nuebar}{\ensuremath{\bar{\nu}_{{\rm e}}}}

\newcommand{\nux}{\ensuremath{\nu_{{\rm x}}}}
\newcommand{\num}{\ensuremath{\nu_{\mu}}}
\newcommand{\numbar}{\ensuremath{\bar{\nu}_{\mu}}}
\newcommand{\nut}{\ensuremath{\nu_{\tau}}}
\newcommand{\nutbar}{\ensuremath{\bar{\nu}_{\tau}}}

\newcommand{\nuesph}{$\nu_{\rm e}$-sphere}

\newcommand{\pderiv}[2]{\frac{\partial{#1}}{\partial{#2}}}

\begin{document}

\title{Fluid Stability Below the Neutrinospheres of Supernova Progenitors and the Dominant Role of Lepto-Entropy Fingers}

\author{S. W. Bruenn\altaffilmark{}, and E. A.  Raley\altaffilmark{}}
\affil{Department of Physics, Florida Atlantic University, Boca Raton, FL 33431--0991}

\and

\author{A. Mezzacappa\altaffilmark{} }
\affil{Physics Division, Oak Ridge National
Laboratory, Oak Ridge, TN 37831--6354}

\begin{abstract}
Instabilities driven by thermal and lepton diffusion (doubly diffusive instabilities) in a Ledoux stable fluid will, if present below the neutrinosphere of the collapsed core of a supernova progenitor (proto-supernova), induce convective-like fluid motions there. These fluid motions may enhance the neutrino emission by advecting neutrinos outward toward the neutrinosphere, and may thus play an important role in the supernova mechanism. ``Neutron fingers,'' in particular, have been suggested as being critical for producing explosions in the sophisticated spherically symmetric supernova simulations by the Livermore group \citep[e.g.,]{wilsonm93}. These have been argued to arise in an extensive region below the neutrinosphere of a proto-supernova where entropy and lepton gradients are stabilizing and destabilizing, respectively, if, as they assert, the rate of neutrino-mediated thermal equilibration greatly exceeds that of neutrino-mediated lepton equilibration. Application of the Livermore group's criteria to models derived from core collapse simulations using both their equation of state and the Lattimer-Swesty equation of state do indeed show a large region below the neutrinosphere unstable to neutron fingers.

Because of the potential importance of fluid instabilities for the supernova mechanism, and the desire to understand the origin of convective-like fluid motions that may arise in upcoming multi-dimensional radiation-hydrodynamical simulations of core collapse, we develop a methodology introduced by \citet{bruennd96} for analyzing the stability of a fluid in the presence of neutrinos of all flavors and in the  presence of a gravitational field. Neutrino-mediated thermal and lepton equilibration between a fluid element and its surroundings (background) is modeled as a linear system characterized by four response functions (i.e., thermal and lepton equilibration driven by entropy and lepton fraction differences between a fluid element and the background), the latter evaluated for a given thermodynamic state and fluid element radius by detailed neutrino transport simulations. These transport simulations employ both traditional and improved neutrino physics.

When applied to an extensive two-dimensional grid of core radii and fluid element sizes for each of several time slices of a number of proto-supernovae, we find no evidence for the neutron finger instability as described by the Livermore group. We find, instead, that the rate of lepton equilibration always exceeds that of thermal equilibration. Furthermore, we find that neither of the ``cross'' response functions, that is, entropy equilibration driven by a lepton fraction difference, and lepton equilibration driven by an entropy difference, is zero and that the first of these tends to be the largest of the four response functions in magnitude. These cross response functions play a critical role in the dynamics of the equilibration of a fluid element with the background. An important consequence of this is the presence of a doubly diffusive instability, which we refer to as ``lepto-entropy fingers,'' in an extensive region below the neutrinosphere where the lepton number, $Y_{\ell}$, is small. This instability is driven by a mechanism very different from that giving rise to neutron fingers, and may play an important role in enhancing the neutrino emission. Deep in the core where the entropy is low and the lepton number higher, our analysis indicates a region unstable to another instability, also involving the cross response functions, which we refer to as ``lepto-entropy semiconvection.'' These instabilities, particularly lepto-entropy fingers, may have already been seen in some multi-dimensional core collapse simulations described in the literature.\end{abstract}


\newpage

\section{Introduction}
\label{sec:DDInstabilities}

It has long been recognized that fluid instabilities may play an important role in the supernova mechanism. It is generally agreed that entropy driven convection between the neutrinosphere and the shock during the postbounce phase of the collapsed core of a supernova progenitor (hereafter "proto-supernova") will contribute turbulent pressure and increase the neutrino energy-deposition efficiency \citep{bethe90, millerwm93, herantbhfc94, burrowshf95, jankam96, mezzacappacbbgsu98b, fryer99, fryerw02, burasrjk02}, but this may not be enough to produce explosions.

More controversial is the nature and the role of fluid instabilities below the neutrinosphere. Here matter and neutrinos are tightly coupled, and convective-like motions induced by instabilities can advect neutrinos. If these convective-like motions occur between the deeper interior and the neutrinosphere, \nue's could be advected from the lepton rich deeper interior to the neutrinosphere, thus adding to the rate of \nue\ diffusion and enhancing the \nue\ luminosity. The analysis of fluid instabilities below the neutrinosphere must take into account the possibilities of thermal and lepton transfer by neutrinos. For example, a displaced fluid element in the presence of an entropy and/or lepton fraction gradient will find that its entropy and/or lepton fraction will differ from that of its surroundings or background. This will induce neutrino mediated thermal and lepton transfer which can modify the hydrodynamic buoyancy forces felt by the fluid element. This in turn can modify the subsequent growth rate of an instability, or the very existence of an instability. In particular, instabilities can arise in a gravitating fluid because of thermal and lepton transport that would otherwise be stable in their absence. These instabilities, involving the transport of two quantities (e.g., energy and leptons), are referred to as doubly diffusive instabilities.

     To introduce the phenomenon, let us consider several classic examples of doubly diffusive instabilities. A doubly diffusive instability will arise if there are two differing ``substances,'' say heat and
leptons, with gradients that can act oppositely in their stabilizing effects. For the sake of
a concrete model, consider the case of heat and salt in water. Let us first imagine the case in which a
region of hot salty water overlies a region of cold fresh water \citep{stern60}. Normally the diffusion
of heat occurs more rapidly than the diffusion of salt, and we will assume this to be the case here. Let
us also assume that the magnitudes of the gradients in heat and salt are such that the fluid is stably stratified gravitationally (i.e., Ledoux stable). Imagine a parcel of hot salty water to be depressed
slightly, as shown in Figure 1. The rapid diffusion of heat will thermally equilibrate this parcel of
water with the background, but the slower diffusion of salt will result in the water remaining salty.
We thus end up with a pocket of cold salty water in a background cold fresh water, and this pocket, being denser than the background, will continue to sink. The result is that ``fingers'' of salty water will poenetrate the fresh water on a thermal diffusion time scale. Note that what creates the instability in this otherwise stable situation is the presence of diffusion, and thus its description as a ``diffusive instability.'' The above described instability is referred to as salt fingers, and is an example of the kind of doubly diffusive instability that occurs if a slowly diffusing substance with a destabilizing gradient is stabilized in the absence of diffusion by the gradient of a rapidly diffusing substance.

\begin{figure}[h!]
\setlength{\unitlength}{1cm}
{\includegraphics[width=\columnwidth]{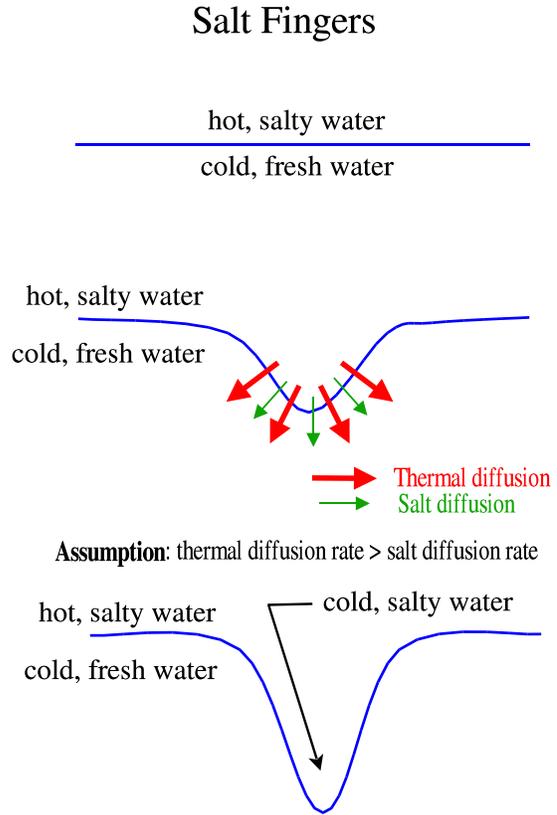}}
\caption{\label{fig:Neutron_Fingers}
Simple model of the salt-finger instability.}
\end{figure}

\begin{figure}[h!]
\setlength{\unitlength}{1cm}
{\includegraphics[width=\columnwidth]{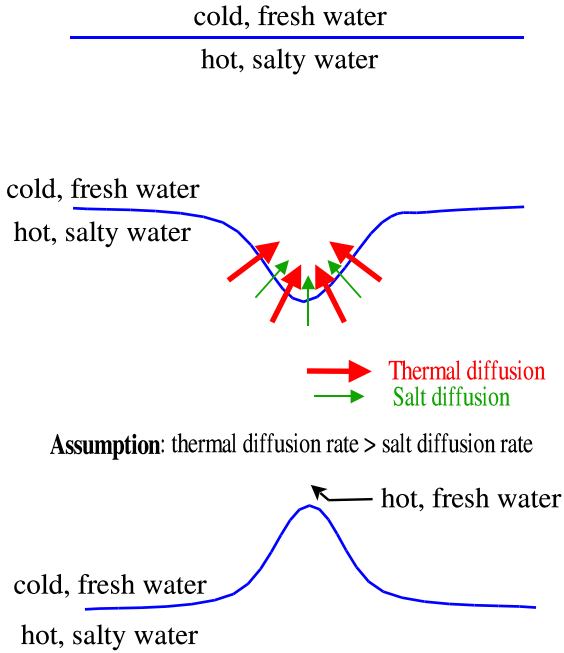}}
\caption{\label{fig:Semi_Convection}
Simple model of the semiconvective instability.}
\end{figure}


     Now imagine that the spatial configuration described above is reversed, namely, that a region of
cold fresh water overlies a region of hot salty water, so that the gradient of salt is now stabilizing,
the gradient of heat dstabilizing, and the magnitudes of the gradients such that the fluid is
Ledoux stable \citep{veronis65}. Imagine further that a parcel of cold fresh water is depressed
slightly into the hot salty water, as shown in Figure 2. Again the rapid diffusion of heat will
thermally equilibrate the parcel with the background, but the slower diffusion of salt will result in
the parcel remaining fresh. We thus end up with a parcel of hot fresh water in a background of hot salty water,  and the parcel, being less dense than originally from the background, will accelerate upwards faster than would have been the case with no diffusion. This can result in an overstable situation, driven by the phase lag between the temperature and velocities of the fluid element which causes net work to be done over each cycle \citep{spiegel72}. In this case the parcel will oscillate with growing amplitude. This instability is referred to as ``semiconvection,'' and is an example of the kind of doubly diffusive instability that occurs if a rapidly diffusing substance with a destabilizing gradient is stabilized in the absence of diffusion by a slowly diffusing substance with a stabilizing gradient.

     Doubly diffusive instabilities such as described above can occur in the collapsed core of a supernova progenitor \citep{smarrwbb81}. Neutron rich (low \ye, where \ye\ is the proton fraction) matter tends to be denser than neutron poor (high \ye) matter at the same temperature and pressure because the latter contains more light particles (electrons and electron neutrinos) that contribute to the pressure. Thus neutron richness in stellar cores can be thought of as playing a role analogous to that of salt in the discussion above. The ramp-up of the bounce shock on core rebound results in an outwardly positive (stabilizing) entropy gradient throughout much of the inner core. On the other hand, a negative lepton gradient formed at and initially confined to the vicinity of the \nuesph\ immediately after shock breakout is advected deeper into the core with time because of the inward matter flow and outward diffusion of \nue's. A destabilizing negative \ye\ gradient stabilized by the outwardly positive entropy gradient can set the stage for a doubly diffusive instability analogous to salt fingers if lepton transport is slower than thermal transport. This instability is referred to as ``neutron fingers.''

     The Livermore group has reported successful explosions in their spherically symmetric supernova simulations employing sophisticated multi-energy neutrino transport, and have emphasized the importance of the neutron-finger instability in powering these explosions \citep{wilsonmww86, wilsonm88, wilsonm93}. The issue of whether or not this instability is operative, and if so whether or not it will power an explosion is important, as no other supernova simulations employing neutrino transport of comparable sophistication but without invoking neutron fingers has produced an explosion \citep{bruenn93, mezzacappalmhtb01, liebendorfermtmhb01, burasrjk02}.
     
     The Livermore group argue that thermal equilibration between a fluid element and the background is driven by the total neutrino number density (i.e., \nue's, \num's, and \nut's and their antiparticles), whereas equilibration in composition (i.e., lepton fraction, \yl) is driven only by the difference in the number densities of \nue's and \nuebar's. The ratio, $R$, of the two equilibration rates is thus given approximately by

\begin{equation}
R = \frac{ | n_{\nue} - n_{\nuebar}  | }{  n_{\nue} + n_{\nuebar} + n_{\num} + n_{\numbar}
+ n_{\nut} + n_{\nutbar} }
\label{eq:DDI1}
\end{equation}

\noindent  where $n_{i}$ is the number density of neutrinos of type $i$. From their simulations they typically find that $R < 0.1$. Based on the magnitude of this inequality the Livermore group tests for the neutron-finger instability by displacing a fluid element at constant composition but equilibrated with the background material in pressure and temperature. If the density of the fluid element relative to its surroundings after displacement is such as to drive the displacement further, the fluid is taken to be neutron-finger unstable. Mathematically, their criterion for neutron-finger instability is given by

\begin{equation}
\left( \pderiv{\rho}{Y_{\rm e} } \right)_{T,P} \frac{d Y_{\rm e} }{dr} > 0
\label{eq:DDI2}
\end{equation}

\noindent where $\frac{d Y_{\rm e} }{dr}$ is the $Y_{\rm e}$  gradient of the background. If a region was established to be neutron-finger unstable, a mixing length algorithm was then employed to simulate the convective fluid motions and mixing. The Livermore group found that the result of the convective-like motions due to the neutron-finger instability in their supernova simulations was the production of a larger \nue\ luminosity, as neutrinos and electrons previously trapped in the core were brought to the vicinity of the neutrinosphere. This increase in \nue\ luminosity was found to be sufficient to power an explosion.

In several papers, \citet{bruennmd95} and \citet{bruennd96} questioned the assumption that the ratio, $R$, of the rates of composition to thermal equilibration is given by equation (\ref{eq:DDI1}). Situations were envisioned in which a fluid element perturbed with respect to the background could drive counter flows of \nue's and \nuebar's so that the numerator of equation (\ref{eq:DDI1}) would be additive in  $n_{\nue}$ and $n_{\nuebar}$ and the denominator subtractive. Detailed simulations, performed by  \citet{bruennd96}, of the equilibration of fluid elements perturbed with respect to the background showed that in many cases counter flows of \nue's and \nuebar's did indeed occur. Furthermore, it was found that lepton transport proceeded quickly by the rapid transport of low energy \nue's and \nuebar's, as the opacity for these was relatively small. On the other hand, thermal equilibration, being more efficiently accomplished by high energy neutrinos, was slowed by the relatively large opacities encountered by these neutrinos. Finally, the \num's, \nut's, and their antiparticles were less effective at mediating thermal equilibration because of there weak thermal coupling with the matter. The result was that the ratio of  composition to thermal equilibration depended on how the fluid element was perturbed with respect to the background, and in many cases this ratio was greater than unity.

Given its potential role in the core collapse supernova mechanism, there is clearly a need to clarify the issue of the neutron-finger instability in a proto-supernova. Additionally there has been the advent during the past decade of multi-dimensional core collapse supernova modeling \citep{millerwm93, herantbhfc94, burrowshf95, jankam96, mezzacappacbbgsu98a, mezzacappacbbgsu98b, fryer99, fryerw02, burasrjk02}, with more simulations of greater sophistication to come. The extant simulations show the presence of fluid instabilities both above and below the neutrinosphere. The fluid instabilities above the neutrinosphere are driven by neutrino heating immediately above the gain radius, which is the surface above the neutrinosphere where neutrino cooling and heating balance. This instability is analogous to that of a fluid in a pot heated from below. The fluid instabilities below the neutrinosphere are more complicated to characterize, as neutrino transport drives both thermal and composition changes and the fluid instabilities can be doubly diffusive.

\citet{bruennd96} developed a procedure for characterizing the stability/instability of a fluid element as a function of the entropy and composition gradient of the background and found that for typical conditions inside the collapsed stellar core of a supernova progenitor the fluid was either unstable to semiconvection or stable. However, they sampled only a small region of the run of thermodynamic conditions inside a supernova progenitor. In this paper we will further develop the methodology of \citet{bruennd96} and use it to examine the stability of proto-supernovae for the entire run of thermodynamic conditions from the center of the core to the neutrinosphere. Futhermore, for each set of thermodynamic conditions we will consider the stability as a function of the size of the fluid element. The latter is important in that it governs the rate of thermal and composition equilibration. In Section  \ref{sec:SPA} we begin our discussion of fluid stability in the ambience of neutrino mediated thermal and lepton transport by presenting the equations of motion of a fluid element that we will use to analyze stability. These equations involve four response functions that describe how thermal and lepton diffusion is driven by an entropy or a lepton fraction difference between a fluid element and the background. In Section \ref{sec:SA} we describe how the solutions of these equations of motion are used to define various types of instability. We also examine the stability of a fluid element as a function of the entropy and lepton gradients of the background for various representative sets of the ``response functions.'' These examples illustrate the different kinds of instabilities that can arise in a proto-supernova. In particular, we find that neutron fingers is unlikely to occur anywhere below the neutrinosphere. However, we discover and describe several new and potentially important modes of doubly diffusive instabilities which we refer to as ``lepto-entropy fingers'' and ``lepto-entropy semiconvection.'' Both of the instabilities involve the cross response functions in essential ways. In Section \ref{sec:Response} we describe how the response functions are computed for a given thermodynamic state and fluid element size. In Section \ref{sec:CCModels} we begin the application of our stability analysis to core collapse models by describing these models, and in Section \ref{sec:Results} we present the results of an extensive set of calculations giving the stability or type of instability and its growth rate as a function both the fluid element size and its location in the core. In Section \ref{sec:Comp} we compare our analyses and results with prior work, and in in Section \ref{sec:Cncl} we present our conclusions.

\section{Single Particle Analysis}
\label{sec:SPA}

We consider a mix of hot, dense material and neutrinos at conditions typical of the region below the neutrinosphere of a proro-supernova ($\rho > 10^{12}$ g cm$^{-3}$, $T > 10$ MeV, $Y_{\ell} < 0.5$. Such a mix can be considered to be equilibrated with respect to strong, electromagnetic, and weak interactions. Its thermodynamic and compositional states can therefore be specified by three independent variables, which we take to be the pressure $p$, entropy $s$, and lepton fraction fraction $Y_{\ell}$.

In a manner analogous to \citet{grossmanna93}, we consider the behavior of an individual fluid element in a background which is assumed to be
time-independent and one-dimensional. The z-axis is taken normal to the plane of symmetry and the
background is assumed to have constant gradients in entropy $\bar{s}$ and lepton fraction $\bar{Y}_{\ell}$. We will assume that the fluid
element is always in pressure equilibrium with the background, i.e., $p = \bar{p}$, where unmarked and
barred variables refer to the fluid element and the background, respectively. However, the fluid
element and the background may differ locally in $s$ and $Y_{\ell}$, and these differences will be denoted by
the quantities $\theta_{s}$ and $\theta_{Y_{\ell}}$, where $\theta_{s} \equiv \theta_{s}(z) = s(z) - \bar{s}(z)$ and
$\theta_{Y_{\ell}} \equiv \theta_{Y_{\ell}}(z) = Y_{\ell}(z) - \bar{Y}_{\ell}(z)$. Given $\bar{p}(z)$,  $\bar{s}(z)$, and $\bar{Y}_{\ell}(z)$, the thermodynamic state of the fluid element is given
by its values of $\theta_{s}$, $\theta_{Y_{\ell}}$, and its vertical location $z$ (positive for
directions away from the center of the core) through

\be
& {\ds     s(z) = \bar{s}(z) + \theta_{s}(z)
} & \nonumber \\ 
& {\ds = \bar{s}(z_{0})
+ \left. \frac{d\bar{s}}{dz} \right|_{z_{0}} (z - z_{0}) + \theta_{s}(z)
} &
\label{eq:eq1}
\ee

\be
& {\ds   Y_{\ell}(z) = \bar{Y_{\ell}}(z) + \theta_{Y_{\ell}}(z)
} & \nonumber \\ 
& {\ds = \bar{Y_{\ell}}(z_{0})
+ \left. \frac{d\bar{Y_{\ell}}}{dz} \right|_{z_{0}} (z - z_{0}) + \theta_{Y_{\ell}}(z)
} &
\label{eq:eq2}
\ee

\begin{equation}
  p = \bar{p}(z).
\label{eq:eq3}
\end{equation}

The quantities $\theta_{s}$, and $\theta_{Y_{\ell}}$, will change with time by the effect of energy and
lepton transport by neutrinos, and by the vertical motion of the fluid element through the gradients of $\bar{s}$ and $\bar{Y}_{\ell}$ of the background. We assume that neutrino energy and lepton transport is linear for small $\theta_{s}$, and $\theta_{Y_{\ell}}$ and describe it by the four ``response functions'' $\Sigma_{s}$,
$\Sigma_{Y_{\ell}}$, $\Upsilon_{s}$, and $\Upsilon_{Y_{\ell}}$, defined by 

\be
& {\ds    \Sigma_{s} = \left. \frac{ \dot{\theta}_{s} }{ \theta_{s} } \right|_{\theta_{Y_{\ell}} = 0}, \quad 
\Sigma_{Y_{\ell}} = \left. \frac{ \dot{\theta}_{s} }{ \theta_{Y_{\ell}} }
\right|_{\theta_{s} = 0},
} & \nonumber \\ 
& {\ds
\Upsilon_{s} = \left. \frac{ \dot{\theta}_{Y_{\ell}} }{ \theta_{s} } \right|_{\theta_{Y_{\ell}} = 0},
\quad 
\Upsilon_{Y_{\ell}} = \left. \frac{ \dot{\theta}_{Y_{\ell}} }{ \theta_{Y_{\ell}} } \right|_{\theta_{s}
= 0}.
} &
\label{eq:eq4}
\ee

\noindent Thus, $\Sigma_{s}$ gives the rate of change of $\theta_{s}$ due to the presence of a nonzero entropy difference, $\theta_{s}$, and a zero lepton fraction difference, $\theta_{Y_{\ell}}$, between the fluid element and the background, $\Sigma_{Y_{\ell}}$ gives the rate of change of $\theta_{s}$ due to the presence of a nonzero lepton fraction difference, $\theta_{Y_{\ell}}$, and a zero entropy difference, $\theta_{s}$, and so forth. With these definitions, it follows that

\begin{equation}
  \left. \dot{\theta}_{s} \right|_{\nu} = \Sigma_{s} \theta_{s} + \Sigma_{Y_{\ell}} \theta_{Y_{\ell}}
\label{eq:eq5}
\end{equation}

\noindent and

\begin{equation}
  \left. \dot{\theta}_{Y_{\ell}} \right|_{\nu} = \Upsilon_{s} \theta_{s} + \Upsilon_{Y_{\ell}}
\theta_{Y_{\ell}}
\label{eq:eq6}
\end{equation}

\noindent where $\left. \dot{\theta}_{s} \right|_{\nu}$ and $\left. \dot{\theta}_{Y_{\ell}}
\right|_{\nu}$ are the rate of change respectively of $\theta_{s}$ and $\theta_{Y_{\ell}}$ due to
neutrino transport. The presence of the ``cross'' response functions $\Sigma_{Y_{\ell}}$ and
$\Upsilon_{s}$ is due to the fact that, (1) entropy transport is brought about by both thermal and lepton
transport, (2) that neutrinos transport both energy and leptons, and (3) that transport is inherently a dissipative
process. Thus,
for example, a difference, $\theta_{Y_{\ell}}$, in $Y_{\ell}$ can force a rate of change of $s$ as well as $Y_{\ell}$, as can a difference, $\theta_{s}$, in $s$. These cross response functions introduce important additional patterns of stability/instability
behavior as a function of the gradients in $s$ and $Y_{\ell}$, as will be described below. Discussions of doubly diffusive instabilities apart from the proto-supernova problem and prior to \citet{bruennd96} did not relate to a context in which cross response functions were important, and these were not considered. 

The equations for $\dot{\theta}_{s}$ and $\dot{\theta}_{Y_{\ell}}$ are
completed by adding to equations (\ref{eq:eq5}) and (\ref{eq:eq6}) the effect of the vertical motion of the fluid element, $v = \dot{z}$, to get

\begin{equation}
 \dot{\theta}_{s} = \Sigma_{s} \theta_{s} + \Sigma_{Y_{\ell}} \theta_{Y_{\ell}}
- \frac{d\bar{s}}{dz} v
\label{eq:eq7}
\end{equation}

\begin{equation}
  \dot{\theta}_{Y_{\ell}} = \Upsilon_{s} \theta_{s} + \Upsilon_{Y_{\ell}} \theta_{Y_{\ell}}
- \frac{d\bar{Y}_{\ell}}{dz} v
\label{eq:eq8}
\end{equation}

\noindent The equation of motion for the fluid element is given by the buoyancy acceleration driven by the difference between its density and that of the background, viz.,

\begin{equation}
  \dot{v} = - \frac{g}{\rho} \left( \pderiv{\rho}{s} \right)_{p,Y_{\ell}} \theta_{s}
- \frac{g}{\rho} \left( \pderiv{\rho}{Y_{\ell}} \right)_{p,s} \theta_{Y_{\ell}}.
\label{eq:eq9}
\end{equation}

\noindent where $g$ is the magnitude of the acceleration of gravity.

\section{Stability Analysis}
\label{sec:SA}

To investigate the stability of a fluid element described by equations (\ref{eq:eq7}) - (\ref{eq:eq9}),
we regard these equations as giving the time evolution of $\theta_{s}$, $\theta_{Y_{\ell}}$, and $v$ and consider in this Section classes of possible time dependent behaviors of a fluid element perturbed from equilibrium as a function of the gradients in $\bar{s}$ and $\bar{Y_{\ell}}$, the radius, $R_{\rm blob}$, of the fluid element, and the gravitational force, $g$. In particular, we look for solutions of the form

\begin{equation}
  \theta_{s} = \theta_{s \, 0} e^{\sigma t}, \qquad
\theta_{Y_{\ell}} = \theta_{Y_{\ell} \, 0} e^{\sigma t}, \qquad
v = v_{0} e^{\sigma t}
\label{eq:st1}
\end{equation}

\noindent where the eigenvalues, $\sigma$, will determine the stability of the fluid, as described below.

In Sections \ref{sec:Ledoux} - \ref{sec:SigmaUpsilon} we look at some particular cases where the response functions take on simplified and special sets of values. The purpose is to relate some classic examples of instability to the solutions of equations equations (\ref{eq:eq7}) - (\ref{eq:eq9}), and to build from the simple to the complex. We delay until Section \ref{sec:GC} an analysis of the general case in which all four response functions have arbitrary nonzero values. The reader who is interested in proceeding directly the the general case is advised to skip directly to Section \ref{sec:GC}.

\subsection{Ledoux Stability Limit}
\label{sec:Ledoux}

We begin with a well-known case by considering the limit in which there is no thermal or lepton transport (i.e., the limit in which all four response functions are zero). Instability in this case is referred to as Ledoux instability, and equations (\ref{eq:eq7}) - (\ref{eq:eq9}) in this case reduce to

\begin{equation}
 \dot{\theta}_{s} = - \frac{d\bar{s}}{dz} v
\label{eq:L1}
\end{equation}

\begin{equation}
  \dot{\theta}_{Y_{\ell}} = - \frac{d\bar{Y}_{\ell}}{dz} v
\label{eq:L2}
\end{equation}

\begin{equation}
  \dot{v} = - \frac{g}{\rho} \left( \pderiv{\rho}{s} \right)_{p,Y_{\ell}} \theta_{s}
- \frac{g}{\rho} \left( \pderiv{\rho}{Y_{\ell}} \right)_{p,s} \theta_{Y_{\ell}}.
\label{eq:L3}
\end{equation}

\noindent The eigenvalues $\sigma$ are given by

\begin{equation}
  0 = \left| \begin{array}{ccc}
\sigma & 0 &  \frac{d\bar{s}}{dz} \\
0 & \sigma &  \frac{d\bar{Y}_{\ell}}{dz} \\
\frac{g}{\rho} \left( \pderiv{\rho}{s} \right)_{p,Y_{\ell}} &
\frac{g}{\rho} \left( \pderiv{\rho}{Y_{\ell}} \right)_{p,s} & \sigma \end{array} \right|
\label{eq:L4}
\end{equation}

\noindent with the solutions

\be
& {\ds   \sigma_{1.2} = \pm  \sqrt{ g }
\sqrt{ \left( \pderiv{ \ln \rho}{ \ln s} \right)_{p,Y_{\ell}} \frac{d \ln \bar{s}}{dz}
+ \left( \pderiv{ \ln \rho}{ \ln Y_{\ell}} \right)_{p,s} \frac{d \ln \bar{Y}_{\ell}}{dz} }
} & \nonumber \\ 
& {\ds = \sqrt{ - \omega_{s}^{2} - \omega_{Y_{\ell}}^{2} },
\qquad \sigma_{3} = 0.
} &
\label{eq:L5}
\ee
\noindent where we define $\omega_{s}$ and  $\omega_{Y_{\ell}}$ by

\be
& {\ds \omega_{s}^{2} = - g  \left( \pderiv{ \ln \rho}{ \ln s} \right)_{p,Y_{\ell}} \frac{d \ln \bar{s}}{dz}, } & \nonumber \\ 
& {\ds \omega_{Y_{\ell}}^{2} = - g  \left( \pderiv{ \ln \rho}{ \ln Y_{\ell}} \right)_{p,s} 
\frac{d \ln \bar{Y}_{\ell}}{dz}.
} &
\label{eq:L6a}
\ee

\noindent Physically, if $\frac{d \ln \bar{Y}_{\ell}}{dz} = 0$ then $\omega_{s}$ is the frequency that a fluid element would oscillate about a stable equilibrium if $\omega_{s}^{2} > 0$, and the growth rate that would characterize the motion of a fluid element away from an unstable equilibrium if $\omega_{s}^{2} < 0$.  $\omega_{Y_{\ell}}$ has an analogous interpretation, e.g., if $\frac{d \ln \bar{s}}{dz} = 0$ then $\omega_{Y_{\ell}}$ is the frequency that a fluid element would oscillate about a stable equilibrium if $\omega_{Y_{\ell}}^{2} > 0$, and the growth rate that would characterize the motion of a fluid element away from an unstable equilibrium if $\omega_{Y_{\ell}}^{2} < 0$. $\omega_{s}$ and $\omega_{Y_{\ell}}$ thus characterize the effect of the background gradients, $\frac{d \ln \bar{s}}{dz}$ and $\frac{d \ln \bar{Y}_{\ell}}{dz}$, respectively, on the dynamics of a fluid element perturbed from equilibrium. In equation (\ref{eq:L5}), the solution $\sigma_{3} = 0$ arises from the fact that a fluid element at rest and having the same thermodynamic conditions as the background (i.e., $\theta_{s} = \theta_{Y_{\ell}} = v = 0$) will remain at rest. Note, however, that the equilibrium implied by the solution $\sigma_{3} = 0$ may not be a stable equilibrium.

The criterion for Ledoux stability is that the eigenvalues $\sigma$ be imaginary, which requires the following inequality be satisfied:

\begin{equation}
  \left( \pderiv{ \ln \rho}{ \ln s} \right)_{p,Y_{\ell}} \frac{d \ln \bar{s}}{dz}
+ \left( \pderiv{ \ln \rho}{ \ln Y_{\ell}} \right)_{p,s} \frac{d \ln \bar{Y}_{\ell}}{dz} < 0.
\label{eq:L6}
\end{equation}

\noindent If this inequality is satisfied, then the two solutions, $\omega_{1.2}$, can be written

\begin{equation}
  \sigma_{1.2} = \pm i \omega_{B-V}
\label{eq:L7}
\end{equation}

\noindent where

\be
& {\ds   \omega_{B-V} = \sqrt{ g }\sqrt{ - \left( \pderiv{ \ln \rho}{ \ln s} \right)_{p,Y_{\ell}} \frac{d \ln \bar{s}}{dz}
- \left( \pderiv{ \ln \rho}{ \ln Y_{\ell}} \right)_{p,s} \frac{d \ln \bar{Y}_{\ell}}{dz} }
} & \nonumber \\ 
& {\ds = \sqrt{ \omega_{s}^{2} + \omega_{Y_{\ell}}^{2} }
} &
\label{eq:L7a}
\ee

\noindent is the Brunt-V\"{a}is\"{a}l\"{a} frequency. In this case the fluid element will oscillate about the equilibrium configuration with the angular frequency $\omega_{B-V}$ and constant amplitude. This situation is therefore stable. On the other hand, if inequality (\ref{eq:L6}) is not satisfied, i.e., if

\begin{equation}
  \left( \pderiv{ \ln \rho}{ \ln s} \right)_{p,Y_{\ell}} \frac{d \ln \bar{s}}{dz}
+ \left( \pderiv{ \ln \rho}{ \ln Y_{\ell}} \right)_{p,s} \frac{d \ln \bar{Y}_{\ell}}{dz} > 0,
\label{eq:L8}
\end{equation}

\noindent then  the solutions, $\sigma_{1}$ and $\sigma_{2}$, can be written $\sigma_{1,2} = \pm |\omega_{B-V}|$, with the positive solution corresponding to an exponentially growing amplitude with growth rate $|\omega_{B-V}|$ given by

\be
& {\ds   |\omega_{B-V}|
} & \nonumber \\ 
& {\ds = \sqrt{ g }
\sqrt{ \left( \pderiv{ \ln \rho}{ \ln s} \right)_{p,Y_{\ell}} \frac{d \ln \bar{s}}{dz}
+ \left( \pderiv{ \ln \rho}{ \ln Y_{\ell}} \right)_{p,s} \frac{d \ln \bar{Y}_{\ell}}{dz} }.
} &
\label{eq:L9}
\ee

\noindent In this case the fluid is said to be convectively unstable. Thus, in the case of no thermal or lepton transport, the ``Ledoux critical line''

\begin{equation}
  \left( \pderiv{\ln \rho}{\ln s} \right)_{p,Y_{\ell}} \frac{d \ln \bar{s}}{dz}
+ \left( \pderiv{\ln \rho}{\ln Y_{\ell}} \right)_{p,s} \frac{d \ln \bar{Y}_{\ell}}{dz} = 0
\label{eq:L10}
\end{equation}

\noindent separates, in the $\frac{d \ln \bar{s}}{dz}$ - $\frac{d \ln \bar{Y}_{\ell}}{dz}$ plane, a
convectively stable region where the left-hand side of equation (\ref{eq:L10}) is $< 0$ from a
convectively unstable region where the left-hand side of equation (\ref{eq:L10}) is $> 0$. 

\begin{figure}[h!]
\setlength{\unitlength}{1cm}
{\includegraphics[width=\columnwidth]{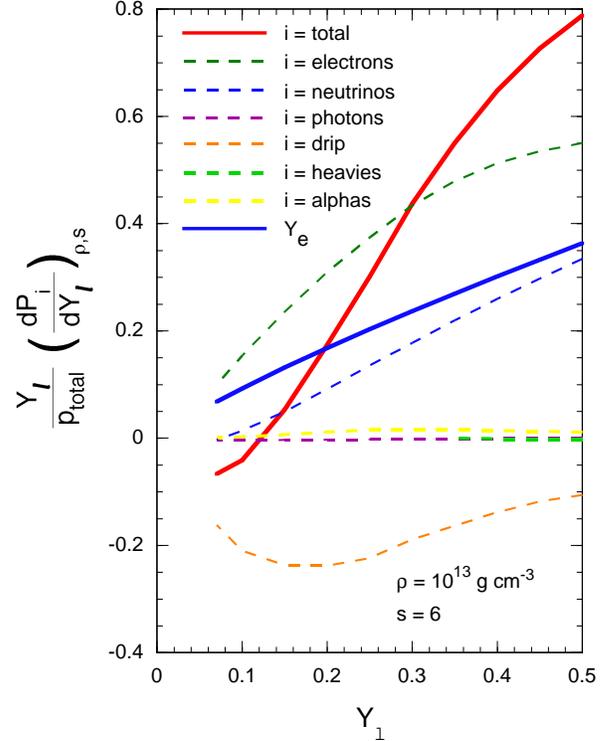}}
\vspace*{-1.0cm}
{\caption{\label{fig:dpdyl_sp}Contributions to the derivative $\left(\pderiv{p}{Y_{l}} \right)_{\rho,s}$. The dashed lines show the contributions of the indicated particles to  this derivative, the solid red line shows this derivative with all contributions, and the solid blue line shows the corresponding values of \ye.}}
\end{figure}

It is interesting to consider the signs of the coefficients of $\frac{d \ln \bar{s}}{dz}$ and $\frac{d \ln \bar{Y}_{\ell}}{dz}$ in equation (\ref{eq:L10}). The
derivative $\left( \pderiv{\ln \rho}{\ln s} \right)_{p,Y_{\ell}}$ is equal to $-Ts \frac{\alpha}{c_{p}}$, where
$c_{p} \equiv T \left( \pderiv{s}{T} \right)_{p,Y_{\ell}}$ is the always positive constant pressure specific heat and
$\alpha \equiv \frac{1}{v} \left( \pderiv{v}{T} \right)_{p,Y_{\ell}}$ is the expansivity. Provided the
latter is positive, which is almost always the case, the derivative $\left( \pderiv{\ln \rho}{\ln s}
\right)_{p,Y_{\ell}}$ is negative, and a positive gradient is $\bar{s}$ is therefore stabilizing. On the other hand, the
derivative $\left( \pderiv{\ln \rho}{\ln Y_{\ell}} \right)_{p,s}$ is equal to $- Y_{\ell} \kappa_{s} 
\left( \pderiv{p}{Y_{\ell}} \right)_{\rho,s}$, where $\kappa_{s} \equiv - \frac{1}{v} \left(
\pderiv{v}{p} \right)_{Y_{\ell},s}$ is the isentropic compressibility (always positive), so the sign of $\left( \pderiv{\ln \rho}{\ln Y_{\ell}} \right)_{p,s}$ can be negative
or positive depending on the sign of $\left( \pderiv{p}{Y_{\ell}} \right)_{\rho,s}$. The latter can be positive or negative depending on the magnitude of $Y_{\ell}$. To explain this curious behavior, note that increasing
$Y_{\ell}$ increases the number of light particles while leaving the baryon number essentially unchanged. The
entropy, which is being held constant in $\left( \pderiv{p}{Y_{\ell}} \right)_{\rho,s}$, depends on both the number of particles and the temperature. Increasing the former will
typically (but not always) require a decrease in the latter in order to maintain constant entropy, and the pressure can therefore increase or decrease.
Whether $p$ increases or decreases depends on which particle specie is dominating the pressure. Figure \ref{fig:dpdyl_sp} shows the contribution of the various particle species to $\left( \pderiv{p}{Y_{\ell}} \right)_{\rho,s}$. It is seen that the light particles (e$^{-}$'s and \nue's) provide a positive  contribution to $\left( \pderiv{p}{Y_{\ell}} \right)_{\rho,s}$ for all values of $Y_{\ell}$ for the given thermodynamic conditions, due to their partial degeneracy, but the baryons are nondegenerate and
provide a negative contribution. For values of $Y_{\ell}$ above 0.11, and for the indicated values of $\rho$ and $s$, the
contribution of e$^{-}$'s and \nue's to $\left( \pderiv{p}{Y_{\ell}} \right)_{\rho,s}$ dominate and
this derivative is positive. But for values of $Y_{\ell}$ below 0.11 the baryons dominate and this
derivative is negative. For values of $Y_{\ell}$ close to 0.11 this derivative, and therefore $\left(
\pderiv{\ln \rho}{\ln Y_{\ell}} \right)_{p,s}$, is almost zero. The stability of a fluid element in
this latter regime is extremely insensitive to the gradient $\frac{d \ln \bar{Y}_{\ell}}{dz}$, and the Ledoux
critical line is almost vertical.

\begin{figure}[!h]
\setlength{\unitlength}{1.0cm}
{\includegraphics[width=\columnwidth]{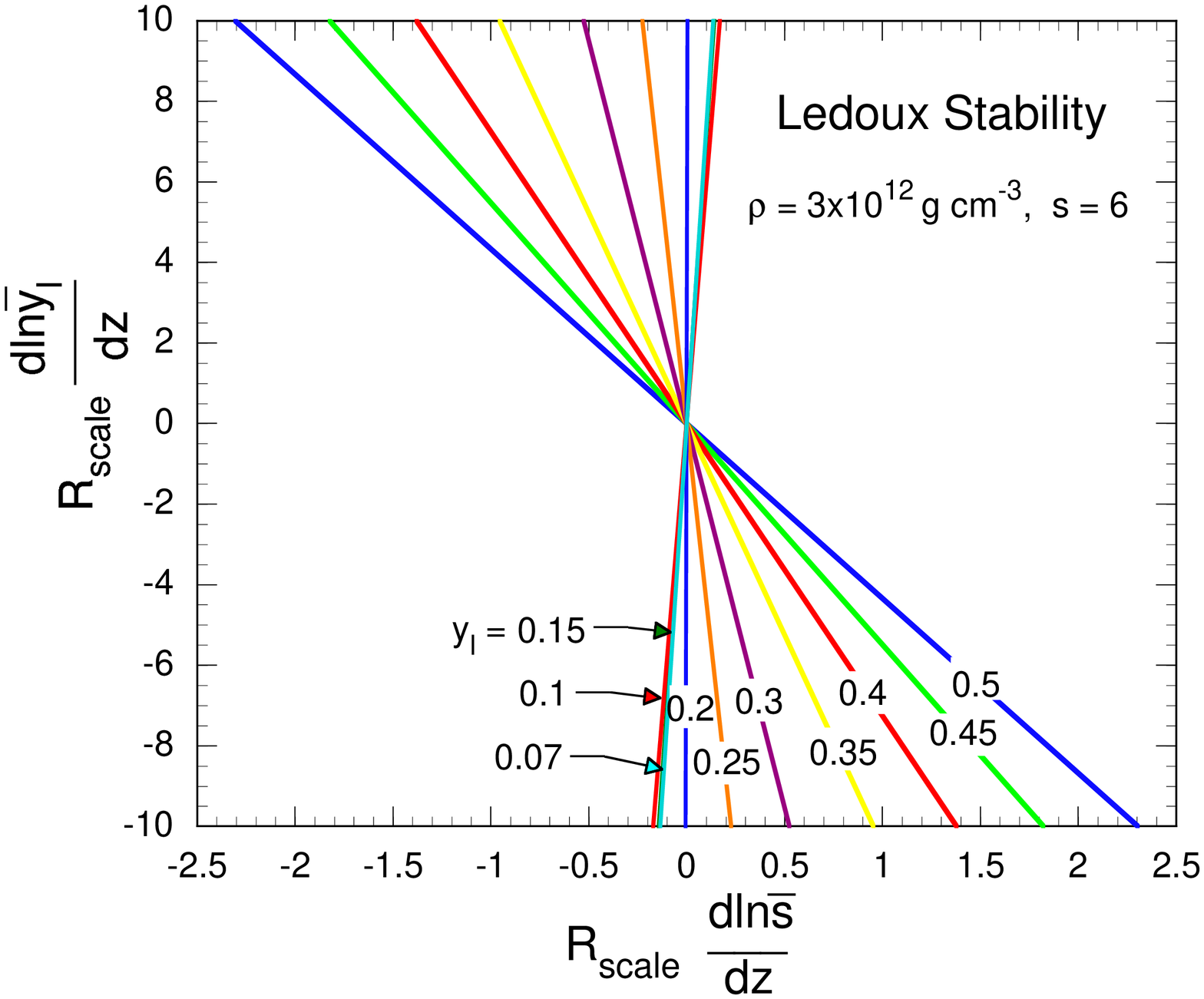}}
\caption{\label{fig:Ledoux_d312_s6}
Ledoux stability lines for $s = 6$ and the indicated values of \yl.}
\end{figure}

\begin{figure}[!h]
\setlength{\unitlength}{1.0cm}
{\includegraphics[width=\columnwidth]{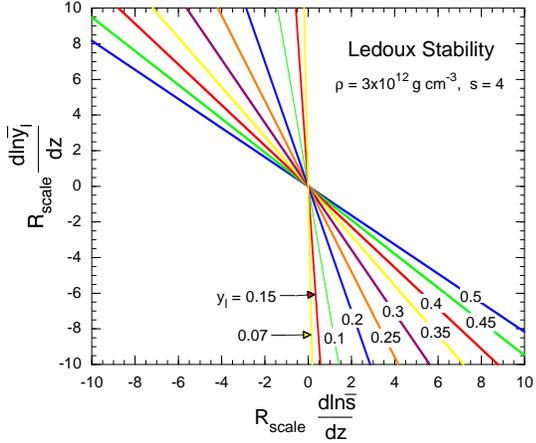}}
\caption{\label{fig:Ledoux_d312_s4}
Ledoux stability lines for $s = 4$ and the indicated values of \yl.}
\end{figure}


To exemplify  the above discussion, Figure \ref{fig:Ledoux_d312_s6} shows the Ledoux critical line as a function of $Y_{\ell}$ for the indicated values of $\rho$ and $s$. As $Y_{\ell}$ decreases it is seen that the slope of the Ledoux critical line goes from
negative through the vertical to positive. The  meaning of a positive slope for the critical line is that a negative (destabilizing) gradient in $\bar{s}$
is now balanced by a negative rather than a positive gradient in $\bar{Y}_{\ell}$. At lower entropies, on the other hand,  such as the case of $s = 4$ shown in Figure \ref{fig:Ledoux_d312_s4}, the baryons do not dominate the derivative $\left( \pderiv{p}{Y_{\ell}} \right)_{\rho,s}$ even for $Y_{\ell}$ as low as 0.07, and the slope of the Ledoux critical line remains negative for all values of $Y_{\ell}$ shown.

The sign change in the slope of the Ledoux critical line as $Y_{\ell}$ falls below a critical value will have important ramifications below for the locations of the instability regions in the $\frac{d \ln \bar{s}}{dz}$ - $\frac{d \ln \bar{Y}_{\ell}}{dz}$ plane when we consider more general sets of values for the response functions.

\subsection{Schwarzschild Stability Limit}
\label{sec:Schwarzschild}

Referring to $\Sigma_{s}$ and $\Upsilon_{Y_{\ell}}$ and to $\Sigma_{Y_{\ell}}$ and $\Upsilon_{s}$ as the direct and cross response functions, respectively, let us consider the case in which the former are nonzero while the latter are still zero. (That is, we take $\Sigma_{s} \ne 0$, $\Upsilon_{Y_{\ell}} \ne 0$, $\Sigma_{Y_{\ell}} = 0$, and $\Upsilon_{s} = 0$.) This corresponds to the case in which an entropy gradient drives an entropy, but not a lepton flow, and a lepton gradient drives a lepton but not an entropy flow. Then equations
(\ref{eq:eq7}) - (\ref{eq:eq9}) become

\begin{equation}
 \dot{\theta}_{s} = \Sigma_{s} \theta_{s} - \frac{d\bar{s}}{dz} v
\label{eq:S1}
\end{equation}

\begin{equation}
  \dot{\theta}_{Y_{\ell}} = \Upsilon_{Y_{\ell}} \theta_{Y_{\ell}} - \frac{d\bar{Y}_{\ell}}{dz} v
\label{eq:S2}
\end{equation}

\begin{equation}
  \dot{v} = - \frac{g}{\rho} \left( \pderiv{\rho}{s} \right)_{p,Y_{\ell}} \theta_{s}
- \frac{g}{\rho} \left( \pderiv{\rho}{Y_{\ell}} \right)_{p,s} \theta_{Y_{\ell}}.
\label{eq:S3}
\end{equation}

\noindent Let us further suppose in this Section that $\Upsilon_{Y_{\ell}} \rightarrow - \infty$, that
is, that the fluid element is instantaneously equilibrated in $Y_{\ell}$ with the background, leaving for the next Section the case in which $\Upsilon_{Y_{\ell}}$ is finite. If $\Upsilon_{Y_{\ell}} \rightarrow - \infty$, it follows that $\theta_{Y_{\ell}} \rightarrow 0$ which implies that $\dot{\theta}_{Y_{\ell}} \rightarrow 0$, so equation (\ref{eq:S2}) must reduce to 

\begin{equation}
  0 = \Upsilon_{Y_{\ell}} \theta_{Y_{\ell}} - \frac{d\bar{Y}_{\ell}}{dz} v
\label{eq:S4}
\end{equation}

\noindent i.e., $\theta_{Y_{\ell}} \rightarrow 0$ and $\Upsilon_{Y_{\ell}} \rightarrow - \infty$ such that $\Upsilon_{Y_{\ell}} \theta_{Y_{\ell}} \rightarrow  \frac{d\bar{Y}_{\ell}}{dz} v$. The equations for the nonzero variables now simplify to

\begin{equation}
 \dot{\theta}_{s} = \Sigma_{s} \theta_{s} - \frac{d\bar{s}}{dz} v
\label{eq:S5}
\end{equation}

\begin{equation}
  \dot{v} = - \frac{g}{\rho} \left( \pderiv{\rho}{s} \right)_{p,Y_{\ell}} \theta_{s}.
\label{eq:S6}
\end{equation}

\noindent Using the first and the last of equations (\ref{eq:st1}), the condition for nontrivial solutions is

\begin{equation}
 0 = \sigma^{2} - \Sigma_{s} \sigma - g \left( \pderiv{\ln \rho}{\ln s} \right)_{p,Y_{\ell}} 
\frac{d \ln \bar{s}}{dz}.
\label{eq:S7}
\end{equation}

Let us consider the case $\Sigma_{s} = 0$ and $\Sigma_{s} \ne 0$ separately. If $\Sigma_{s} = 0$, then

\be
& {\ds \sigma_{\pm} = \pm \sqrt{ g \left( \pderiv{\ln \rho}{\ln s} \right)_{p,Y_{\ell}} 
\frac{d \ln \bar{s}}{dz} }
} & \nonumber \\ 
& {\ds = \pm \sqrt{ -\omega_{s}^{2} }
=  \pm \sqrt{ - g \left| \left( \pderiv{\ln \rho}{\ln s} \right)_{p,Y_{\ell}} \right|
\frac{d \ln \bar{s}}{dz} }
} &
\label{eq:S8}
\ee

\noindent where in the last expression we have explicitly taken into account the fact that $ \left( \pderiv{\ln \rho}{\ln s} \right)_{p,Y_{\ell}} < 0$. Equation (\ref{eq:S8}) expresses the Schwarzschild stability criterion, viz., if $\frac{d\bar{s}}{dz} > 0$ then the roots are imaginary and the fluid is oscillatory stable, the oscillation frequency being $\omega_{s}$ defined in equation (\ref{eq:L6a}); if $\frac{d\bar{s}}{dz} < 0$ then there is one positive and one negative real root and the fluid is convectively unstable with a growth rate, $|\omega_{s}|$, given by the magnitude of the positive root, e.g.,

\begin{equation}
|\omega_{s}| = \sqrt{ g \left| \left( \pderiv{\ln \rho}{\ln s} \right)_{p,Y_{\ell}} \right|
\left| \frac{ d \ln \bar{s} }{dz}  \right| }
\label{eq:S9}
\end{equation}

\noindent The critical stability line, i.e., the line in the $\frac{d \ln Y_{\ell} }{dz}$-$\frac{d \ln s}{dz}$ plane that separates real from imaginary roots (instability from stability) is the vertical line $\frac{d \ln \bar{s}}{dz} = 0$.

Assume now that $\Upsilon_{Y_{\ell}}$ is as above (viz., $\Upsilon_{Y_{\ell}} \rightarrow - \infty$) but $\Sigma_{s} \ne 0$, and that $\Sigma_{s} < 0$, i.e., that neutrino transport will drive an entropy perturbation of a fluid element back towards equilibrium with the background. In this case the solutions of equations (\ref{eq:S1} - (\ref{eq:S3}) are

\be
& {\ds \sigma_{\pm} = \frac{\Sigma_{s}}{2} 
 \pm \sqrt{ \frac{\Sigma^{2}_{s}}{4} - \omega_{s}^{2} }
} & \nonumber \\ 
& {\ds =  \frac{- | \Sigma_{s} | }{2} 
 \pm \sqrt{ \frac{\Sigma^{2}_{s}}{4}
 - g \left| \left( \pderiv{\ln \rho}{\ln s} \right)_{p,Y_{\ell}} \right| \frac{d \ln \bar{s}}{dz} }.
} &
\label{eq:S10}
\ee

\begin{figure}[h!]
\setlength{\unitlength}{1.0cm}
{\includegraphics[width=\columnwidth]{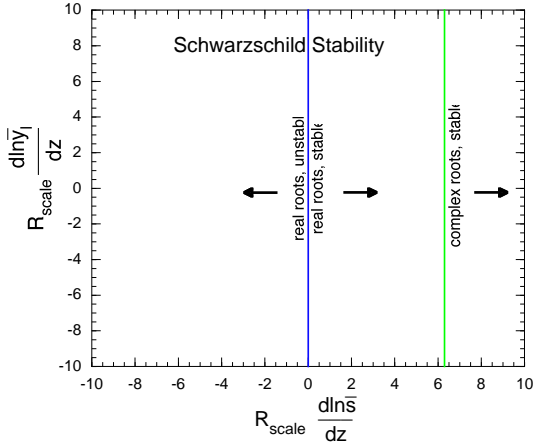}}
{\caption{\label{fig:Schwarzschild}The two critical lines  characterizing Scwarzschild stability.}}
\end{figure}

\noindent In contrast to the case in which $\Sigma_{s} = 0$ and for which the critical stability line $\frac{d \ln \bar{s}}{dz} = 0$ separates both the real and imaginary solutions and the unstable and stable regimes, with $\Sigma_{s} < 0$ the critical line bifurcates into two parallel critical lines. The situation is shown schematically in Figure \ref{fig:Schwarzschild}. The critical line $\frac{d \ln \bar{s}}{dz} = 0$, which we will refer to as critical line 1, is the line along which one real root is zero, and so changes sign as this line is crossed. It still separates stable from unstable solutions. However a second critical line $\frac{d \ln \bar{s}}{dz} = \frac{ \Sigma_{s}^{2}}{4g} \left| \left( \pderiv{\ln \rho}{\ln s} \right)_{p,Y_{\ell}} \right|^{-1}$ now appears, which we refer to as critical line 2. Along this line two real roots switch to a complex conjugate pair, so this critical line separates real from complex solutions. To the right of the critical line 2 the fluid is stable; if perturbed it will oscillate with decreasing amplitude. Along critical 2 line the neutrino transport critically damps the fluid. In the region between the two critical lines the fluid is stable but non-oscillatory---neutrino transport overdamps the fluid. To the left ($\frac{d \ln s }{dz} < 0$) of the critical line 1 the fluid is unstable, as in the $\Sigma_{s} = 0$ case, and equation (\ref{eq:S10}) can be written

\begin{equation}
\sigma_{\pm} =  \frac{- | \Sigma_{s} | }{2} 
 \pm \sqrt{ \frac{\Sigma^{2}_{s}}{4}
 - \omega_{s}^{2} }
\label{eq:S11}
\end{equation}

\noindent Close but to the left of critical line 1 separating stable from unstable solutions, where $\left| \frac{d \ln \bar{s}}{dz} \right| << \frac{ \Sigma_{s}^{2}}{4g} \left| \left( \pderiv{\ln \rho}{\ln s} \right)_{p,Y_{\ell}} \right|^{-1}$, or equivalently where $|\omega_{s}| << \Sigma_{s}/2$, the growth rate, $|\omega_{S}|$, of the instability is 

\begin{equation}
|\omega_{S}| = |\omega_{s}| \times \frac{  |\omega_{s}| }{ \Sigma_{s} }.
\label{eq:S12}
\end{equation}

\noindent The growth rate is thus reduced from the $\Sigma_{s} = 0$ case by a factor of $\frac{  |\omega_{s}| }{ \Sigma_{s} }$. This factor arises from the reduction in $\theta_{s}$ (and thus in the buoyancy force) due to the tendency of neutrino transport (as modeled by a negative $\Sigma_{s}$) to thermally equilibrate the fluid element with the background. Finally, far to the left of the first critical line, where $\left| \frac{d \ln \bar{s}}{dz} \right| >> \frac{ \Sigma_{s}^{2}}{4g} \left| \left( \pderiv{\ln \rho}{\ln s} \right)_{p,Y_{\ell}} \right|^{-1}$, or equivalently where $|\omega_{s}| >> \Sigma_{s}/2$, the growth rate of the instability is

\begin{equation}
|\omega_{S}| = |\omega_{s}| - \frac{  \Sigma_{s} }{ 2 },
\label{eq:S13}
\end{equation}

\noindent i.e., just slightly less than the $\Sigma_{s} = 0$ case. Here there is also a reduction in $\theta_{s}$ (and thus in the buoyancy force) due to the tendency of neutrino transport to thermally equilibrate the fluid element with the background, but this effect is now small compared with the growth rate, $|\omega_{s}|$, the fluid element would experience in the absence of transport.

Now imagine for a the moment that $\Sigma_{s} > 0$, i.e., that neutrino transport will drive an entropy  perturbation of a fluid element further away from equilibrium. Then

\be
& {\ds \sigma_{\pm} =  \frac{| \Sigma_{s} | }{2} 
 \pm \sqrt{ \frac{\Sigma^{2}_{s}}{4}
 - g \left| \left( \pderiv{\ln \rho}{\ln s} \right)_{p,Y_{\ell}} \right| \frac{d \ln \bar{s}}{dz} }
} & \nonumber \\ 
& {\ds  = \frac{| \Sigma_{s} | }{2} \pm \sqrt{ \frac{\Sigma^{2}_{s}}{4} - \omega_{s}^{2} } .
} &
\label{eq:S14}
\ee

\noindent In this case there is no stability. Critical line 1 ($\frac{d \ln \bar{s}}{dz} = 0$) separates the region to its left where equation (\ref{eq:S14}) gives one positive and one negative root from the region to the right but to the left of critical line 2 ($\frac{d \ln \bar{s}}{dz} = \frac{ \Sigma_{s}^{2}}{4g} \left| \left( \pderiv{\ln \rho}{\ln s} \right)_{p,Y_{\ell}} \right|^{-1}$) where equation (\ref{eq:S14}) gives two positive roots. To the right of critical line 2 equation (\ref{eq:S14}) gives complex solutions with a positive real part --- the motion of a perturbed fluid element in this case is oscillatory with a growing amplitude, i.e., semiconvective.

\subsection{$\Sigma_{s} \ne 0, \Upsilon_{Y_{\ell} } \ne 0$}
\label{sec:SigmaUpsilon}

We generalize the case considered in the preceding Section by allowing $\Upsilon_{Y_{\ell} }$ to be finite rather than - $\infty$; that is we consider the case in which the two ``direct'' response functions, $\Sigma_{s}$ and $\Upsilon_{Y_{\ell} }$, are nonzero and arbitrary, but the two ``cross'' response functions, $\Sigma_{Y_{\ell}}$ and $\Upsilon_{s} $, are still zero. Then equations (\ref{eq:eq7}) - (\ref{eq:eq9}) become equations (\ref{eq:S1}) - (\ref{eq:S3}), which are written here again as

\begin{equation}
 \dot{\theta}_{s} = \Sigma_{s} \theta_{s} - \frac{d\bar{s}}{dz} v,
\label{eq:SU1}
\end{equation}

\begin{equation}
  \dot{\theta}_{Y_{\ell}} = \Upsilon_{Y_{\ell}} \theta_{Y_{\ell}} - \frac{d\bar{Y}_{\ell}}{dz} v,
\label{eq:SU2}
\end{equation}

\noindent and

\begin{equation}
  \dot{v} = - \frac{g}{\rho} \left( \pderiv{\rho}{s} \right)_{p,Y_{\ell}} \theta_{s}
- \frac{g}{\rho} \left( \pderiv{\rho}{Y_{\ell}} \right)_{p,s} \theta_{Y_{\ell}}.
\label{eq:SU3}
\end{equation}

\noindent Solutions have the form given by equations (\ref{eq:st1}), and exist for values of $\sigma$ satisfying

\begin{equation}
\sigma^{3} + A \sigma^{2} + B \sigma^{1} + C = 0,
\label{eq:SU4}
\end{equation}

\noindent where

\begin{equation}
A = - \left( \Sigma_{s} + \Upsilon_{Y_{\ell}} \right),
\label{eq:SU5}
\end{equation}

\be
& {\ds B = \Sigma_{s} \Upsilon_{Y_{\ell}}
- g \left( \pderiv{ \ln \rho }{ \ln s } \right)_{p,Y_{\ell}} \frac{ d \ln \bar{s} }{dz}
} & \nonumber \\ 
& {\ds - g \left( \pderiv{ \ln \rho }{ \ln Y_{\ell} } \right)_{p,s} \frac{ d \ln \bar{Y}_{\ell} }{dz},
} &
\label{eq:SU6}
\ee

\noindent and

\be
& {\ds C = \Sigma_{s} g \left( \pderiv{ \ln \rho }{ \ln Y_{\ell} } \right)_{p,s}
\frac{ d \ln \bar{Y}_{\ell} }{dz}
} & \nonumber \\ 
& {\ds + \Upsilon_{Y_{\ell}} g \left( \pderiv{ \ln \rho }{ \ln s } \right)_{p,Y_{\ell}}
 \frac{ d \ln \bar{s} }{dz}.
} &
\label{eq:SU7}
\ee

The eigenvalues, $\sigma$, in this case are given by the solution of a cubic equation, and so display a richer variety. In particular, there are now three roots rather than two. The three roots are either all real, or consist of one real root and a complex conjugate pair. In the above discussion of Schwarzschild stability, critical line 1 in the $\frac{d \ln Y_{\ell} }{dz} -  \frac{d \ln s}{dz}$ plane was defined as the line along which a root is zero, so that a root changes sign as this line is crossed (i.e., a stable/unstable mode switches to a unstable/stable mode). The analog of that line here is obtained by setting $C$ in equation (\ref{eq:SU7}) to zero. This gives

\begin{equation}
\frac{ d \ln \bar{Y}_{\ell} }{dz} = - \frac{ \Upsilon_{Y_{\ell}} }{ \Sigma_{s}  }
\frac{ \left( \pderiv{ \ln \rho }{ \ln s } \right)_{p,Y_{\ell}} }
{ \left( \pderiv{ \ln \rho }{ \ln Y_{\ell} } \right)_{p,s} }  \frac{ d \ln \bar{s} }{dz},
\label{eq:SU8}
\end{equation}

\noindent which we will also refer to as critical line 1. As in the above case of Schwarzschild stability, critical line 1 here is the line across which a root changes sign.

The second critical line in the $\frac{d \ln Y_{\ell} }{dz} -  \frac{d \ln s}{dz}$ plane introduced in our discussion of Schwarzschild stability was the line across which a pair of real roots switch to a complex conjugate pair (growing/decaying modes switch to growing/decaying oscillatory modes). The analog here is the line across which two real roots switch to a complex conjugate pair, and is obtained from the equation \citep{abramowitzs65}

\begin{equation}
q^{3} + r^{2} = 0,
\label{eq:SU9}
\end{equation}

\noindent where

\begin{equation}
q= \frac{1}{3} B - \frac{1}{9} A^{2}, \qquad r = \frac{1}{6} ( AB - 3C ) - \frac{1}{27} A^{2}.
\label{eq:SU10}
\end{equation}

\noindent Thus, critical line 2 is given by equations (\ref{eq:SU5}) - (\ref{eq:SU7}), (\ref{eq:SU9}) and (\ref{eq:SU10}). This is a fairly messy nonlinear equation in $\frac{d \ln Y_{\ell} }{dz}$ and $\frac{d \ln s}{dz}$ and will not be written down here. It will be solved numerically for a number of illustrative examples below.

Finally, and unlike the case of Schwarzschild stability, there is the possibility that the real part of a complex conjugate pair of roots can change sign. This will happen if there is a complex conjugate pair of roots, and if they sum to zero. To determine the condition for this note \citep{abramowitzs65} that the roots of cubic equation (\ref{eq:SU4}) satisfy 

\be
& {\ds \sigma_{1} + \sigma_{2} + \sigma_{3} = - A,
} & \nonumber \\ 
& {\ds \sigma_{1} \sigma_{2} + \sigma_{1} \sigma_{3} + \sigma_{2} \sigma_{3} = B,
} & \nonumber \\ 
& {\ds \sigma_{1} \sigma_{2} \sigma_{3} = - C.
} &
\label{eq:SU11}
\ee

\noindent Then $0 = AB - C$ gives

\be
& {\ds 0 = \sigma_{1} \sigma_{2} ( \sigma_{1} + \sigma_{2}) 
+  \sigma_{1} \sigma_{3} ( \sigma_{1} + \sigma_{3}) 
} & \nonumber \\ 
& {\ds +  \sigma_{2} \sigma_{3} ( \sigma_{2} + \sigma_{3})
+ 2 \sigma_{1} \sigma_{2} \sigma_{3},
} & \nonumber \\ 
& {\ds = ( \sigma_{1} + \sigma_{2}) ( \sigma_{1} \sigma_{2} + \sigma_{1} \sigma_{3}
+ \sigma_{2} \sigma_{3} + \sigma_{3}^{2} )
} &
\label{eq:SU12}
\ee

\noindent where the subscripts 1, 2, and 3 can be cyclically permuted in the above expressions. Thus equation (\ref{eq:SU12}) is satisfied if any two roots sum to zero. This is the desired condition. Using equations (\ref{eq:SU5}) - (\ref{eq:SU7}) equation $0 = AB - C$ can be written

\begin{equation}
\frac{ d \ln \bar{Y}_{\ell} }{dz} = - \frac{ \Sigma_{s} }{ \Upsilon_{Y_{\ell}}  }
\frac{ \left( \pderiv{ \ln \rho }{ \ln s } \right)_{p,Y_{\ell}} }
{ \left( \pderiv{ \ln \rho }{ \ln Y_{\ell} } \right)_{p,s} }  \frac{ d \ln \bar{s} }{dz}
+ \frac{ \Sigma_{s} ( \Sigma_{s} + \Upsilon_{Y_{\ell}} ) }
{ g \left( \pderiv{ \ln \rho }{ \ln Y_{\ell} } \right)_{p,s} },
\label{eq:SU13}
\end{equation}

\noindent and we will refer to this line as critical line 3.

\begin{figure}[!h]
\setlength{\unitlength}{1.0cm}
{\includegraphics[width=\columnwidth]{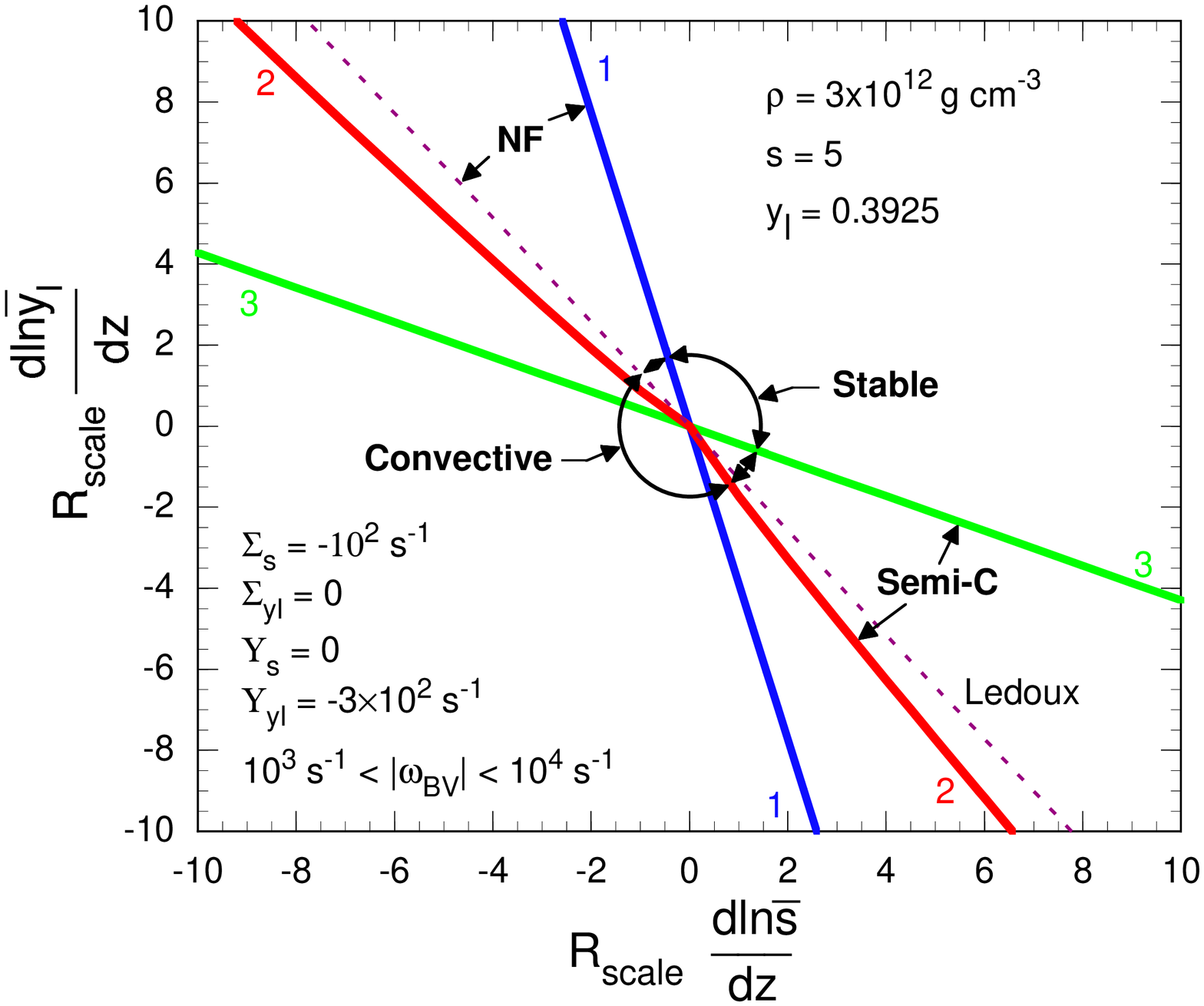}}
\caption{\label{fig:CRF0_2}
Fluid stability as a function of the $\ln s$ and $\ln Y_{\ell}$ gradients for the case in which $|\Upsilon_{Y_{\ell}}| = 3|\Sigma_{s}|$ for the indicated thermodynamic state and values of $\Sigma_{s}$, $\Upsilon_{Y_{\ell}}$, and $\omega_{BV}$. The value of R$_{\rm scale}$ is 30 km, a typical value of the nascent neutron star radius a few hundred milliseconds after core bounce.}
\end{figure}

\begin{figure}[!h]
\setlength{\unitlength}{1.0cm}
{\includegraphics[width=\columnwidth]{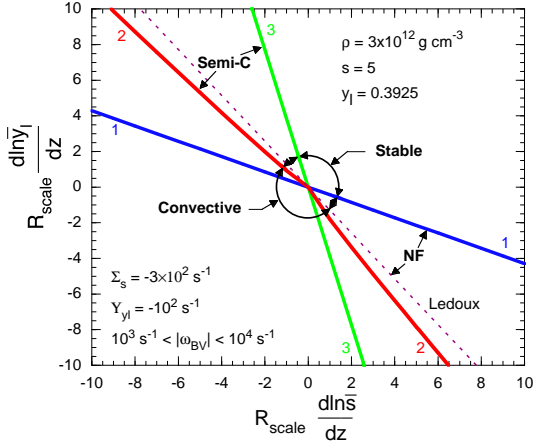}}
\caption{\label{fig:CRF0_2rv}
Fluid stability as a function of the $\ln s$ and $\ln Y_{\ell}$ gradients for the case in which $|\Upsilon_{Y_{\ell}}| = 3|\Sigma_{s}|$ for the indicated thermodynamic state and values of $\Sigma_{s}$, $\Upsilon_{Y_{\ell}}$, and $\omega_{BV}$. The value of R$_{\rm scale}$ is 30 km, a typical value of the nascent neutron star radius a few hundred milliseconds after core bounce.}
\end{figure}


Figures \ref{fig:CRF0_2} and \ref{fig:CRF0_2rv} show these three critical lines along with the Ledoux line in the $\frac{d \ln Y_{\ell} }{dz} -  \frac{d \ln s}{dz}$ plane for the indicated thermodynamic state and values of $\Sigma_{s}$, $\Upsilon_{Y_{\ell}}$, and $|\omega_{BV}|$. The absolute value of the Brunt-V\"{a}is\"{a}l\"{a} frequency, $|\omega_{BV}|$, where $\omega_{BV}$ is given by equation (\ref{eq:L7a}), is an indicator of the buoyancy growth rate in the absence of neutrino transport. Its value, however, depends on the gradients of $\ln s$ and $\ln Y_{\ell}$ and therefore varies over the $\frac{d \ln Y_{\ell} }{dz} -  \frac{d \ln s}{dz}$ plane. A range of typical values for $\omega_{BV}$ is given in the figures. A comparison of these values with  the absolute values, $|\Sigma_{s}|$ and $|\Upsilon_{Y_{\ell}}|$, of the direct response functions will indicate the relative importance of buoyancy versus neutrino transport on the dynamics of the fluid element.

We have divided the $\frac{d \ln Y_{\ell} }{dz} -  \frac{d \ln s}{dz}$ plane into four distinct regions delimiting different types of fluid stability/instability, the boundary between any two adjacent regions being one of the critical lines. In the region denoted ``stable'' all roots of equation (\ref{eq:SU4}) are negative, if real, or have a negative real part if a complex conjugate pair. A perturbed fluid element in this region will return to its unperturbed state, and the fluid is therefore stable. In the region marked ``Semi-C'' the largest growth rate arises from the positive real part of a complex conjugate pair of roots, and the dominate unstable mode is therefore oscillatory and growing. In this region the fluid is unstable to semiconvection. In the region marked ``NF'' the largest growth rate arises from a positive real root, but the region resides on the stable side of the Ledoux critical line and would therefore be stable in the absence of thermal and lepton transport. The instability in this region is therefore driven by thermal and lepton transport. We refer to this instability for the moment as ``neutron fingers,'' although we will show below that for the conditions in the collapsed stellar cores of supernova progenitors that we have analyzed this instability is not neutron fingers per se but something quite different. Finally, in the region marked ``convection'' the largest growth rate arises from a positive real root, as in the NF case, but the region resides on  the unstable side of the Ledoux critical line and would therefore be unstable even in the absence of thermal and lepton diffusion. We summarize our stability/instability classification in Table \ref{tab:SC}.

\begin{table*}[t]
\begin{center}
\caption{ \label{tab:SC}{\bfseries{Stability Classification}}} 
\begin{tabular}{|c|c|c|} \hline
{\bf Classification} & {\bf Roots } & {\bf Subsidiary Conditions} \\ \hline
 Stable & $r_{1} < 0$ &  \\ 
Semi-C & $r_{1} > 0, s_{1} \ne 0$ &   \\ 
NF & $r_{1} > 0, s_{1} = 0$ & Ledoux stable in the absence of transport  \\ 
Convection & $r_{1} > 0$ & Ledoux unstable in the absence of transport  \\ \hline
 \end{tabular}
\end{center}
$r_{1}$ is the largest real root, or the real part of a complex conjugate pair. $s_{1}$ is the imaginary part of the root of which $r_{1}$ is the real part.\\
\end{table*}

To characterize in more detail the stability/instability of the fluid in the $\frac{d \ln Y_{\ell} }{dz} -  \frac{d \ln s}{dz}$ plane for the conditions shown in Figure \ref{fig:CRF0_2}, and to delineate the role of the critical stability lines, we will start in the upper right quadrant of that figure where the gradients of $\ln s$ and $\ln Y_{\ell}$ are both positive. The condition that the fluid is stable, i.e., that all  the roots of equation (\ref{eq:SU4}) are negative, if real, or that the real root is negative and the real part of a complex conjugate pair is also negative, is that all of the following inequalities be fulfilled \citep{aleksandrovkl63}: $A > 0$, $C > 0$, and $AB > C$, where $A$, $B$, and $C$ are given by equations (\ref{eq:SU5}) - (\ref{eq:SU7}). $A$, given by equation (\ref{eq:SU5}), is clearly positive, as both $\Sigma_{s}$ and $\Upsilon_{Y_{\ell}}$ are negative. For the given thermodynamic conditions the logarithmic derivatives $\left( \pderiv{ \ln \rho }{ \ln Y_{\ell} } \right)_{p,s}$ and $\left( \pderiv{ \ln \rho }{ \ln s } \right)_{p,Y_{\ell}}$ are both negative, therefore $C$, given by equation (\ref{eq:SU7}), is also positive. Finally,

\be
& {\ds AB - C =
} & \nonumber \\ 
& {\ds = - \Sigma_{s}^{2} \Upsilon_{Y_{\ell}} - \Sigma_{s} \Upsilon_{Y_{\ell}}^{2}
+ \Sigma_{s} g \left( \pderiv{ \ln \rho }{\ln s } \right)_{p, Y_{\ell}} \frac{d \ln \bar{s} }{dz}
} & \nonumber \\ 
& {\ds + \Upsilon_{Y_{\ell}} g \left( \pderiv{ \ln \rho }{\ln Y_{\ell} } \right)_{p, s} \frac{d \ln \bar{Y}_{\ell} }{dz}
> 0
} &
\label{eq:SU14}
\ee

\noindent since each term to the left of the inequality sign of equation (\ref{eq:SU14}) is positive. Thus all three of the above inequalities are satisfied in the upper right quadrant of Figure \ref{fig:CRF0_2} where the gradients of $\ln s$ and $\ln Y_{\ell}$ are both positive, and the fluid is therefore stable in this region. For the particular case shown in Figure \ref{fig:CRF0_2}, in the upper right-hand quadrant there is a negative real root and a complex conjugate pair with a negative real part.

To characterize the stability/instability of the fluid in the remaining parts of the $\frac{d \ln Y_{\ell} }{dz} -  \frac{d \ln s}{dz}$ plane, let us move clockwise around the figure from the  upper right-hand quadrant. The character of the roots (and therefore the stability of the fluid) does not change until critical line 3 is crossed, at which point the real part of the complex conjugate pair of roots changes sign from negative to positive. The fluid, if perturbed, will now oscillate with growing amplitude, and according to our taxonomy is therefore unstable to semiconvection. Continuing clockwise, critical line 2 is crossed and the complex conjugate pair of roots with positive real part become two real positive roots. The fluid, if perturbed, will move away from equilibrium with exponentially increasing amplitude. The fluid is therefore unstable in this region, and since it is would be Ledoux unstable in the absence of transport, it is therefore convective. Crossing line 1 at the bottom of the figure, one of the positive real roots changes sign, so there are now one positive real root and two negative real roots. Because of the positive real root the fluid is again convective.  Continuing clockwise to the left of the figure and up, critical line 3 is crossed and the real part of a complex conjugate pair of roots, if present, would change sign. There not being a complex conjugate pair of roots in this region, there is no change in the character of the roots and the fluid remains convective. Continuing clockwise and crossing critical line 2 at the upper left-hand side of the figure, two negative real roots change to a complex conjugate pair with a negative real part, leaving one real positive root. The fluid is still convective. On crossing the Ledoux line the character of the roots remains unchanged but the fluid would be stable in the absence of thermal and lepton transport. The fluid is now unstable to neutron fingers. Finally, crossing critical line 1 from left to right at the top of the figure the positive real root changes sign and we are left with a negative real root and a complex conjugate pair with a negative real part. We are back in the region where the fluid is stable.

The above taxonomical sectoring of the $\frac{d \ln Y_{\ell} }{dz} -  \frac{d \ln s}{dz}$ plane in Figure \ref{fig:CRF0_2} into regions of stability, convection, semiconvection, and neutrino fingers is a general characterization of a gravitating fluid with thermal and lepton transport represented by negative values for $\Sigma_{s}$ and $\Upsilon_{Y_{\ell}}$. The location of these regions will depend, however, on the magnitudes of $\Sigma_{s}$ and $\Upsilon_{Y_{\ell}}$, and on the thermodynamic state of the fluid and the force of gravity. Figure \ref{fig:CRF0_2} has illustrated a case in which $\Upsilon_{Y_{\ell}} = 3\Sigma_{s}$ (lepton equilibration more rapid than thermal equilibration). Interchanging the values of $\Sigma_{s}$ and $\Upsilon_{Y_{\ell}}$, so that $\Upsilon_{Y_{\ell}} = \frac{1}{3} \Sigma_{s}$ (thermal equilibration more rapid than lepton equilibration), and keeping the same thermodynamic state and gravity gives regions of stability and instability shown in Figure \ref{fig:CRF0_2rv}. Note, as  is apparent from equations (\ref{eq:SU8}) and (\ref{eq:SU13}), that interchanging the values of $\Sigma_{s}$ and $\Upsilon_{Y_{\ell}}$ interchanges the slopes of critical lines 1 and 3 (and also changes the intercept of critical line 3). Also, the location of the NF and Semi-C regions are interchanged. Referring back to the discussion in Section \ref{sec:DDInstabilities}, Figures \ref{fig:CRF0_2} and  \ref{fig:CRF0_2rv} show, respectively, that a destabilizing (negative) gradient in $Y_{\ell}$ stabilized by a stable (positive) gradient in $s$ leads to semiconvection if $\Upsilon_{Y_{\ell}} > \Sigma_{s}$ (lepton transport more rapid than thermal transport) and neutron fingers if $\Sigma_{s} > \Upsilon_{Y_{\ell}}$ (thermal transport more rapid than lepton transport). The opposite is true if the destabilizing (negative) gradient is $s$ and the stabilizing (positive) gradient is $Y_{\ell}$ (tops of Figures (\ref{fig:CRF0_2}) and  (\ref{fig:CRF0_2rv}). 

\begin{figure}[!h]
\setlength{\unitlength}{1.0cm}
{\includegraphics[width=\columnwidth]{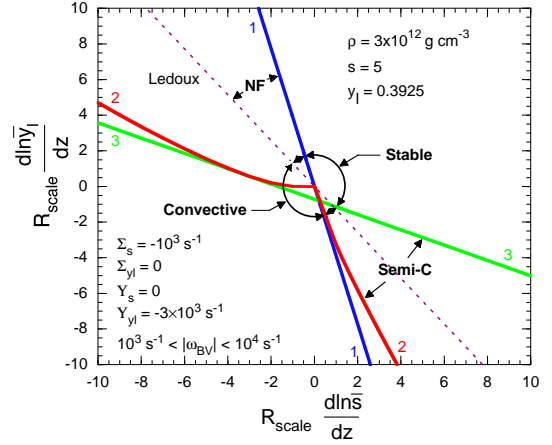}}
\caption{\label{fig:CRF0_3}
Fluid stability as a function of the $\ln s$ and $\ln Y_{\ell}$ gradients for the case in which $|\Upsilon_{Y_{\ell}}| = 3|\Sigma_{s}|$, but larger in magnitude than for the  case shown in Figure \ref{fig:CRF0_2}, for the same thermodynamic state and values of $\Sigma_{s}$, $\Upsilon_{Y_{\ell}}$, 
and $\omega_{BV}$.}
\end{figure}

\begin{figure}[!h]
\setlength{\unitlength}{1.0cm}
{\includegraphics[width=\columnwidth]{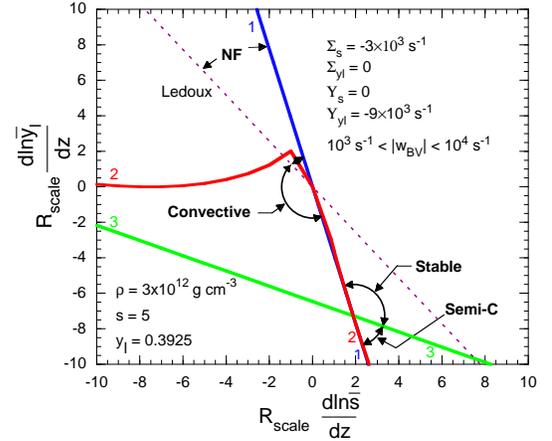}}
\caption{\label{fig:CRF0_4}
Fluid stability as a function of the $\ln s$ and $\ln Y_{\ell}$ gradients for the case in which $|\Upsilon_{Y_{\ell}}| = 3|\Sigma_{s}|$, but still larger in magnitude than for the  case shown in Figure \ref{fig:CRF0_2}, for the same thermodynamic state and values of $\Sigma_{s}$, $\Upsilon_{Y_{\ell}}$, and $\omega_{BV}$.}
\end{figure}

Figures \ref{fig:CRF0_3} and  \ref{fig:CRF0_4} show the effect of successively increasing the values of $\Sigma_{s}$ and $\Upsilon_{Y_{\ell}}$, but keeping their ratio the same, and keeping the thermodynamic state and gravitational acceleration the same (all equal to the case shown in Figure \ref{fig:CRF0_2}). It is apparent from equations (\ref{eq:SU8}) and (\ref{eq:SU13}) that the slopes of critical lines 1 and 3 are unaffected by the magnitudes of  $\Sigma_{s}$ and $\Upsilon_{Y_{\ell}}$ provided that their ratio remains the same. However, the intercept of critical line 3 with the vertical axis becomes more negative (recall that  $\left( \pderiv{ \ln \rho }{ \ln Y_{\ell} } \right)_{p,s}$ is negative for the case being considered). The latter effect is to extend the stable region with increasing magnitudes of $\Sigma_{s}$ and $\Upsilon_{Y_{\ell}}$ into the region that was formerly semiconvective.

An effect not evident from the figures is the dependence of the instability growth rates on the magnitudes of $\Sigma_{s}$ and $\Upsilon_{Y_{\ell}}$ for given values of the gradients $d \ln \bar{Y}_{\ell}/dz$ and  $d \ln \bar{s}/dz$. Qualitativelly, the growth rates for instabilities that are driven by thermal or lepton transport (semiconvection or neutron fingers) increase with the magnitudes of $\Sigma_{s}$ and $\Upsilon_{Y_{\ell}}$, while those that are buoyancy driven (convection) decrease with the magnitudes of $\Sigma_{s}$ and $\Upsilon_{Y_{\ell}}$. The latter effect arises because thermal and lepton transport tend to equilibrate the perturbed fluid element with its surroundings, thus reducing the buoyancy forces driving the instability. More quantitative results for instability growth rates will be presented for the surveys of proto-supernovae in Section \ref{sec:Results}.

\begin{figure}[!h]
\setlength{\unitlength}{1.0cm}
{\includegraphics[width=\columnwidth]{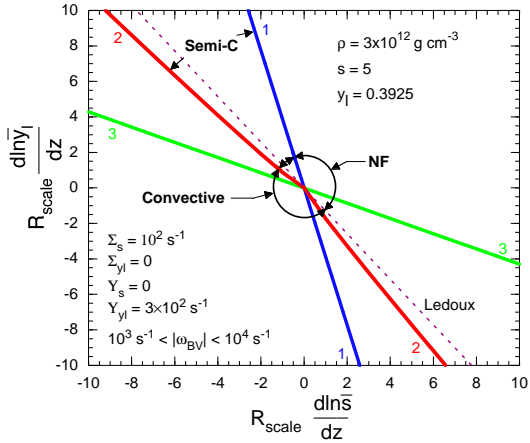}}
\caption{\label{fig:CRF0_m2}
Fluid stability as a function of the $\ln s$ and $\ln Y_{\ell}$ gradients for the case in which $\Sigma_{s}$ and $\Upsilon_{Y_{\ell}}$ have the same magnitudes as in the case shown in Figure \ref{fig:CRF0_2} but opposite (positive) signs. The thermodynamic state and values of $\omega_{BV}$ are the same as in the case shown in Figure \ref{fig:CRF0_2}.}
\end{figure}


Finally, we show in Figure (\ref{fig:CRF0_m2}) the effect on fluid stability of positive values of  $\Sigma_{s}$ and $\Upsilon_{Y_{\ell}}$. In this figure, the thermodynamic state and gravitational acceleration are the same as that shown in Figure (\ref{fig:CRF0_2}), but both the values of $\Sigma_{s}$ and $\Upsilon_{Y_{\ell}}$ have the same magnitudes but positive signs. This causes the fluid to be destabilized for all values of $d \ln Y_{\ell}/dz$ and $d \ln s/dz$, and is analogous to the case discussed at the end of Section \ref{sec:Schwarzschild}. While we have not encountered positive values of $\Sigma_{s}$ and $\Upsilon_{Y_{\ell}}$ in the proto-supernovae we have surveyed, we do encounter positive values of the cross response functions $\Sigma_{Y_{\ell}}$ and $\Upsilon_{s}$, as will be discussed in the next Section.

\subsection{General Case - Lepto-Entropy Fingers and Lepto-Entropy Semiconvection}
\label{sec:GC}

In the general case all of the response functions, $\Sigma_{s}$, $\Sigma_{Y_{\ell}}$, $\Upsilon_{s}$, and $\Upsilon_{Y_{\ell}}$, given by equations (\ref{eq:eq4}) are nonzero, and the equations describing the motion of a perturbed fluid element are equations (\ref{eq:eq7}) - (\ref{eq:eq9}). Solutions for the motion of a perturbed fluid element again have the form given by equations (\ref{eq:st1}), and exist for values of $\sigma$ satisfying

\begin{equation}
\sigma^{3} + A \sigma^{2} + B \sigma^{1} + C = 0,
\label{eq:GC1}
\end{equation}

\noindent where $A$, $B$, and $C$ are now given by

\begin{equation}
A = - \left( \Sigma_{s} + \Upsilon_{Y_{\ell}} \right),
\label{eq:GC2}
\end{equation}

\be
& {\ds B = \Sigma_{s} \Upsilon_{Y_{\ell}} - \Sigma_{Y_{\ell}} \Upsilon_{s}
} & \nonumber \\ 
& {\ds - g \left( \pderiv{ \ln \rho }{ \ln s } \right)_{p,Y_{\ell}} \frac{ d \ln \bar{s} }{dz}
- g \left( \pderiv{ \ln \rho }{ \ln Y_{\ell} } \right)_{p,s} \frac{ d \ln \bar{Y}_{\ell} }{dz},
} &
\label{eq:GC3}
\ee

\be
& {\ds C = \Sigma_{s} g \left( \pderiv{ \ln \rho }{ \ln Y_{\ell} } \right)_{p,s}
\frac{ d \ln \bar{Y}_{\ell} }{dz}
} & \nonumber \\ 
& {\ds + \Upsilon_{Y_{\ell}} g \left( \pderiv{ \ln \rho }{ \ln s } \right)_{p,Y_{\ell}}
 \frac{ d \ln \bar{s} }{dz}
} & \nonumber \\ 
& {\ds - \Sigma_{Y_{\ell}} g \frac{ Y_{\ell} }{s} \left( \pderiv{ \ln \rho }{ \ln s } \right)_{p,Y_{\ell}}
 \frac{ d \ln \bar{Y}_{\ell} }{dz}
} & \nonumber \\ 
& {\ds -  \Upsilon_{s} g \frac{s}{ Y_{\ell} } \left( \pderiv{ \ln \rho }{ \ln Y_{\ell} } \right)_{p,s}
\frac{ d \ln \bar{s} }{dz}.
} &
\label{eq:GC4}
\ee

\noindent Critical lines 1 (across which a root changes sign) and 3 (across which the real part of a complex conjugate pair changes sign) are now given, respectively, by

\be
& {\ds \frac{ d \ln \bar{Y}_{\ell} }{dz} =
} & \nonumber \\ 
& {\ds = - \frac{ \frac{ \bar{s} }{ \bar{Y}_{\ell} } \Upsilon_{s} \left( \pderiv{ \ln \rho }{ \ln Y_{\ell} } \right)_{p,s}
- \Upsilon_{Y_{\ell}} \left( \pderiv{ \ln \rho }{ \ln s } \right)_{p,Y_{\ell}} }
{ \frac{ \bar{Y}_{\ell} }{ \bar{s} } \Sigma_{Y_{\ell}} \left( \pderiv{ \ln \rho }{ \ln s } \right)_{p,Y_{\ell}}
- \Sigma_{s} \left( \pderiv{ \ln \rho }{ \ln Y_{\ell} } \right)_{p,s} }  \frac{ d \ln \bar{s} }{dz},
} &
\label{eq:GC5}
\ee

\noindent and

\be
& {\ds \frac{ d \ln \bar{Y}_{\ell} }{dz} = - 
\frac{ \Sigma_{s} \left( \pderiv{ \ln \rho }{ \ln s } \right)_{p,Y_{\ell}}
-  \frac{ \bar{s} }{ \bar{Y}_{\ell} } \Upsilon_{s} \left( \pderiv{ \ln \rho }{ \ln Y_{\ell} } \right)_{p,s} }
{ \frac{ \bar{Y}_{\ell} }{ \bar{s} } \Sigma_{Y_{\ell}} \left( \pderiv{ \ln \rho }{ \ln s } \right)_{p,Y_{\ell}}
+ \Upsilon_{Y_{\ell}} \left( \pderiv{ \ln \rho }{ \ln Y_{\ell} } \right)_{p,s} }
 \frac{ d \ln \bar{s} }{dz},
} & \nonumber \\ 
& {\ds - \frac{1}{g}  \frac{ ( \Upsilon_{s} \Sigma_{Y_{\ell}} - \Upsilon_{Y_{\ell}} \Sigma_{s} )
( \Sigma_{s} + \Upsilon_{Y_{\ell}} ) }
{ \frac{ \bar{Y}_{\ell} }{ \bar{s} } \Sigma_{Y_{\ell}} 
\left( \pderiv{ \ln \rho }{ \ln s } \right)_{p,Y_{\ell}}
+ \Upsilon_{Y_{\ell}} \left( \pderiv{ \ln \rho }{ \ln Y_{\ell} } \right)_{p,s} }
} &
\label{eq:GC6}
\ee

\begin{figure}[!h]
\setlength{\unitlength}{1.0cm}
{\includegraphics[width=\columnwidth]{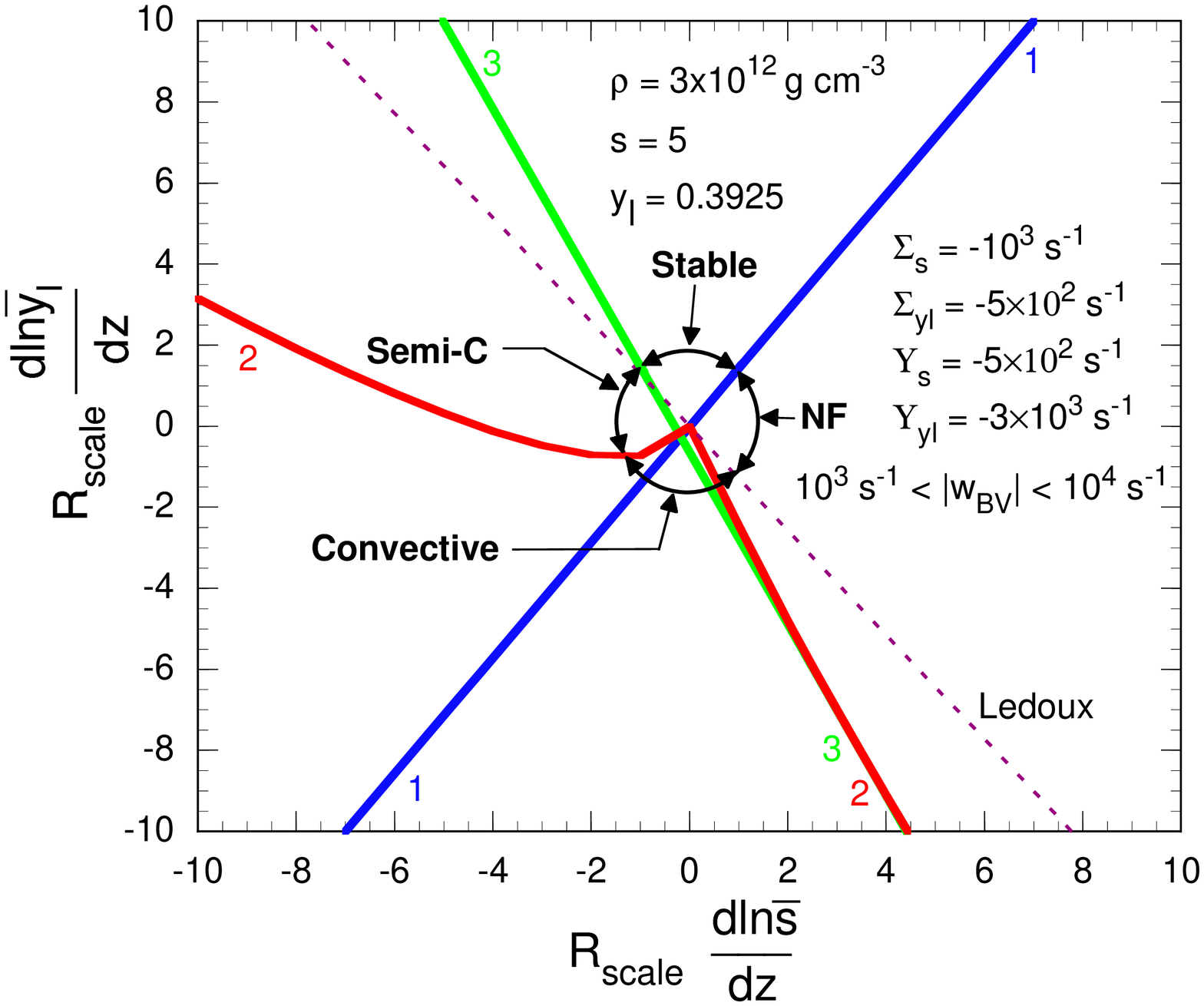}}
\caption{\label{fig:CRF0_3_mh}
Fluid stability as a function of the $\ln s$ and $\ln Y_{\ell}$ gradients for the case shown in Figure \ref{fig:CRF0_3}, but with $\Sigma_{Y_{\ell}} = -5 \times 10^{2}$ s$^{-1}$ and $\Upsilon_{s} = -5 \times 10^{2}$ s$^{-1}$.}
\end{figure}

\begin{figure}[!h]
\setlength{\unitlength}{1.0cm}
{\includegraphics[width=\columnwidth]{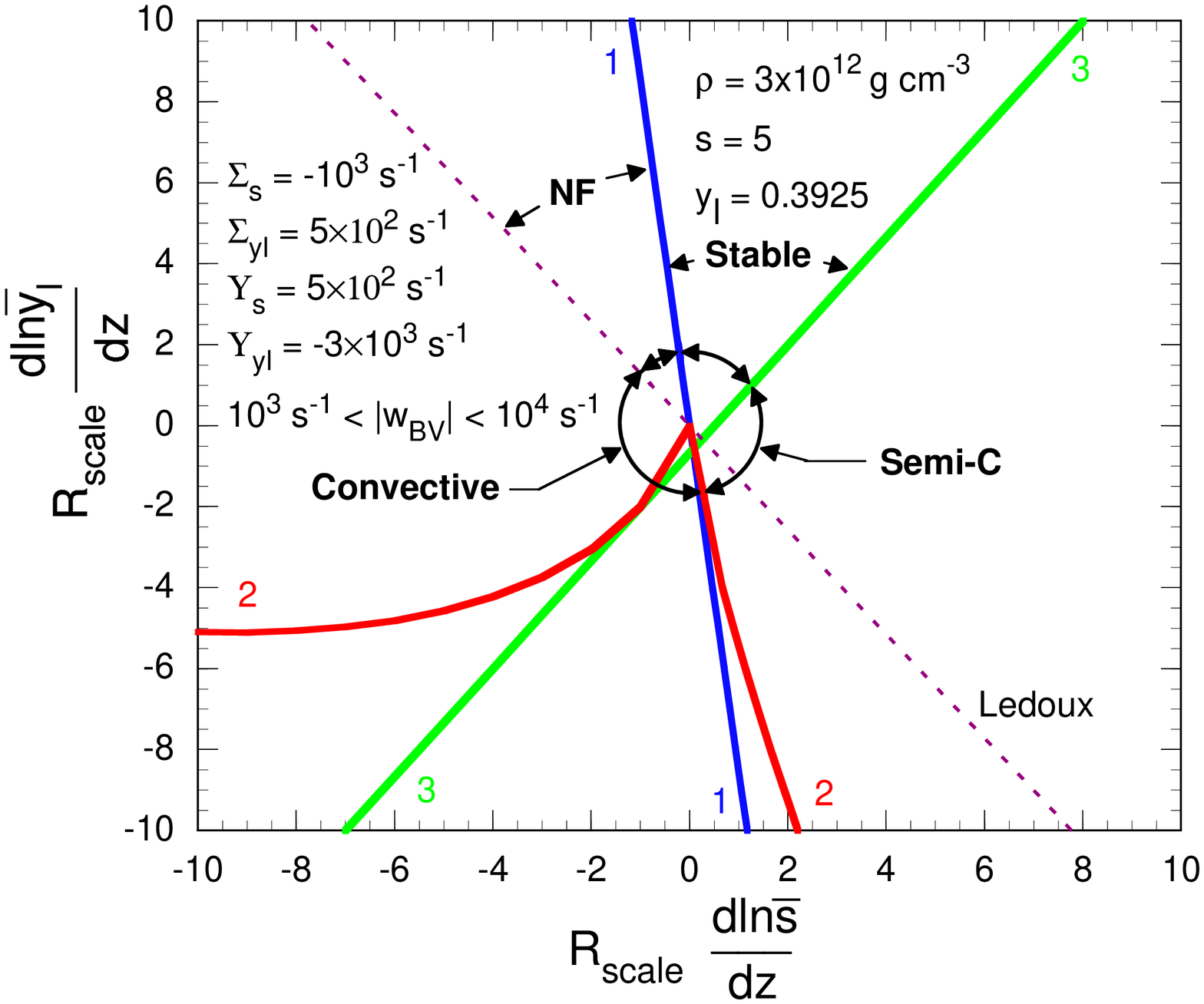}}
\caption{\label{fig:CRF0_3_h}
Fluid stability as a function of the $\ln s$ and $\ln Y_{\ell}$ gradients for the case shown in Figure \ref{fig:CRF0_3}, but with $\Sigma_{Y_{\ell}} = 5 \times 10^{2}$ s$^{-1}$ and $\Upsilon_{s} = 5 \times 10^{2}$ s$^{-1}$.}
\end{figure}

\begin{figure}[!h]
\setlength{\unitlength}{1.0cm}
{\includegraphics[width=\columnwidth]{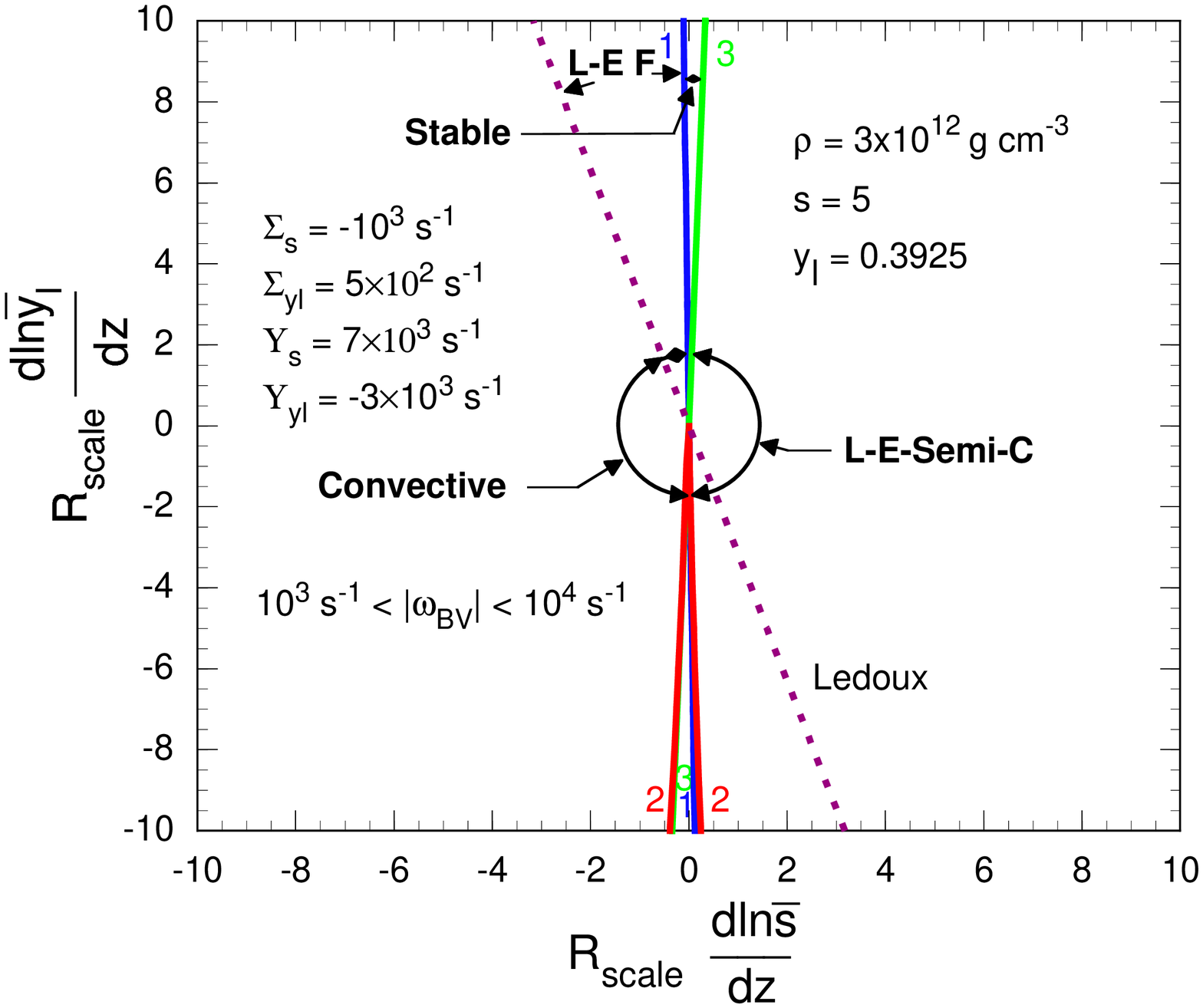}}
\caption{\label{fig:CRF0_3_t}
Fluid stability as a function of the $\ln s$ and $\ln Y_{\ell}$ gradients for the case shown in Figure \ref{fig:CRF0_3}, but with $\Sigma_{Y_{\ell}} = 5 \times 10^{2}$ s$^{-1}$ and $\Upsilon_{s} = 7 \times 10^{3}$ s$^{-1}$.}
\end{figure}

Nonzero values of the cross response functions, $\Sigma_{Y_{\ell}}$ and $\Upsilon_{s}$, if non-negligible, can significantly affect the existence and location of the stable region, and the location and nature of the three unstable regions in the $\frac{d \ln \bar{s}}{dz}$ - $\frac{d \ln \bar{Y}_{\ell}}{dz}$ plane. Figures \ref{fig:CRF0_3_mh} - \ref{fig:CRF0_3_t} show cases in which the thermodynamic state, the value of the gravitational acceleration, and the two direct response functions, $\Sigma_{s}$ and $\Upsilon_{Y_{\ell}}$, are the same as for the case shown in Figure \ref{fig:CRF0_3}, but the cross response functions have  various nonzero values. The relatively small negative values given to $\Sigma_{Y_{\ell}}$ and $\Upsilon_{s}$ in Figure \ref{fig:CRF0_3_mh} cause the regions of stability and instability to be substantially shifted relative to that shown in Figure  \ref{fig:CRF0_3}. Changing the sign of the cross response functions from negative to positive causes further shifts in the regions of stability and instability, as evident from a comparison of Figures \ref{fig:CRF0_3_h} and \ref{fig:CRF0_3_mh}. 

\begin{figure}[!h]
\setlength{\unitlength}{1.0cm}
{\includegraphics[width=\columnwidth]{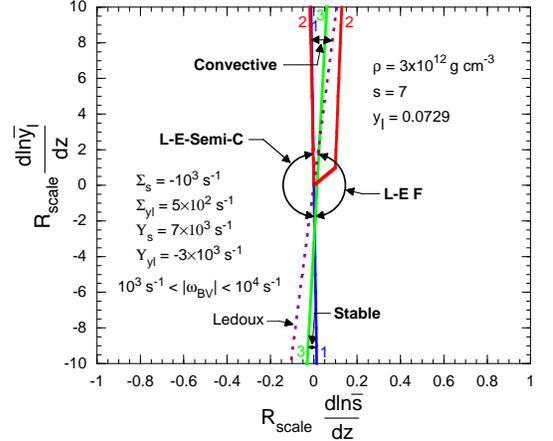}}
\caption{\label{fig:CRF0_3_yt}
Fluid stability as a function of the $\ln s$ and $\ln Y_{\ell}$ gradients for the values of the gravitational acceleration and the response functions shown in Figure \ref{fig:CRF0_3_t}, but for a thermodynamic state such that $\left( \pderiv{ \ln \rho }{ \ln Y_{\ell} } \right)_{p,s} > 0$.}
\end{figure}

Increasing the value of $\Upsilon_{s}$ by an order of magnitude, which is more consistent with the numerical results for the response functions, leads to some interesting results. The stable region collapses to a very small sector about vertical $d \ln s/dz$ axis, as shown by Figure  \ref{fig:CRF0_3_t}, and several new modes of instability appear. Figure \ref{fig:CRF0_3_yt} shows a case for which the values of the response functions and the gravitational acceleration are the same as shown in Figure \ref{fig:CRF0_3_t}, but the thermodynamic state causes the derivative $\left( \pderiv{ \ln \rho }{ \ln Y_{\ell} } \right)_{p,s}$ to be positive rather than negative. (This can arise for small values of $Y_{\ell}$, as discussed in Section \ref{sec:Ledoux}.) It is seen that in this case the region of stability is reflected about the origin, and that again new modes of instability appear. As will be described below, these instabilities arise because of the substantial magnitudes of the cross response functions together the large ratio of $\Upsilon_{s}$ to $\Sigma_{Y_{\ell}}$. These cross response functions cause a $\theta_{Y_{\ell}}$ (a difference in $Y_{\ell}$ between the fluid element and the background) to develop primarily in response to a $\theta_{s}$ (a difference in $s$ between the fluid element and the background). This is in contradistinction to the Ledoux case in which a $\theta_{Y_{\ell}}$ develops from a displacement of the fluid element through a gradient in the background $Y_{\ell}$. 

\subsubsection{Lepto-Entropy Fingers (L-E-F)}
\label{sec:LEF}

We  will discuss the instabilities that that are exhibited in Figure \ref{fig:CRF0_3_yt} here and in the next Section, and return to the instabilities exhibited in Figure \ref{fig:CRF0_3_t} in Section \ref{sec:LEFS}. The instability on the right-hand side of Figure \ref{fig:CRF0_3_yt} denoted by L-E-F is new in that it has not been described or analyzed in the literature up to now. Moreover, it is likely to play the dominate role in producing convective-like fluid motions below the neutrinosphere of a proto-supernova. We begin by noting that the relative values of the response functions chosen for Figure \ref{fig:CRF0_3_yt} as well as the fact that $\left( \pderiv{ \ln \rho }{ \ln Y_{\ell} } \right)_{p,s} > 0$ are representative of these parameters in the region below the neutrinosphere and above the cold, unshocked inner core of a supernova progenitor. The relative values of the response functions seem to be generic, and will be discussed in more detail below. The positive value of  $\left( \pderiv{ \ln \rho }{ \ln Y_{\ell} } \right)_{p,s}$ is due to low values of $Y_{\ell}$ which come about because the material in this region, unlike that in the cold inner core, encountered the shock and did so while still outside the neutrinosphere. This material therefore suffered extensive subsequent deleptonization when passing through the neutrinosphere by electron captures on the shock-dissociated free protons. The low values of the lepton fraction ($Y_{ell} \simless 0.08$) and the moderately high values of the entropy ($s \sim 6$) characteristic of this shocked material is the reason that $\left( \pderiv{ \ln \rho }{ \ln Y_{\ell} } \right)_{p,s} > 0$, in accordance with Figure (\ref{fig:dpdyl_sp}) and the discussion at the end of Section \ref{sec:Ledoux}. Finally, the material in this region tends to have a positive entropy gradient, a relic of the shock ramping up through this region.

A positive gradient in $\bar{s}$, positive cross response functions, and a positive derivative $\left( \pderiv{ \ln \rho }{ \ln Y_{\ell} } \right)_{p,s}$ can, under circumstances to be described, subject the region to an instability that we will dub ``lepto-entropy fingers'' (L-E F). This instability is thermal and lepton transport driven, as in the case of neutron fingers, but is very different from the latter. Recall that the neutron finger mechanism,  to operate in the presence of a positive (stabilizing) entropy gradient and a destabilizing lepton fraction gradient, requires that thermal transport be much more rapid than lepton transport. The argument given for the presence of the neutron finger instability in proto-supernovae by the Livermore group, in fact, is based on the premise that thermal transport is rapid and lepton transport is negligibly slow. However, our results, which will be presented in detail in Section \ref{sec:Results} and which span an extensive grid of thermodynamic states and fluid element radii show that typically $|\Upsilon_{Y_{\ell}}| \simeq 10 \times |\Sigma_{s}|$. From the definitions of $\Sigma_{s}$ and $\Upsilon_{Y_{\ell}}$, given by equations (\ref{eq:eq4}), this indicates that lepton transport is {\sl considerably more rapid} than thermal transport, a fact noted by \citet{bruennmd95} and \citet{bruennd96}. The reason for this was pointed out in \citet{bruennd96} and is that the energy, $\epsilon_{\nu}$, transported by a neutrino is limited by the rapid rise in neutrino opacity with $\epsilon_{\nu}$, whereas a \nue\ or a \nuebar\ transports the same lepton number (viz., $y_{\ell_{\nu}} = 1$ and $y_{\ell_{\nu}} = -1$ for a \nue\ and a \nuebar\, respectively) independently of its energy. Furthermore, the \nue\ and \nuebar\ flows between a perturbed fluid element and the background can be oppositely directed, that is, additive in lepton number and subtractive in energy, depending on the temperature and the \nue\ and \nuebar\ chemical potential of the fluid element relative to the background.

Of critical importance in understanding the instability exhibited on the right-hand side of Figure \ref{fig:CRF0_3_yt}, the lepto-entropy finger instability, is the fact that both $\left( \pderiv{ \ln \rho }{ \ln Y_{\ell} } \right)_{p,s} > 0$ and the cross response function $\Upsilon_{s}$ are large and positive in magnitude, typically as large or larger than $|\Upsilon_{Y_{\ell}}|$, and that the following inequalities are satisfied

\begin{equation}
  |\Upsilon_{i}| >>\omega_{j} >> |\Sigma_{k}|
\label{eq:GC7a}
\end{equation}

\noindent where $i$, $j$, and $k$ can each denote either $s$ or $Y_{\ell}$. Thus, to operate in the presence of a positive (stabilizing) entropy gradient, lepto-entropy fingers requires lepton transport to be much more rapid than thermal transport, the opposite of the neutron finger requirement, and additionally requires that $\left( \pderiv{ \ln \rho }{ \ln Y_{\ell} } \right)_{p,s} > 0$ and that $\Upsilon_{s}$ be large and positive.

The nature of the lepto-entropy finger instability can be illustrated by using inequalities (\ref{eq:GC7a}) to simplify equations (\ref{eq:eq7}) - (\ref{eq:eq9}) to give

\begin{equation}
 \dot{\theta}_{s} = - \frac{d\bar{s}}{dz} v
\label{eq:GC7b}
\end{equation}

\begin{equation}
  0 = \Upsilon_{s} \theta_{s} + \Upsilon_{Y_{\ell}} \theta_{Y_{\ell}}
\label{eq:GC7c}
\end{equation}

\noindent and

\begin{equation}
  \dot{v} = - \frac{g}{\rho} \left( \pderiv{\rho}{s} \right)_{p,Y_{\ell}} \theta_{s}
- \frac{g}{\rho} \left( \pderiv{\rho}{Y_{\ell}} \right)_{p,s} \theta_{Y_{\ell}}.
\label{eq:GC7d}
\end{equation}

\noindent We have dropped the two small transport terms in equation (\ref{eq:eq7}) to get equation (\ref{eq:GC7b}), and we have dropped $\dot{\theta}_{Y_{\ell}}$ and the last term on the right-hand side of equation (\ref{eq:eq8}) in comparison with the large and oppositely signed transport terms to get equation (\ref{eq:GC7c}). Solving equation (\ref{eq:GC7c}) for $\theta_{Y_{\ell}}$ gives

\begin{equation}
   \theta_{Y_{\ell}}(t) = \left| \frac{ \Upsilon_{s} }{ \Upsilon_{Y_{\ell}} } \right| \theta_{s}
\label{eq:GC7}
\end{equation}

\noindent where we have written equation (\ref{eq:GC7}) explicitly for the case in which $\Upsilon_{s} > 0$ and $\Upsilon_{Y_{\ell}} < 0$. Equation (\ref{eq:GC7}) together with equation (\ref{eq:GC7b}) says essentially that the change of $\theta_{s}$ given by equation (\ref{eq:GC7b}) is quasistatic compared with the rate at which the large transport terms $\Upsilon_{s} \theta_{s}$ and $\Upsilon_{Y_{\ell}} \theta_{Y_{\ell}}$ can adjust $\theta_{Y_{\ell}}$, and that $\theta_{Y_{\ell}}$ will therefore adjust to the instantaneous value of $\theta_{s}$ as dictated by equation (\ref{eq:GC7}).

Equation (\ref{eq:GC7}) is critical to understanding the lepto-entropy finger instability, and the lepto-entropy semiconvective instability described in the next  Section. While $\theta_{s}$ arises in response to a displacement of the fluid element through a gradient in the background entropy, as dictated by equation (\ref{eq:GC7b}) and by the identical equation (\ref{eq:L1}) for the familiar Ledoux case, $\theta_{Y_{\ell}}$ arises in response to $\theta_{s}$, as dictated by equation (\ref{eq:GC7}), rather than by a displacement of the fluid element through a gradient in the background lepton fraction, as dictated by equation (\ref{eq:L2}) for the Ledoux case.

Plugging equation (\ref{eq:GC7}) into equation (\ref{eq:GC7d}) gives

\begin{equation}
  \dot{v} = - g \left[ \left( \pderiv{\ln \rho}{s} \right)_{p,Y_{\ell}}
+ \left( \pderiv{\ln \rho}{Y_{\ell}} \right)_{p,s}
\left| \frac{ \Upsilon_{s} }{ \Upsilon_{Y_{\ell}} } \right| \right] \theta_{s}.
\label{eq:GC9}
\end{equation}

\noindent The derivative $\left( \pderiv{\ln \rho}{s} \right)_{p,Y_{\ell}}$ in equation (\ref{eq:GC9}) is always $< 0$, but if the derivative $\left( \pderiv{\ln \rho}{Y_{\ell}} \right)_{p,s} > 0$, which occurs in the outer region of the core where $Y_{\ell}$ is low as discussed above (e.g., Figure \ref{fig:dpdyl_sp} and Figures \ref{fig:w_ls_15_100} - \ref{fig:w_ls_25_200}), and if

\begin{equation}
 \left[ \left( \pderiv{\ln \rho}{s} \right)_{p,Y_{\ell}}
+ \left( \pderiv{\ln \rho}{Y_{\ell}} \right)_{p,s}
\left| \frac{ \Upsilon_{s} }{ \Upsilon_{Y_{\ell}} } \right| \right] > 0
\label{eq:GC10}
\end{equation}

\noindent then the fluid is unstable. For suppose, as an example, that $d\bar{s}/dz > 0$. In the absence of transport this is a stabilizing gradient, for an outward displacement of a fluid element at constant entropy will result in $\theta_{s} < 0$ which will make it more dense than the background and drive it back. Now, however, equation (\ref{eq:GC9}) with equation (\ref{eq:GC10}) satisfied asserts that $\theta_{s} < 0$ will lead to an outwards acceleration of the fluid element, i.e., the fluid is unstable. A more mathematical statement of the above can be obtained by using equation (\ref{eq:GC7b}) for $\theta_{s}$ in the derivative of equation (\ref{eq:GC9}) to get

\begin{equation}
  \ddot{v} = - g \left[ \left( \pderiv{\ln \rho}{s} \right)_{p,Y_{\ell}}
+ \left( \pderiv{\ln \rho}{Y_{\ell}} \right)_{p,s}
\left| \frac{ \Upsilon_{s} }{ \Upsilon_{Y_{\ell}} } \right| \right] \left( - \frac{d\bar{s}}{dz} v \right),
\label{eq:GC10a}
\end{equation}

\noindent which has growing solutions if $\frac{d\bar{s}}{dz} > 0$ and if equation (\ref{eq:GC10}) is satisfied.

How does this curious instability come about? A displacement of the fluid element through an entropy gradient results in a difference, $\theta_{s}$, between the entropy of the fluid element and the background. The large value of $\Upsilon_{s}$ means that the appearance of the entropy difference $\theta_{s}$ will quickly {\sl induce} via equation (\ref{eq:GC7}) a lepton fraction difference, $\theta_{Y_{\ell}}$, between the lepton fraction of the fluid element and the background. The $\theta_{Y_{\ell}}$ so induced is dependent only on $\theta_{s}$, has the same sign as  $\theta_{s}$, and is independent of the $Y_{\ell}$ gradient. Because $\left( \pderiv{\ln \rho}{s} \right)_{p,Y_{\ell}}$ and $\left( \pderiv{\ln \rho}{Y_{\ell}} \right)_{p,s}$ are of opposite sign, if the perturbation $\theta_{s}$ results in a restorative force, the induced perturbation $\theta_{Y_{\ell}}$ will result in an anti-restorative force. If the induced perturbation $\theta_{Y_{\ell}}$ is large enough relative to $\theta_{s}$, the anti-restorative force wins out, and the perturbation grows, i.e., a ``lepton finger'' sustained by the entropy difference between the fluid element and the background will penetrate into the background. Thus, a Ledoux stable region can be destabilized by this diffusive lepto-entropy finger instability. Crucial to the emergence of the lepto-entropy finger instability is the large positive value of $\Upsilon_{s}$, and we shall illustrate by an example in Section \ref{sec:Response} how this large positive value arises.

Despite the rather extreme assumptions that went into the derivation of inequality (\ref{eq:GC10}) (viz., $|\Upsilon_{s}| >> \omega_{Y_{\ell}}$, $|\Upsilon_{Y_{\ell}}| >> \omega_{Y_{\ell}}$, $|\Sigma_{s}| << \omega_{s}$, $|\Sigma_{Y_{\ell}}| << \omega_{s}$), inequality (\ref{eq:GC10}) proves to be remarkably robust for predicting the lepto-entropy finger instability in proto-supernovae. In Section \ref{sec:Results} we test this criterion for lepto-entropy fingers for one of the models we analyze by reducing $\Upsilon_{s}$ to violate the inequality in equation (\ref{eq:GC10}) by 5 percent for every core radius and fluid element size for which it was satisfied, and find that the region previously found unstable to lepto-entropy fingers almost completely disappears.

The above analysis of the lepto-entropy finger instability breaks down when $|d\bar{s}/dz| << |d  \bar{Y}_{\ell}/dz|$. An outward displacement will then result in a very small magnitude of $\theta_{s}$ and the last term in equation (\ref{eq:eq8}) involving the relatively large $d  \bar{Y}_{\ell}/dz$ cannot now be neglected. Thus, there is a small sector about the vertical axis where $|d\bar{s}/dz| << |d  \bar{Y}_{\ell}/dz|$ which bounds the lepto-entropy finger region in Figure \ref{fig:CRF0_3_yt} where the fluid is either stable or convective.

\subsubsection{Lepto-Entropy Semiconvection (L-E-Semi-C)}
\label{sec:LES}

Consider now the case for which  $\frac{d\bar{s}}{dz} < 0$ (Ledoux destabilizing) but for which the conditions leading to equation (\ref{eq:GC9}) are again satisfied (viz., $|\Upsilon_{s}| >> \omega_{Y_{\ell}}$, $|\Upsilon_{Y_{\ell}}| >> \omega_{Y_{\ell}}$, $|\Sigma_{s}| << \omega_{s}$, $|\Sigma_{Y_{\ell}}| << \omega_{s}$). Then equation (\ref{eq:GC10a}) as it stands gives stable oscillatory solutions if equation (\ref{eq:GC10}) is satisfied. However, the approximations leading to equation (\ref{eq:GC10a}) are now inadequate, as the addition of even a small term in $\dot{v}$, due for example to the samll but nonzero values of $\Sigma_{s}$ and $\Sigma_{Y_{\ell}}$ in equation (\ref{eq:eq7}), can, depending on its sign, turn the oscillatory solution of equation (\ref{eq:GC10a}) into a damped oscillatory solution (stable) or a growing oscillatory solution (semiconvective). Equation (\ref{eq:GC10a}) was obtained by approximating equations (\ref{eq:eq7}) - (\ref{eq:eq9}) by equations (\ref{eq:GC7b}) - (\ref{eq:GC7d}). To obtain a better approximation, we approximate equations (\ref{eq:eq7}) - (\ref{eq:eq9}) instead by

\begin{equation}
 \dot{\theta}_{s} = \Sigma_{s} \theta_{s} + \Sigma_{Y_{\ell}} \theta_{Y_{\ell}}
- \frac{d\bar{s}}{dz} v
\label{eq:GC11a}
\end{equation}

\begin{equation}
0 = \Upsilon_{s} \theta_{s} + \Upsilon_{Y_{\ell}} \theta_{Y_{\ell}}
\label{eq:GC11b}
\end{equation}

\begin{equation}
  \dot{v} = - \frac{g}{\rho} \left( \pderiv{\rho}{s} \right)_{p,Y_{\ell}} \theta_{s}
- \frac{g}{\rho} \left( \pderiv{\rho}{Y_{\ell}} \right)_{p,s} \theta_{Y_{\ell}}.
\label{eq:GC11c}
\end{equation}

\noindent Here we have assumed, as in our analysis of lepto-entopy fingers above, that equation (\ref{eq:eq8}) is dominated by the transport terms and have dropped both the $\dot{\theta}_{Y_{\ell}}$ and the $\frac{d\bar{Y}_{\ell}}{dz} v$ terms to get equation (\ref{eq:GC11b}). Unlike our analysis of lepto-entopy fingers, however, we have retained the transport terms in equation (\ref{eq:eq7}) to get equation (\ref{eq:GC11a}). Solving equation (\ref{eq:GC11b}) gives as before

\begin{equation}
\theta_{Y_{\ell}} = - \frac{ \Upsilon_{s} }{ \Upsilon_{Y_{\ell}} } \theta_{s}
= \left| \frac{ \Upsilon_{s} }{ \Upsilon_{Y_{\ell}} } \right| \theta_{s}
\label{eq:GC11}
\end{equation}

\noindent where we have written the last expression in equation (\ref{eq:GC11}) to explicitly account for the fact that $\Upsilon_{s} > 0$ and $ \Upsilon_{Y_{\ell}} < 0$. Here as before the large values of the response functions $ \Upsilon_{s}$ and $\Upsilon_{Y_{\ell}}$ in comparison with the terms we have dropped guarantees that the value of $\theta_{Y_{\ell}}$ adjusts immediately to the instantaneous value of $\theta_{s}$. Substituting equation (\ref{eq:GC11}) into equation (\ref{eq:GC11c}) gives equation (\ref{eq:GC9}) again, i,e.,

\be
& {\ds   \dot{v} = - g \left[ \left( \pderiv{\ln \rho}{s} \right)_{p,Y_{\ell}}
+ \left( \pderiv{\ln \rho}{Y_{\ell}} \right)_{p,s}
\left| \frac{ \Upsilon_{s} }{ \Upsilon_{Y_{\ell}} } \right| \right] \theta_{s}.
} &
\label{eq:GC12}
\ee

\noindent Taking the time derivative of equation (\ref{eq:GC12}) and using equation (\ref{eq:GC11a}) for $\dot{\theta}_{s}$, we obtain in place of equation (\ref{eq:GC10a})

\be
& {\ds  \ddot{v} = - g \left[ \left( \pderiv{\ln \rho}{s} \right)_{p,Y_{\ell}}
+ \left( \pderiv{\ln \rho}{Y_{\ell}} \right)_{p,s}
\left| \frac{ \Upsilon_{s} }{ \Upsilon_{Y_{\ell}} } \right| \right] \times
} & \nonumber \\ 
& {\ds \times \left( \Sigma_{s} \theta_{s} + \Sigma_{Y_{\ell}} \theta_{Y_{\ell}} - \frac{d\bar{s}}{dz} v \right)
} & \nonumber \\ 
& {\ds = - g \left[ \left( \pderiv{\ln \rho}{s} \right)_{p,Y_{\ell}}
+ \left( \pderiv{\ln \rho}{Y_{\ell}} \right)_{p,s}
\left| \frac{ \Upsilon_{s} }{ \Upsilon_{Y_{\ell}} } \right| \right] \times
} & \nonumber \\ 
& {\ds \times \left[ \left( \Sigma_{s} + \Sigma_{Y_{\ell}} 
\left| \frac{ \Upsilon_{s} }{ \Upsilon_{Y_{\ell}} }\right|  \right) \theta_{s} 
- \frac{d\bar{s}}{dz} v \right]
} & \nonumber \\ 
& {\ds = \left( \Sigma_{s} + \Sigma_{Y_{\ell}} 
\left| \frac{ \Upsilon_{s} }{ \Upsilon_{Y_{\ell}} }\right|  \right) \dot{v}
} & \nonumber \\ 
& {\ds - g \left[ \left( \pderiv{\ln \rho}{s} \right)_{p,Y_{\ell}}
+ \left( \pderiv{\ln \rho}{Y_{\ell}} \right)_{p,s}
\left| \frac{ \Upsilon_{s} }{ \Upsilon_{Y_{\ell}} } \right| \right] 
\left( - \frac{d\bar{s}}{dz} \right) v
} &
\label{eq:GC13}
\ee

\noindent where we have used the equation (\ref{eq:GC11}) in obtaining the second expression of equation (\ref{eq:GC13}), and equation (\ref{eq:GC12}) in obtaining the last. Equation (\ref{eq:GC13}) is a more accurate version of equation (\ref{eq:GC10a}) and differs from the latter in that some of the terms multiplying $\dot{v}$ have been included. It will be used in place of equation (\ref{eq:GC10a}) when the coefficient of $\dot{v}$ are needed to decide between growing or decaying oscillatory solutions.

To continue our consideration of the case in which $\frac{d\bar{s}}{dz} < 0$, rather than examining equation (\ref{eq:GC13}) it is conceptually easier to examine its first integral

\be
& {\ds  \dot{v} =  \left( \Sigma_{s} + \Sigma_{Y_{\ell}} 
\left| \frac{ \Upsilon_{s} }{ \Upsilon_{Y_{\ell}} }\right|  \right) v
} & \nonumber \\ 
& {\ds - g \left[ \left( \pderiv{\ln \rho}{s} \right)_{p,Y_{\ell}}
+ \left( \pderiv{\ln \rho}{Y_{\ell}} \right)_{p,s}
\left| \frac{ \Upsilon_{s} }{ \Upsilon_{Y_{\ell}} } \right| \right] \times
} & \nonumber \\ 
& {\ds \times \left( - \frac{d\bar{s}}{dz} \right) ( x - x_{0}).
} &
\label{eq:GC14}
\ee

\noindent We note first from equation (\ref{eq:GC14}) that if the coefficient of $x - x_{0}$ is greater than zero (i.e., $\left[ \left( \pderiv{\ln \rho}{s} \right)_{p,Y_{\ell}} + \left( \pderiv{\ln \rho}{Y_{\ell}} \right)_{p,s}
\left| \frac{ \Upsilon_{s} }{ \Upsilon_{Y_{\ell}} } \right| \right] > 0$, which, of course, requires that $\left( \pderiv{\ln \rho}{Y_{\ell}} \right)_{p,s} > 0$ since $\left( \pderiv{\ln \rho}{s} \right)_{p,Y_{\ell}} < 0$), then the solutions are oscillatory. The scenario here is analogous to that giving rise to lepto-entropy fingers discussed above. In fact, the same mechanism that destabilizes a fluid element in the presence of a stabilizing entropy gradient in the case above of lepto-entopy fingers, here tends to stabilize a fluid element in the presence of a destabilizing entropy gradient. Thus, an outward displacement of a fluid element through a negative entropy gradient results in the entropy of  the fluid element becoming greater than that of the background, that is, a $\theta_{s} > 0$ develops. This by itself would lead to a positive buoyancy force that would tend to force the fluid element farther outward. However, a $\theta_{s} > 0$ induces a $\theta_{Y_{\ell}} > 0$, in accordance with equation (\ref{eq:GC11}). If $\theta_{Y_{\ell}}$ is large enough, i.e., if $\left[ \left( \pderiv{\ln \rho}{s} \right)_{p,Y_{\ell}} + \left( \pderiv{\ln \rho}{Y_{\ell}} \right)_{p,s}
\left| \frac{ \Upsilon_{s} }{ \Upsilon_{Y_{\ell}} } \right| \right] > 0$, then the negative buoyancy arising from $\theta_{Y_{\ell}} > 0$ together with a positive $\left( \pderiv{\ln \rho}{Y_{\ell}} \right)_{p,s}$ will overcome the positive buoyancy arising from $\theta_{s} > 0$, and the fluid element will be forced back. The fluid element will therefore oscillate.

Whether the oscillatory motion of the fluid element is growing or decaying depends on the sign of the coefficient of $v$ in equation (\ref{eq:GC14}). If this coefficient is less than zero, i.e., if $\left( \Sigma_{s} + \Sigma_{Y_{\ell}} \left| \frac{ \Upsilon_{s} }{ \Upsilon_{Y_{\ell}} }\right|
\right) < 0$, the solutions of equation (\ref{eq:GC14}) are decaying oscillatory and the fluid is stable. If $\left( \Sigma_{s} + \Sigma_{Y_{\ell}} \left| \frac{ \Upsilon_{s} }{ \Upsilon_{Y_{\ell}} }\right|
\right) > 0$, on the other hand, the solution of equation (\ref{eq:GC14}) is growing oscillatory and the fluid is unstable to semiconvection.

To examine the sign of $\left( \Sigma_{s} + \Sigma_{Y_{\ell}} \left| \frac{ \Upsilon_{s} }{ \Upsilon_{Y_{\ell}} }\right| \right)$ further, we note that the response function $\Sigma_{s}$ is always negative. If the cross response function $\Sigma_{Y_{\ell}}$ were zero, the coefficient of $v$ in equation (\ref{eq:GC14}) would be negative, the solutions would be decaying oscillatory and the fluid would be stable. This is analogous to the Schwarzschild stability described in Section \ref{sec:Schwarzschild} with  $\Sigma_{s} < 0$. As pointed out there, an entropy perturbation, $\theta_{s}$, and the buoyancy force it induces will be diminished by the tendency of neutrino transport (modeled by a negative $\Sigma_{s}$) to thermally equilibrate the fluid element with the background. Here the situation is similar except that the buoyancy force arising from an entropy perturbation, $\theta_{s}$, comes about through the $\theta_{Y_{\ell}}$ that it induces. In either case the effect of a $\Sigma_{s} < 0$ is to reduce the buoyancy force with a phase such that a net work is extracted from the motion of the fluid element during the course of each oscillation.

However, if $\Sigma_{Y_{\ell}} > 0$ and additionally satisfies the inequality 

\begin{equation}
 \Sigma_{Y_{\ell}} >  \frac{ \Upsilon_{s} }{ \Upsilon_{Y_{\ell}} } \left| \Sigma_{s} \right|,
\label{eq:GC17}
\end{equation}

\noindent then the coefficient of $v$ in equation (\ref{eq:GC14}) is $ > 0$, the solutions are growing oscillatory, and fluid is semiconvectively unstable. In this case, where both $\Sigma_{s}$ and $\Sigma_{Y_{\ell}}$ are nonzero, there are two competing effects governing the growth or decay of the amplitde of oscillation of the fluid element. These are embodied by the two terms in the coefficient of $v$ in equation (\ref{eq:GC14}). The $\Sigma_{s}$ term, as described above, models the effect of an entropy difference $\theta_{s}$ on the thermal transport between the fluid element and the background viz., for $\Sigma_{s} < 0$ the induced thermal transport drives a $\dot{\theta}_{s}$ tending to reduce the magnitude of $\theta_{s}$, and with it the buoyancy force. The effect is that a net work is extracted from an oscillating fluid element during the course of each oscillation. The term involving $\Sigma_{Y_{\ell}}$ affects the motion more indirectly. It begins with the fact that an entropy difference, $\theta_{s}$, induces a lepton difference, $\theta_{Y_{\ell}}$, in accordance with equation (\ref{eq:GC11}). This lepton difference, being, in turn, another driver of thermal transport between the fluid element and the background, as modeled by equation (\ref{eq:GC11a}), provides a contribution to $\dot{\theta}_{s}$. This contribution has the same sign as $\theta_{s}$ if $\Sigma_{Y_{\ell}} > 0$, which is usually the case. If inequality (\ref{eq:GC17}) is satisfied, the contribution to $\dot{\theta}_{s}$ arising from the term involving $\Sigma_{Y_{\ell}}$ is larger in magnitude than the contribution to $\dot{\theta}_{s}$ arising from the term $\Sigma_{s}$, and the net effect is a  $\dot{\theta}_{s}$ having the same sign as $\theta_{s}$. With this, the oscillation amplitude of the fluid element grows. We thus have a case in which a lepton difference, $\theta_{Y_{\ell}}$, induced via equation (\ref{eq:GC11}) by an entropy difference, $\theta_{s}$, does two things. It causes a growing mode to become oscillatory (the coefficient of $x - x_{0}$ in equation (\ref{eq:GC14})), and it causes this oscillatory solution to grow (the coefficient of $v$ in equation (\ref{eq:GC14})). We refer to this instability as ``lepto-entropy semiconvection,'' and denote it by ``L-E-Semi-C'' in the figures.

\begin{figure}[!h]
\setlength{\unitlength}{1.0cm}
{\includegraphics[width=\columnwidth]{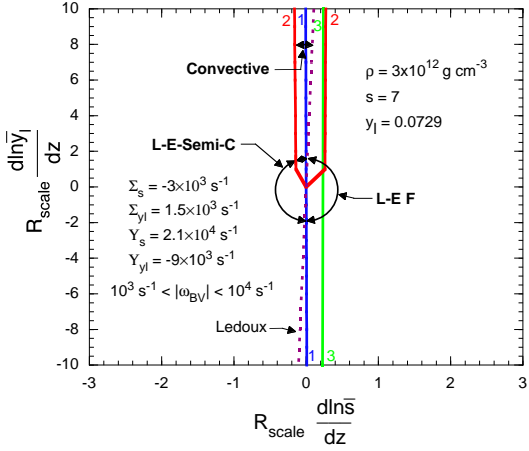}}
\caption{\label{fig:CRF0_3_yt_large}
Fluid stability as a function of the $\ln s$ and $\ln Y_{\ell}$ gradients for the case shown in Figure \ref{fig:CRF0_3_yt}, but with each of the response functions increased by a factor of three.}
\end{figure}

\begin{figure}[!h]
\setlength{\unitlength}{1.0cm}
{\includegraphics[width=\columnwidth]{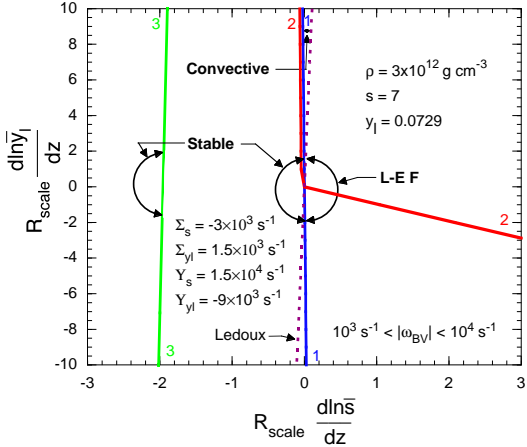}}
\caption{\label{fig:CRF0_3_yt_small}
Fluid stability as a function of the $\ln s$ and $\ln Y_{\ell}$ gradients for the values of the gravitational acceleration and the response functions shown in Figure \ref{fig:CRF0_3_yt_large}, but with $\Upsilon_{s}$ reduced so that inequality (\ref{eq:GC17}) is no longer satisfied.}
\end{figure}

     The condition for lepto-entropy semiconvection expressed by inequality (\ref{eq:GC17}) can be tested by reducing $\Upsilon_{s}$ so that inequality (\ref{eq:GC17}) is no longer satisfied. Doing so for the response functions shown in Figure \ref{fig:CRF0_3_yt} required a substantial reduction in $\Upsilon_{s}$ before the lepto-entropy semiconvective instability disappeared. However, the magnitude of response functions in Figure \ref{fig:CRF0_3_yt} are not large enough for the velocity term in equation (\ref{eq:eq8}) to be neglected in comparison with the diffusion terms. Increasing all of the response functions by a factor of three (equivalent to reducing the size of the fluid element) gives the results shown in Figures \ref{fig:CRF0_3_yt_large} and \ref{fig:CRF0_3_yt_small}. Comparison of these two figures shows that lepto-entropy semiconvection is present for the case shown in Figure \ref{fig:CRF0_3_yt_large} in which inequality (\ref{eq:GC17}) is satisfied, but not in Figure \ref{fig:CRF0_3_yt_small} in which the inequality is not satisfied.

\subsubsection{Lepto-Entropy Fingers and Lepto-Entropy Semiconvection in regimes where $\left( \pderiv{ \ln \rho }{ \ln Y_{\ell} } \right)_{p,s} < 0$ }
\label{sec:LEFS}

\begin{figure}[!h]
\setlength{\unitlength}{1.0cm}
{\includegraphics[width=\columnwidth]{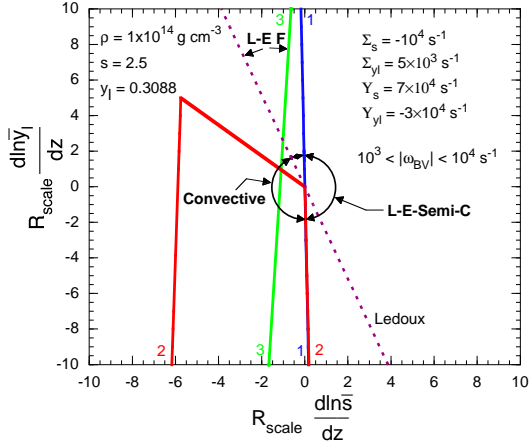}}
\caption{\label{fig:CRF0_3_d14_semi}
Fluid stability as a function of the $\ln s$ and $\ln Y_{\ell}$ gradients for the specified thermodynamic state for a case in which the response functions satisfy inequality (\ref{eq:GC17}).}
\end{figure}

Another important case for proto-supernova is the case in which the conditions leading to equation (\ref{eq:GC9}) are again satisfied (viz., $|\Upsilon_{s}| >> \omega_{Y_{\ell}}$, $|\Upsilon_{Y_{\ell}}| >> \omega_{Y_{\ell}}$, $|\Sigma_{s}| << \omega_{s}$, $|\Sigma_{Y_{\ell}}| << \omega_{s}$), but now $\left( \pderiv{\ln \rho}{Y_{\ell}} \right)_{p,s} < 0$, which occurs in the cold inner core where $Y_{\ell}$ is high. Figure \ref{fig:CRF0_3_d14_semi} shows the stability/instability regions in the $\frac{d \ln \bar{s}}{dz}$ - $\frac{d \ln \bar{Y}_{\ell}}{dz}$ plane for a thermodynamic state representative of the colder and less highly deleptonized inner core where the above conditions are satisfied. The instabilities encountered here are similar to those exhibited in Figure \ref{fig:CRF0_3_t}, and the discussion here will therefore be applicable to the conditions shown in Figure \ref{fig:CRF0_3_t}.

To examine this case, we note that the conditions stated above imply that the approximations leading to equation (\ref{eq:GC14}) are satisfied here, and we rewrite this equation as

\be
& {\ds  \dot{v} =  \left( \Sigma_{s} + \Sigma_{Y_{\ell}} 
\left| \frac{ \Upsilon_{s} }{ \Upsilon_{Y_{\ell}} }\right|  \right) v
} & \nonumber \\ 
& {\ds - g \left[ \left| \left( \pderiv{\ln \rho}{s} \right)_{p,Y_{\ell}} \right|
+ \left| \left( \pderiv{\ln \rho}{Y_{\ell}} \right)_{p,s}
\frac{ \Upsilon_{s} }{ \Upsilon_{Y_{\ell}} } \right| \right] \times
} & \nonumber \\ 
& {\ds \times \left( \frac{d\bar{s}}{dz} \right) ( x - x_{0}).
} &
\label{eq:GC18}
\ee

\noindent where we have explicitly exhibited the fact that now both $\left( \pderiv{\ln \rho}{s} \right)_{p,Y_{\ell}} < 0$ and $\left( \pderiv{\ln \rho}{Y_{\ell}} \right)_{p,s} < 0$. Equation (\ref{eq:GC18})
gives a stability criterion similar to the Schwarzschild stability criterion of equation (\ref{eq:S8}), viz., stable oscillatory solutions if $\frac{d\bar{s}}{dz} > 0$ and unstable solutions if $\frac{d\bar{s}}{dz} < 0$. In the latter case, we have convection if we are in a region which is Ledoux unstable (most of the left-hand side of Figure \ref{fig:CRF0_3_d14_semi}), and lepto-entropy fingers in the region which is Ledoux stable (upper wedge on the left-hand side of Figure \ref{fig:CRF0_3_d14_semi}). The explanation for the presence of lepto-entropy fingers here is similar in origin to the case discussed in Section \ref{sec:LEF}. Here a negative (destabilizing) gradient in $\bar{s}$ is stabilized, in the absence of transport, by a sufficiently large positive gradient in $\bar{Y}_{\ell}$ depending on the relative magnitude of $\left( \pderiv{\ln \rho}{Y_{\ell}} \right)_{p,s}$ and $\left( \pderiv{\ln \rho}{s} \right)_{p,Y_{\ell}}$. With transport, however, the lepton fraction difference, $\theta_{Y_{\ell}}$, between the fluid element and the background no longer arises from the displacement of the fluid element in a background gradient of $\bar{Y}_{\ell}$, but in response to the entropy difference $\theta_{s}$ in accordance with equation (\ref{eq:GC11}), as in the case discussed in Section \ref{sec:LEF}. Rather than stabilizing the fluid element, the induced $\theta_{Y_{\ell}}$ with the same sign as $\theta_{s}$, together with the negative value of $\left( \pderiv{\ln \rho}{Y_{\ell}} \right)_{p,s}$, further destabilizes the fluid element, and we have a case which we will again refer to as lepto-entropy fingers.

The case in which $\frac{d\bar{s}}{dz} > 0$ is encountered frequently throughout much of the extent of the cold inner core of a proto-supernova. This positive entropy gradient arises during infall because the electron captures that take place in matter with increasing radius are more out of equilibrium due to a faster infall velocity of this matter. (Recall that the inner core collapses approximately homologously with infall velocity proportional to radius.) This causes an increased entropy production with radius.

\begin{figure}[!h]
\setlength{\unitlength}{1.0cm}
{\includegraphics[width=\columnwidth]{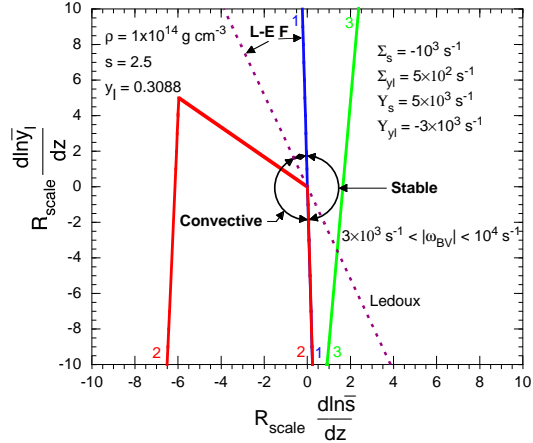}}
\caption{\label{fig:CRF0_3_d14_stbl}
Fluid stability as a function of the $\ln s$ and $\ln Y_{\ell}$ gradients for the same thermodynamic state as shown in Figure \ref{fig:CRF0_3_d14_semi} but for a case in which the response functions do not satisfy inequality (\ref{eq:GC17}).}
\end{figure}


A positive entropy gradient in equation (\ref{eq:GC18}) gives oscillatory solutions. In this case the entropy gradient is stabilizing, i.e., an outward displacement of the fluid element gives rise to a negative entropy difference, $\theta_{s}$, which together with a negative $\left( \pderiv{\ln \rho}{s} \right)_{p,Y_{\ell}}$ gives rise to a negative buoyancy forcing the fluid element inward. A negative $\theta_{Y_{\ell}}$ arises in response to the negative $\theta_{s}$, in accordance with equation (\ref{eq:GC11}), and this together with the negative $\left( \pderiv{\ln \rho}{Y_{\ell}} \right)_{p,s}$ adds to the restoring force.
 
These oscillatory solutions are stable or semiconvective depending on whether the sign of the coefficient of $v$ in equation (\ref{eq:GC18}) is negative or positive, respectively. If the sign is positive, then the discussion at the end of Section \ref{sec:LES} applies. In particular, an entropy difference $\theta_{s}$ drives a thermal diffusion which tends to reduce the magnitude of $\theta_{s}$, damping the oscillation of the fluid element. But the $\theta_{Y_{\ell}}$ that arises in response to $\theta_{s}$, in accordance with equation (\ref{eq:GC11}), feeds back by driving a thermal diffusion tending to increase the magnitude of $\theta_{s}$. If this latter thermal diffusion is large enough (i.e., if inequality (\ref{eq:GC17}) is satisfied) it dominates and causes the oscillation amplitude of the fluid element to grow. We then get a semiconvection driven by a $\theta_{Y_{\ell}}$ induced by a $\theta_{s}$, and we will refer to this case again as lepto-entropy semiconvection.

Referring back to Figure \ref{fig:CRF0_3_d14_semi}, the values of the response functions in this figure were chosen to satisfy inequality (\ref{eq:GC17}) but to otherwise be representative. Note that the region on the right (where $\frac{d\bar{s}}{dz} > 0$) is unstable to lepto-entropy semiconvection, in accordance with our discussion discussed above. As a test of the condition for lepto-entropy semiconvection as prescribed by inequality (\ref{eq:GC17}), Figure \ref{fig:CRF0_3_d14_stbl} shows the $\frac{d \ln \bar{s}}{dz}$ - $\frac{d \ln \bar{Y}_{\ell}}{dz}$ plane for the same thermodynamic state and response function values as Figure \ref{fig:CRF0_3_d14_semi} except that the value of the  response function $\Upsilon_{s}$ has been decreased from $7 \times 10^{4}$ to $5 \times 10^{4}$ so that inequality (\ref{eq:GC17}) is now violated. The region in the $\frac{d \ln \bar{s}}{dz}$ - $\frac{d \ln \bar{Y}_{\ell}}{dz}$ plane unstable to lepto-entropy semiconvection in Figure \ref{fig:CRF0_3_d14_semi} is now stable, validating the prescription given by inequality (\ref{eq:GC17}) for this case. 

\section{Computation of the Response Functions}
\label{sec:Response}

In the preceding Section we discussed the effect of different sets of values of the response functions, $\Sigma_{s}, \Sigma_{ Y_{\ell} }$, $\Upsilon_{s}$, and $\Upsilon_{ Y_{\ell} }$, on the locations of the stability and instability regions in the $\frac{d \ln \bar{s}}{dz}$ - $\frac{d \ln \bar{Y}_{\ell}}{dz}$ plane, and on the modes of instability. In this Section we discuss how we obtain these response functions for a given thermodynamic state and fluid element size. Because the response functions depend on details of neutrino transport, and may be sensitive to these details, our approach is to compute the response functions by sophisiticated radiation-hydrodynamical equilibration simulations. The thermodynamic conditions (i.e., the values of the density $\rho$, temperature $T$, and lepton fraction $Y_{\ell}$) of spherical fluid element are specified, and the size (radius) of the fluid element is chosen. The fluid element is then represented numerically by 20 spherical mass shells of equal width, and the background by 60 additional mass shells of the same thickness surrounding the fluid element. Initially  the fluid element and the background are at the same thermodynamic conditions. The fluid element is then subjected to a perturbation in its thermodynamic conditions relative to the background, and the neutrino mediated re-equilibration of the fluid element with the background is then numerically computed. The radiation-hydrodynamical code used to perform these equilibration simulations is the same as that used to model core collapse and will be described in the next Section.

Two procedures, ``isothermal'' and ``adiabatic,'' for perturbing a fluid element relative to the background were tried. An isothermal perturbation proceeds as follows. Prior to performing the perturbation, the thermodynamic conditions of the fluid are frozen, neutrino interactions with the matter, but not transport, are turned on, and neutrinos are allowed to equilibrate locally with the fluid element and the background. The result is a uniform neutrino sea of all flavors in equilibrium with both the fluid element and the background. To complete the isothermal perturbation, the neutrino distributions are kept constant and the temperature or electron fraction or both of the fluid element are perturbed relative to the background, with the density of the fluid element adjusted so that pressure equilibrium between the fluid element and the background is maintained. In summary, an isothermal perturbation is a perturbation of the thermodynamic state of the fluid element relative to both the background and a uniform neutrino ``sea.'' To complete the simulation, neutrino transport is turned on. As the neutrinos re-equilibrate the fluid element with the background the radius, and therefore density, of the fluid is continually adjusted so that pressure equilibrium is maintained at all times. The results of this simulation are the time evolutions of the quantities $\langle s \rangle - \bar{s} = \theta_{s}$ and $\langle Y_{\ell} \rangle - \bar{Y}_{\ell} = \theta_{ Y_{\ell} }$, where $\langle s \rangle$ and $\langle Y_{\ell} \rangle$ are the entropy and lepton fraction mass averaged over the fluid element, $\bar{s}$ and $\bar{Y}_{\ell}$ are the entropy and lepton fraction of the background, and $\theta_{s}$ and $\theta_{ Y_{\ell} }$ are defined at the beginning of Section \ref{sec:SPA}.

An adiabatic perturbation begins with a perturbation of the temperature or electron fraction or both of the fluid element relative to the background. To complete the adiabatic perturbation, neutrino interactions with the matter, but not transport, are turned on, and neutrinos are allowed to equilibrate locally with the fluid element and the background. The result is that both the fluid element and the neutrinos within it are perturbed with respect to the background and the background neutrinos. Thus, an adiabatic perturbation is a perturbation of both the matter and neutrinos such that local thermodynamic equilibrium between the matter and neutrinos is maintained. Once the neutrinos have achieved local equilibration with the matter, neutrino transport is turned on and the re-equilibration of the fluid element with the background is followed. Pressure equilibrium between the fluid element and the background is maintained continuously by adjusting the radius (and therefore density) of the fluid element. As before, the results of this simulation are the time evolutions of the quantities $\langle s \rangle - \bar{s} = \theta_{s}$ and $\langle Y_{\ell} \rangle - \bar{Y}_{\ell} = \theta_{ Y_{\ell} }$.

From a given simulation, time derivatives of the quantities $\theta_{s}$ and $\theta_{ Y_{\ell} }$ are computed by taking finite differences, i.e.,

\be
& {\ds   \dot{\theta}_{s} = \frac{ \theta_{s}(\Delta_{s} t) - \theta_{s}(0) } { \Delta_{s} t },
} & \nonumber \\ 
& {\ds \dot{\theta}_{Y_{\ell}} = \frac{ \theta_{Y_{\ell}}(\Delta_{Y_{\ell}} t)
- \theta_{Y_{\ell}}(0) } { \Delta_{Y_{\ell}} t }
} &
\label{eq:R1}
\ee

\noindent where we have defined $\Delta_{s} t$ and $\Delta_{Y_{\ell}} t$ to be the times required for $\theta_{s}$ and $\theta_{Y_{\ell}}$ to be reduced to $1/e$ of their original perturbed values, respectively. Experiments showed that the choice of $1/e$ as a measure of the approach to equilibration in the determination of $\dot{\theta}_{s}$ and $\dot{\theta}_{Y_{\ell}}$ was not critical, although the use of shorter time intervals gave slightly higher values for $\dot{\theta}_{s}$ and $\dot{\theta}_{Y_{\ell}}$, as would be expected.

To determine values of the response functions for a given thermodynamic state and fluid element radius, we performed two equilibration simulations beginning with different linearly independent sets of values of $\theta_{s}$ and $\theta_{ Y_{\ell} }$ for the perturbation of the fluid element with respect to the background. Call these two sets of perturbations $\theta_{s}^{(1)}, \theta_{ Y_{\ell} }^{(1)}$ and $\theta_{s}^{(2)}, \theta_{ Y_{\ell} }^{(2)}$. We then took a linear combination of the two perturbations, choosing the coefficients $a$ and $b$ such that

\begin{equation}
\theta_{s} = a \theta_{s}^{(1)} + b \theta_{s}^{(2)} \ne 0, 
\theta_{ Y_{\ell} } = a \theta_{ Y_{\ell} }^{(1)} + b \theta_{ Y_{\ell} }^{(2)} = 0,
\label{eq:R2}
\end{equation}

\noindent and another linear combination, choosing $c$ and $d$ such that

\begin{equation}
\theta_{s} = c \theta_{s}^{(1)} + d \theta_{s}^{(2)} = 0, \qquad
\theta_{ Y_{\ell} } = c \theta_{ Y_{\ell} }^{(1)} + d \theta_{ Y_{\ell} }^{(2)} \ne 0.
\label{eq:R3}
\end{equation}

\noindent From these linear combinations we compute the response functions. For example,
 
\be
& {\ds \Sigma_{s} = \left. \frac{ \dot{\theta}_{s} }{ \theta_{s} } \right|_{\theta_{Y_{\ell}} = 0}
= \frac{ a \dot{\theta}_{s}^{(1)} + b \dot{\theta}_{s}^{(2)} }
{ a \theta_{s}^{(1)} + b \theta_{s}^{(2)} },
} & \nonumber \\ 
& {\ds \Sigma_{Y_{\ell}} = \left. \frac{ \dot{\theta}_{s} }{ \theta_{Y_{\ell}} }
\right|_{\theta_{s} = 0}
= \frac{ c \dot{\theta}_{s}^{(1)} + d \dot{\theta}_{s}^{(2)} }
{ c \theta_{Y_{\ell}}^{(1)} + d \theta_{Y_{\ell}}^{(2)} },
} &
\label{eq:R4}
\ee

\be
& {\ds \Upsilon_{s} = \left. \frac{ \dot{\theta}_{Y_{\ell}} }{ \theta_{s} } \right|_{\theta_{Y_{\ell}} = 0}
= \frac{ a \dot{\theta}_{Y_{\ell}}^{(1)} + b \dot{\theta}_{Y_{\ell}}^{(2)} }
{ a \theta_{s}^{(1)} + b \theta_{s}^{(2)} },
} & \nonumber \\ 
& {\ds \Upsilon_{Y_{\ell}} = \left. \frac{ \dot{\theta}_{Y_{\ell}} }{ \theta_{Y_{\ell}} } \right|_{\theta_{s}
= 0}
= \frac{ c \dot{\theta}_{Y_{\ell}}^{(1)} + d \dot{\theta}_{Y_{\ell}}^{(2)} }
{ c \theta_{Y_{\ell}}^{(1)} + d \theta_{Y_{\ell}}^{(2)} }.
} &
\label{eq:R5}
\ee

(Naturally it would have been possible to perturb $\theta_{s}$ at $\theta_{Y_{\ell}} = 0$, and vice versa, and this would have made it unnecessary to chose linear combinations of the results. However, the procedure we adopted proved to be simpler to implement.)

\begin{figure}[!h]
\setlength{\unitlength}{1.0cm}
{\includegraphics[width=\columnwidth]{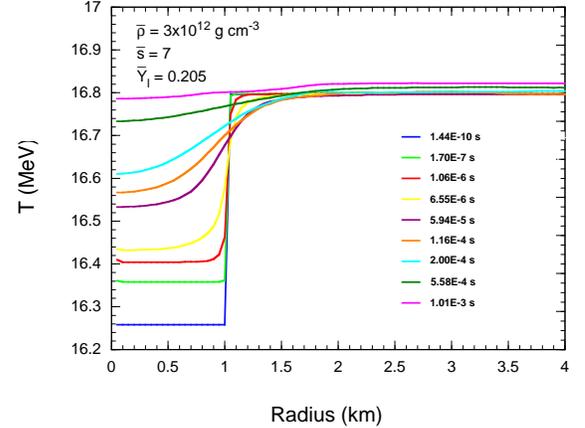}}
\caption{\label{fig:t_r_time}
Temperature as a function of radius at selected times for an equilibration simulation with improved neutrino rates. The radius of the fluid element is 1 km, and the thermodynamic state of the background and the selected times are indicated.}
\end{figure}

\begin{figure}[!h]
\setlength{\unitlength}{1.0cm}
{\includegraphics[width=\columnwidth]{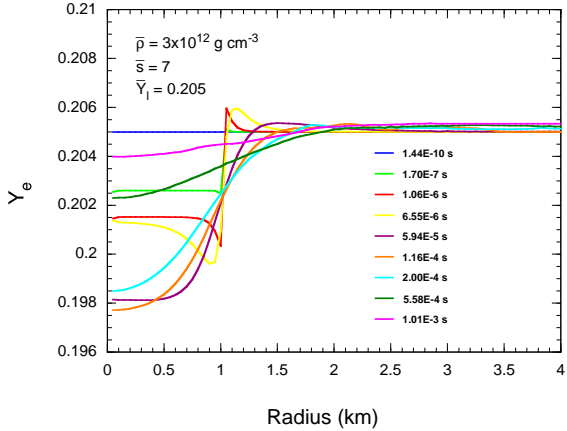}}
\caption{\label{fig:ye_r_time}
Same as Figure \ref{fig:t_r_time} but showing the electron fraction as a function of radius at selected times.}
\end{figure}

\begin{figure}[!h]
\setlength{\unitlength}{1.0cm}
{\includegraphics[width=\columnwidth]{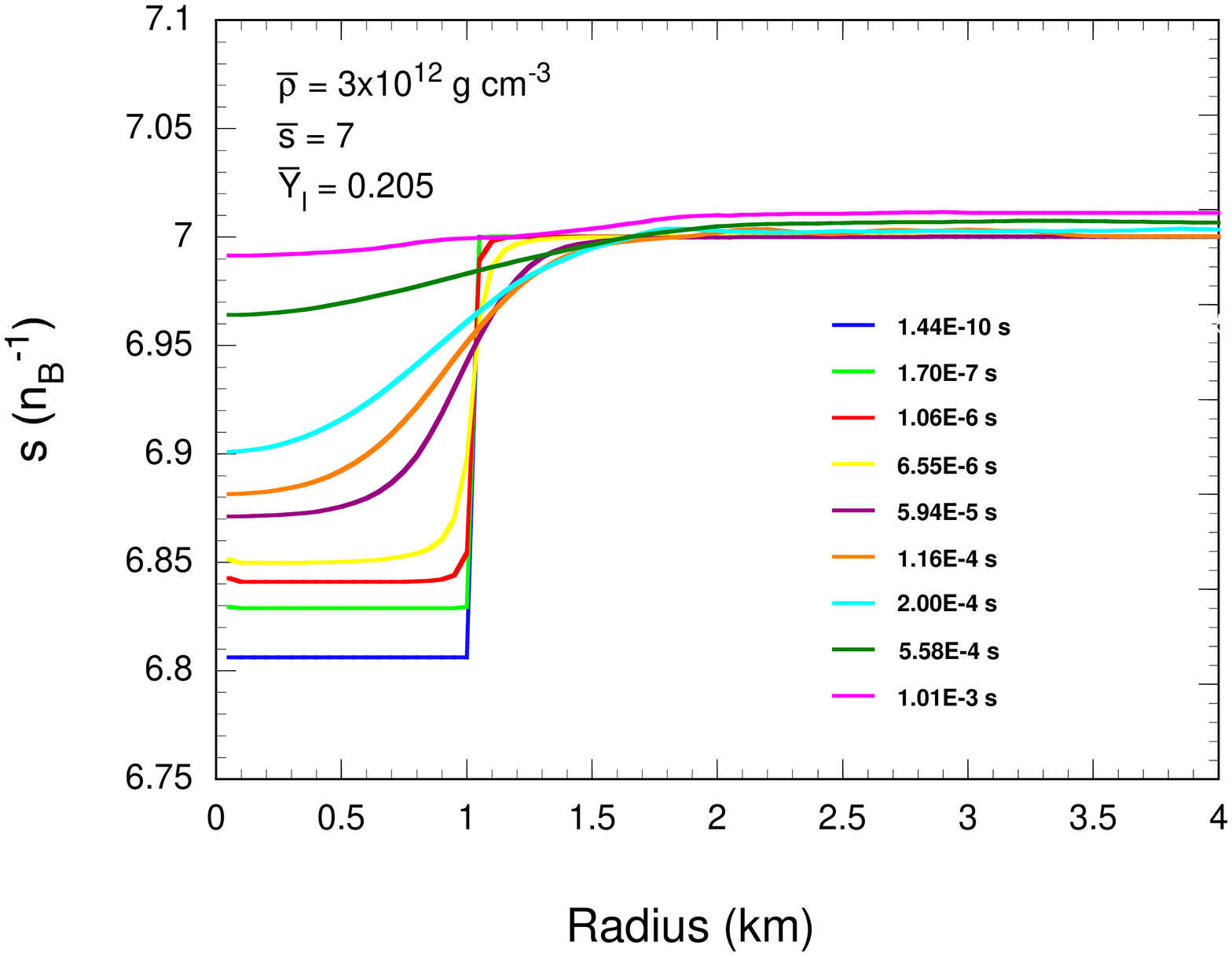}}
\caption{\label{fig:s_r_time}
Same as Figure \ref{fig:t_r_time} but showing the entropy as a function of radius at selected times.}
\end{figure}
\begin{figure}[!h]
\setlength{\unitlength}{1.0cm}
{\includegraphics[width=\columnwidth]{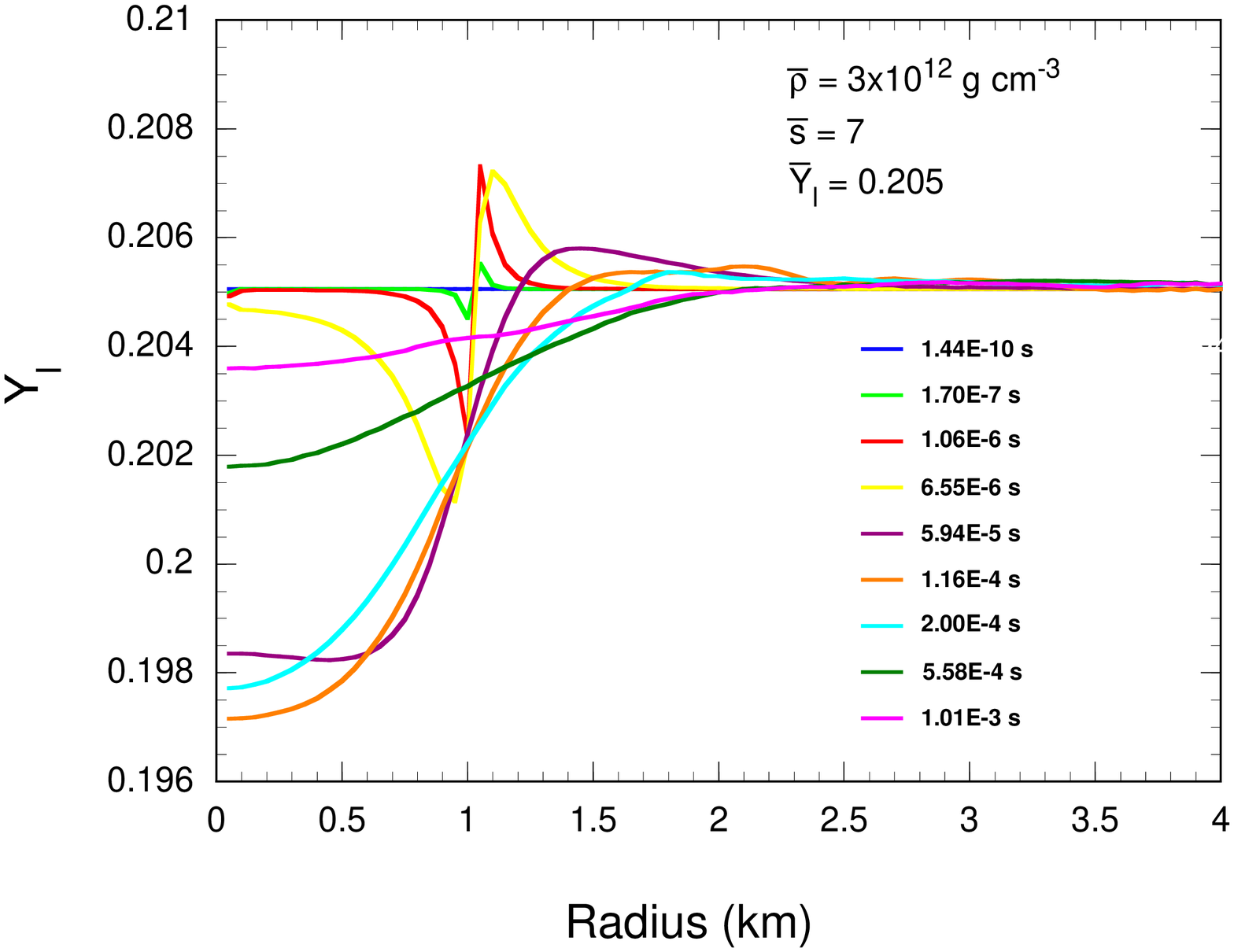}}
\caption{\label{fig:yl_r_time}
Same as Figure \ref{fig:t_r_time} but showing the entropy as a function of radius at selected times.}
\end{figure}
\begin{figure}[!h]
\setlength{\unitlength}{1.0cm}
{\includegraphics[width=\columnwidth]{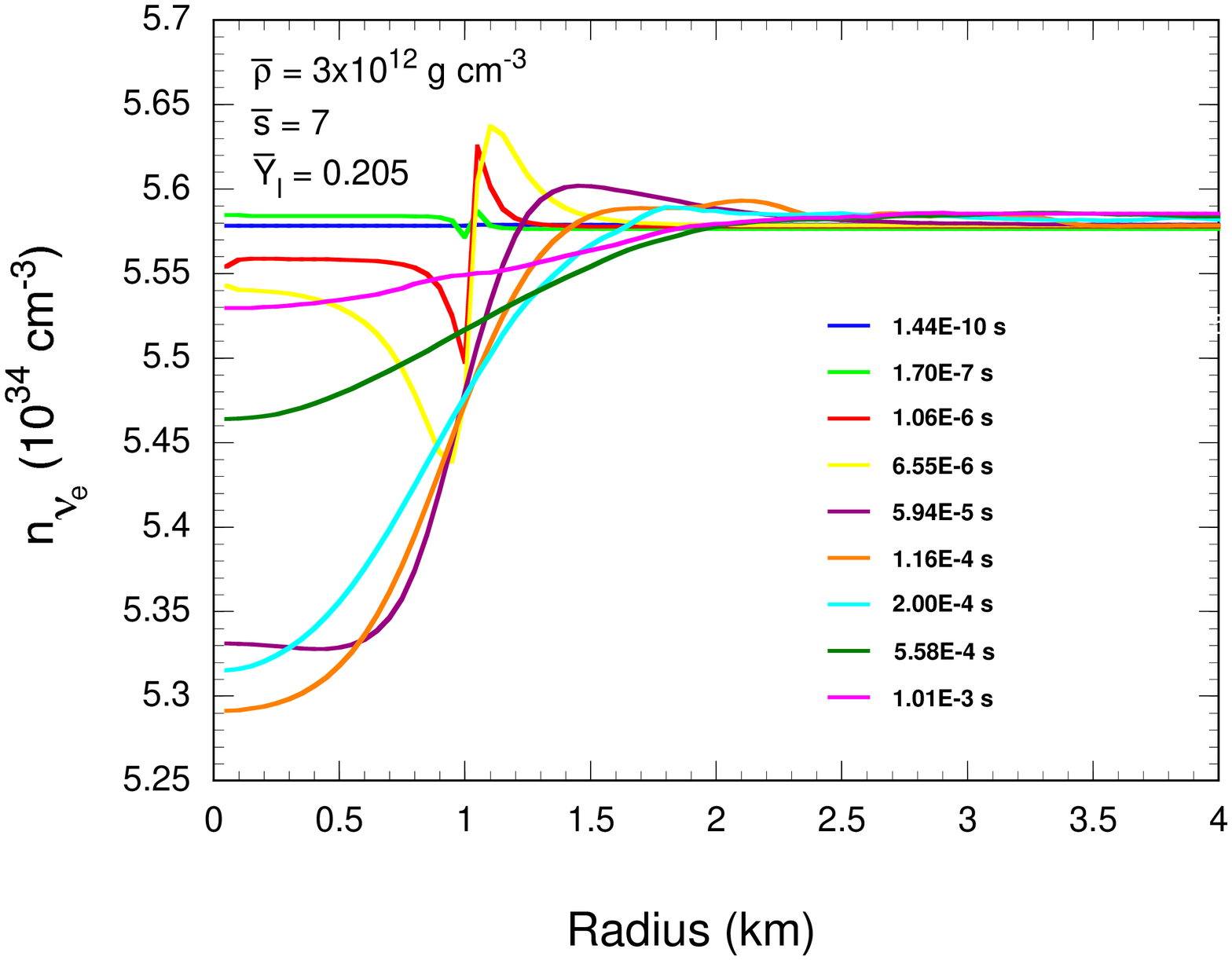}}
\caption{\label{fig:nue_r_time}
Same as Figure \ref{fig:t_r_time} but showing the \nue\ number density as a function of radius at selected times.}
\end{figure}
\begin{figure}[!h]
\setlength{\unitlength}{1.0cm}
{\includegraphics[width=\columnwidth]{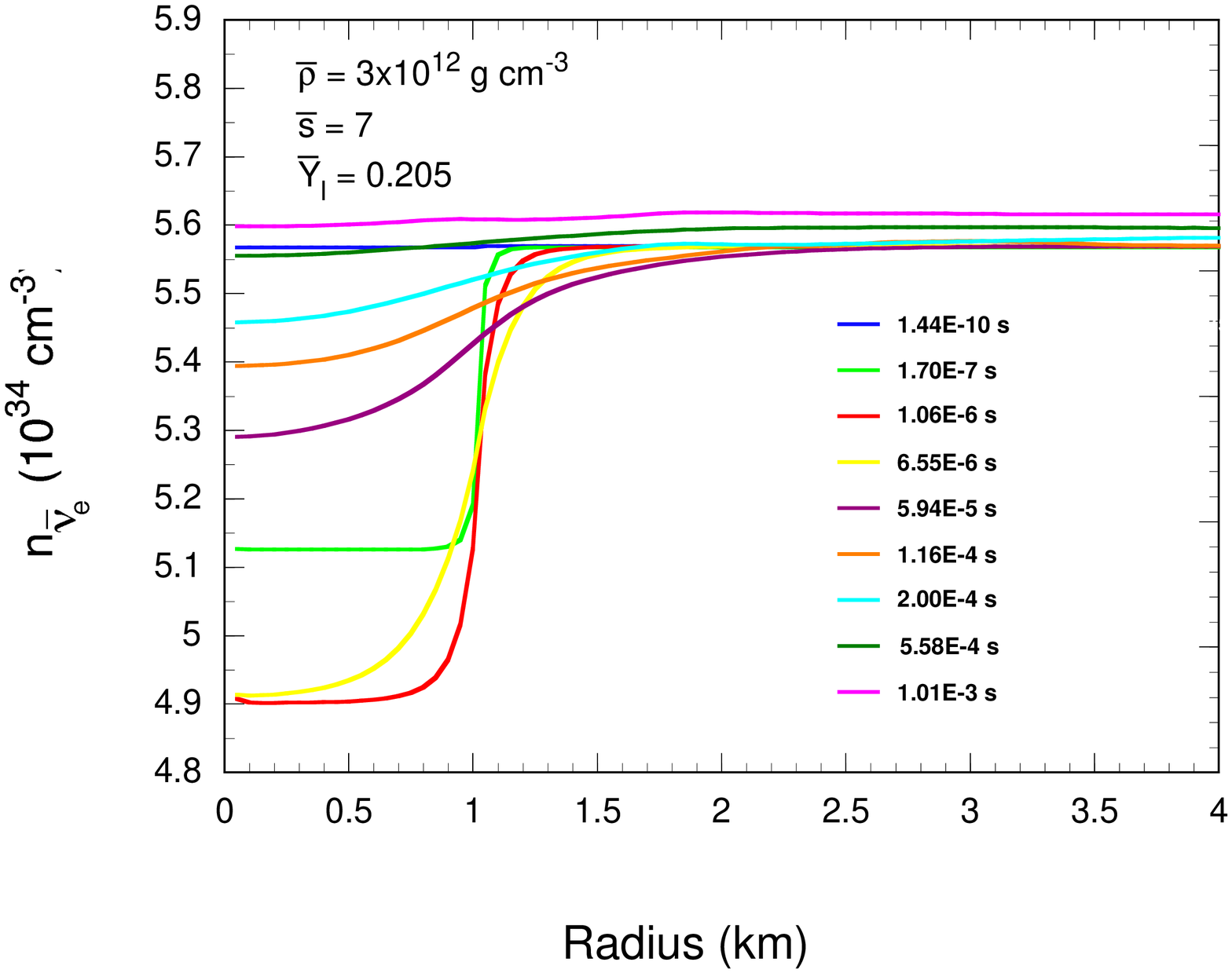}}
\caption{\label{fig:nuebar_r_time}
Same as Figure \ref{fig:t_r_time} but showing the \nuebar\ number density as a function of radius at selected times.}
\end{figure}

As we discussed above in Section \ref{sec:GC}, a large positive value of $\Upsilon_{s}$ is critical to the emergence of the lepto-entropy finger instability. To illustrate how this comes about with a specific example, we show in Figures \ref{fig:t_r_time} - \ref{fig:nuebar_r_time} some of the details of an equilibration simulation for an isothermal perturbation. The simulation was performed with improved neutrino rates. For ease of illustration, the state of the background that we chose, $\bar{\rho} = 3 \times 10^{12}$ g cm$^{-3}$, $\bar{s} = 7$, $\bar{Y}_{\ell}$, is a thermodynamic state for which the \nue\ (and therefore \nuebar) chemical potential vanishes. The equilibrium \nue\ and \nuebar\ number densities in the background are therefore equal to each other. Because of this, it was possible to perturb the temperature of the fluid element and adjust its density to keep the pressure the same as the background while maintaining $Y_{\ell}$ at its unperturbed value. (Different values of \nue\ and \nuebar\ number densities would have resulted in a perturbed value of $Y_{\ell}$ in the fluid element after its density was adjusted to maintain pressure equilibrium, even if its value of  $Y_{\rm e}$ was kept equal to that of the background.) While the choice of any other thermodynamic state would have illustrated our points, a choice which makes it possible perturb the temperature (and entropy) of the fluid element, maintain pressure equilibrium, and keep the value of $Y_{\ell}$ at its unperturbed value provides the simplest and cleanest illustration of how a lepton fraction perturbation arises in response to an entropy perturbation, which is the basis of both the lepto-entropy fingers and lepto-entropy semiconvection instabilities.

Figures \ref{fig:t_r_time} and \ref{fig:ye_r_time} show the matter temperature and electron fraction profiles at selected times. The fluid element was perturbed relative to the background by decreasing its temperature, leaving the electron fraction unchanged, and adjusting the density to restore pressure equilibrium with  the background. This initial perturbation is shown in the figures by the temperature and electron fraction profiles at the smallest time, $t = 1.44 \times 10^{-10}$. The above perturbation in $T$ results in the entropy perturbation shown in Figure \ref{fig:s_r_time} by the profile at $t = 1.44 \times 10^{-10}$. Because of our choice of a thermodynamic state with zero \nue\ chemical potential, there is no net perturbation in the lepton fraction, as shown in Figure \ref{fig:yl_r_time} by the profile at $t = 1.44 \times 10^{-10}$. As explained earlier in this Section, an isothermal perturbation is effected by first allowing the neutrinos to equilibrate locally with the fluid element and the background, both at the same thermodynamic state. The matter, but not the neutrinos, of the fluid is then perturbed. Thus, in particular, the electron neutrino and antineutrino number densities, $n_{\nu_{\rm e}}$ and $n_{\bar{\nu}_{\rm e}}$, are initially uniform when the matter is perturbed, as shown by the $n_{\nu_{\rm e}}$ and $n_{\bar{\nu}_{\rm e}}$ profiles in Figures \ref{fig:nue_r_time} and \ref{fig:nuebar_r_time}, respectively, at $t = 1.44 \times 10^{-10}$.

\begin{figure}[!h]
\setlength{\unitlength}{1.0cm}
{\includegraphics[width=\columnwidth]{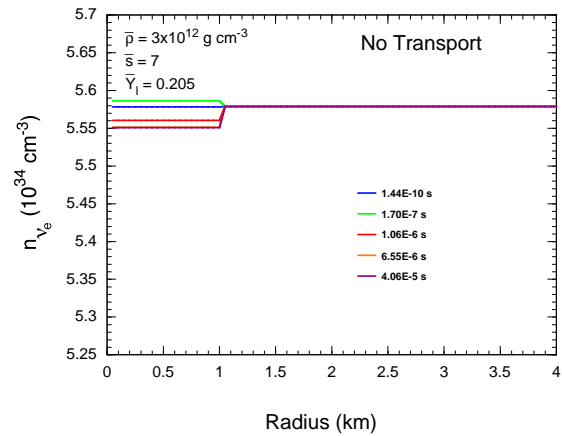}}
\caption{\label{fig:nue_r_time_notrans}
Same as Figure \ref{fig:nue_r_time} but with zone to zone transport turned off.}
\end{figure}

\begin{figure}[!h]
\setlength{\unitlength}{1.0cm}
{\includegraphics[width=\columnwidth]{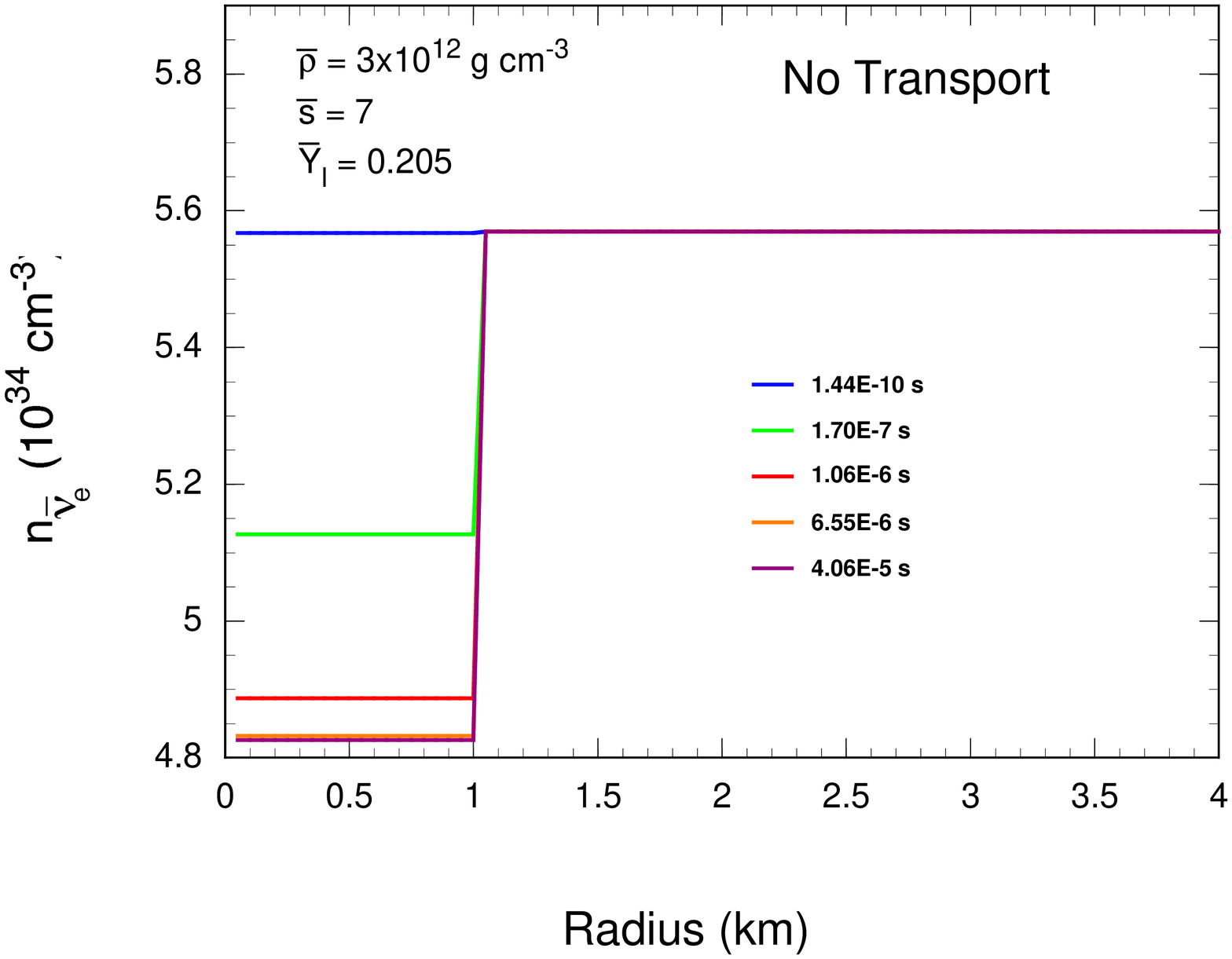}}
\caption{\label{fig:nuebar_r_time_notrans}
Same as Figure \ref{fig:nuebar_r_time} but with zone to zone transport turned off..}
\end{figure}

What happens almost immediately after the perturbation is that the \nue's and \nuebar's in the perturbed fluid element equilibrate locally with the matter, essentially establishing, thereby, conditions equivalent to an adiabatic perturbation. This is evident by comparing the \nue\  and \nuebar\  profiles at times $1.7 \times 10^{-7}$ s and $1.06 \times 10^{-6}$ s in Figures \ref{fig:nue_r_time} and \ref{fig:nuebar_r_time} with the corresponding profiles in Figures \ref{fig:nue_r_time_notrans} and \ref{fig:nuebar_r_time_notrans},. These latter figures show the time evolution of local the neutrino-matter equilibration with transport turned off so that only local equilibration can occur. It is seen that the profiles with transport turned on (Figures \ref{fig:nue_r_time} and \ref{fig:nuebar_r_time}) and the corresponding profiles with transport turned off (Figures \ref{fig:nue_r_time_notrans} and \ref{fig:nuebar_r_time_notrans}) follow each other closely, and that local equilibration is achieved in about 1 $\mu$ s. After this time the effects of transport begin to be seen, and the subsequent evolution is the same whether the initial perturbation is isothermal or adiabatic. This is why there is very little difference in the values of the response functions whether the perturbation is isothermal or adiabatic. 

The local equilibration is very asymmetric, with $n_{\bar{\nu}_{\rm e}}$ decreasing by $\sim 12.5$ \% versus a $\sim 0.36$ \% decrease in $n_{\nu_{\rm e}}$, and this asymmetry is responsible for much of what follows. The asymmetry is due to the fact that the equilibrated values of $n_{\nu_{\rm e}}$ and  $n_{\bar{\nu}_{\rm e}}$ depend on electron number density, $n_{\rm e}^{-}$, and positron number density, $n_{\rm e^{+}}$, respectively, through the reactions ${\rm e}^{-} + {\rm p} \rightleftharpoons {\rm n} + \nu_{\rm e}$ and ${\rm e}^{+} + {\rm n} \rightleftharpoons {\rm p} + \bar{\nu}_{\rm e}$. The positrons are thermally produced as electron-positron pairs, so the value of $n_{\rm e^{+}}$ is very sensitive to temperature and drops significantly with the temperature perturbation, whereas $n_{\rm e}^{-}$ consists largely of fixed electrons (for the entropies of interest here), and is much less sensitive to temperature. Thus, while the perturbation of the fluid element reduces both $n_{\rm e}^{-}$ and $n_{\rm e^{+}}$ equally, the relative reduction in $n_{\rm e^{+}}$ is much greater. The result is that the local equilibration of the \nue's and \nuebar's with the matter entails more $\bar{\nu}_{\rm e}$ absorption on protons (Figure \ref{fig:nuebar_r_time}, profiles at $1.7 \times 10^{-7}$ s and $1.06 \times 10^{-6}$ s) than $\nu_{\rm e}$ absorption on neutrons (Figure \ref{fig:nue_r_time}, profiles at $1.7 \times 10^{-7}$ s and $1.06 \times 10^{-6}$ s), and this depresses $Y_{\rm e}$ (Figure \ref{fig:ye_r_time}, profiles at $1.7 \times 10^{-7}$ s and $1.06 \times 10^{-6}$ s).

After local equilibration, rapid lepton transport between the fluid element and the background adjusts the $n_{\nu_{\rm e}}$ and $n_{\bar{\nu}_{\rm e}}$ distributions toward values for which there is a quasi net lepton flow equilibrium (i.e., a small ratio of net lepton flow to the flow in each direction) of \nue's and \nuebar's. Since the local matter-neutrino equilibration reduced the value of $n_{\bar{\nu}_{\rm e}}$ much more than that of $n_{\nu_{\rm e}}$, the difference between these two quantities drives a net lepton flow out of the fluid element, raising the value of $n_{\bar{\nu}_{\rm e}}$ and reducing that of $n_{\nu_{\rm e}}$ (Figures \ref{fig:nuebar_r_time} and \ref{fig:nue_r_time}, respectively, profiles at $6.55 \times 10^{-6}$ s, $5.94 \times 10^{-5}$ s, and $1.16 \times 10^{-4}$ s) until the establishment of a quasi net lepton flow equilibrium. The net flow of leptons out of the fluid element as a quasi net lepton flow is established reduces the value of $Y_{\ell}$ in the fluid element (Figure \ref{fig:yl_r_time}, profiles at $6.55 \times 10^{-6}$ s, $5.94 \times 10^{-5}$ s, and $1.16 \times 10^{-4}$ s). At this point, then, there is a rapid reduction in the value of $\theta_{Y_{\ell}}$ in response to the initial negative perturbation, $\theta_{s}$, of the fluid element.  This is the origin of the large positive value of $\Upsilon_{s}$, in accordance with its definition by equation \ref{eq:eq4}, which is so crucial for the appearance of lepto-entropy fingers and lepto-entropy semiconvection.

After the establishment of a quasi net lepton flow equilibrium, the differences, $\Delta n_{\nu_{\rm e}}$ and $\Delta n_{\bar{\nu}_{\rm e}}$, between the values of $n_{\nu_{\rm e}}$ and $n_{\bar{\nu}_{\rm e}}$, respectively, between the fluid element and the background adjust as energy diffuses between the fluid element and the background, and slowly approach zero as the temperature of the fluid element equilibrates with that of the background on a relatively long timescale (Figure \ref{fig:t_r_time}, profiles at $1.16 \times 10^{-4}$ s, $2.00 \times 10^{-4}$ s, $5.58 \times 10^{-4}$ s, and $1.01 \times 10^{-3}$ s).

In summary, an isothermal perturbation of the fluid element is followed by a local equilibration of matter and neutrinos on the shortest timescale, producing conditions equivalent to that of an adiabatic perturbation. Rapid lepton transport then adjusts the values of $n_{\nu_{\rm e}}$ and $n_{\bar{\nu}_{\rm e}}$ on the next shortest time scale, establishing a net lepton flow equilibrium and producing thereby a lepton fraction difference, $\theta_{Y_{\ell}}$, in response to the initial entropy difference, $\theta_{s}$. The slower thermal diffusion rate equilibrates the temperature of the fluid with the background on the longest time scale, while the lepton flow continually adjusts the values of $n_{\nu_{\rm e}}$ and $n_{\bar{\nu}_{\rm e}}$ to maintain a quasi net lepton flow equilibrium between the fluid element and the background.

\section{Core Collapse Models}
\label{sec:CCModels}

As representative of the cores of supernova progenitors, we have chosen to examine  at 100 ms and 200 ms after bounce the cores of a 15 \msol (S15s7b) and a 25 \msol (S25s7b) stellar progenitor. These were provided by \citet{woosley95} and evolved from the main sequence to the onset of core collapse as described in \citet{woosleyw95}. Precollapse model S15s7b has an iron core mass of 1.278 \msol, representative of a small iron core mass progenitor, while model precollapse model S25s7b has an iron core mass of 1.770 \msol, representative of a large iron core mass progenitor. Each progenitor was evolved through core collapse, bounce, and the shock reheating epoch by the code described in  \citet{bruenn85} and \citet{bruennh91}, and extensively revised as described in  \citet{bruenndm01} and \citet{bruenn03}. The code is fully general relativistic. Neutrino transport is multigroup, three flavor and uses a flux-limited diffusion approximation. Twenty energy zones were used in these simulations spanning in geometric progression the neutrino energy range from 4 MeV to 400 MeV. The neutrino microphysics described in \citet{bruenn85} and \citet{bruennh91} was used for the simulations, as well as some revised and new neutrino microphysics described below. Also used were the ion screening corrections given by \citet{horowitz96} and implemented as described in detail in \citet{bruennm97}.

In some of the simulations, designated by ``LS,'' the Lattimer-Swesty equation of state (Lattimer \& Swesty 1991) was used when the following three conditions were satisfied locally: (1)
$n_{{\rm B}} > 10^{-8}$ fm$^{-3}$  ($\rho > 1.67 \times 10^{7}$ \gcm), where $n_{{\rm B}}$
is the number density of nucleons (free  and bound) per cubic Fermi, (2) $T > 0.05$ MeV,
and (3) the matter was assumed to be in nuclear statistical equilibrium (NSE). The 
Cooperstein-BCK equation of state (\citealp{cooperstein85a}; \citealp*{baronck85}) was used
when the second and third condition was satisfied, but not the first. For nuclei not in
NSE, which comprised the silicon layer, oxygen layer, and other exterior layers until they
encountered the shock, seven alpha particle nuclei (plus neutrons and protons) were tracked explicitly. The nuclei were treated as an ideal gas (with excited states), and a
nine-species nuclear reaction network was used to follow the nuclear transmutations. We will refer to this equation of state as the ``non-NSE'' equation of state. A zone not in NSE were flashed to NSE if its temperature reached 0.44 MeV \citep*{thielemannnh96}. In order to compare our results more closely with those of the Livermore group, we also ran some of the simulations. designated by ``W,''  using the Livermore equation of state \citep{wilson03}.  This equation of state was provided in tabular form. This W equation of state was joined smoothly to the non-NSE equation of state, and the latter was used when nuclei were not in NSE. Table \ref{tab:CCS} summarizes the core collapse simulations that were performed to produce the core configurations that will be analyzed in the next Section. 

\begin{table}[h]
\begin{center}
\caption{ \label{tab:CCS}{\bfseries{Core Collapse Simulations}}} 
\begin{tabular}{|c|c|c|} \hline
{\bf Simulation} & {\bf Progenitor } & {\bf Equation of State} \\ \hline
 M15\_LS & S15s7b & Lattimer-Swesty \\ 
 M15\_W & \ S15s7b & Livermore  \\ 
M25\_LS & \ S25s7b & Lattimer-Swesty  \\ 
 M25\_W & \ S25s7b & Livermore  \\ \hline
 \end{tabular}
\end{center}
\end{table}

For each of the models included in Table \ref{tab:CCS} the the stability of the fluid from the core center to the region just beneath the neutrinosphere was analyzed for two time slices, 100 ms and 200 ms after core bounce. The response functions, $\Sigma_{s}, \Sigma_{ Y_{\ell} }$, $\Upsilon_{s}$, and $\Upsilon_{ Y_{\ell} }$, were obtained by performing equilibration simulations as described in Section \ref{sec:Response} by the same code described above for the core collapse simulation. For most of the equilibration simulations, the ``standard'' neutrino microphysics described in \citet{bruenn85} and \citet{bruennh91} was used, i.e., the same neutrino microphysics as used in the core collapse simulations. We observe, however, that neutrino physics has advanced beyond this standard. Recently, analytic expressions have been derived for neutrino absorption, emission, and elastic scattering on nucleons which take into account the nucleon recoil and blocking factors for arbitrary degeneracy \citep{reddypl98, burrowss98, burrowss99}. These new rates become smaller at high densities than the corresponding rates without nucleon blocking, and the recoil effects call into question the assumption that the scattering rates are isoenergetic. The latter will predominantly affect the thermalization of the \nux's (by ``\nux's'' we refer to  \num's and \nut's and their antiparticles), as these particles lack the energy exchange channels of the charged-current processes on baryons which dominate the thermalization of the \nue's and \nuebar's. Thermalization simulations \citep{thompsonbh00}, for example, show that \nux-nucleon scattering is an important thermalization process for \nux's with energies exceeding 15 MeV. Other neutrino processes that may be important for equilibrating a fluid element with the background are neutrino-nucleon bremsstrahlung and the related neutrino-nucleon inelastic scattering \citep{hannestadr98, thompsonbh00}.

We have performed equilibration simulations for several of the models with the improved \citep{reddypl98} absorption, emission, and elastic scattering on nucleons, and with the inclusion of bremsstrahlung and neutrino-nucleon inelastic scattering as formulated by \citet{hannestadr98}. Models analyzed for stability using this improved neutrino microphysics are denoted by ``Impr'', while those using the standard neutrino microphysics are denoted by ``Stnd.'' Table \ref{tab:S} lists the models analyzed for stability. We did not deem it necessary to include models obtained from core collapse simulations employing the improved neutrino physics because they differ very little below the neutrinosphere from those computed with the standard neutrino physics up to 200 ms after bounce. The improved neutrino physics mainly affects the luminosities and {\small RMS} energies of the \nux's near and above the neutrinosphere. Furthermore, we wanted to perform the core collapse simulations with the Livermore equation of state and comparison models using the Lattimer-Swesty equation of state with neutrino microphysics as similar to that used by the Livermore group as possible.

\begin{table*}[t]
\begin{center} 
\caption{\label{tab:S}{\bfseries{Models Analyzed for Stability}}}
\begin{tabular}{|c|c|c|c|c|} \hline
{\bf Model} & {\bf Progenitor } & {\bf Time from Bounce } & {\bf EOS} & {\bf $\nu$-Microphysics} \\ \hline
 M15\_LS\_100\_Stnd & S15s7b & 100 ms & Lattimer-Swesty & Standard \\ 
 M15\_LS\_100\_Impr & S15s7b & 100 ms & Lattimer-Swesty & Improved \\ 
 M15\_W\_100\_Stnd & \ S15s7b & 100 ms & Livermore & Standard \\ 
 M15\_LS\_200\_Stnd & S15s7b & 200 ms & Lattimer-Swesty & Standard \\ 
 M15\_W\_200\_Stnd & \ S15s7b & 200 ms & Livermore & Standard \\ 
 M25\_LS\_100\_Stnd & S25s7b & 100 ms & Lattimer-Swesty & Standard \\ 
 M25\_LS\_100\_Impr & S25s7b & 100 ms & Lattimer-Swesty & Improved \\ 
 M25\_W\_100\_Stnd & \ S25s7b & 100 ms & Livermore & Standard \\ 
 M25\_LS\_200\_Stnd & S25s7b & 200 ms & Lattimer-Swesty & Standard \\ 
 M25\_W\_200\_Stnd & \ S25s7b & 200 ms & Livermore & Standard \\ \hline
\end{tabular}
\end{center}
\end{table*}

\nopagebreak

\section{The Stability  of Collapsed Cores below the Neutrinosphere}
\label{sec:Results}

\begin{figure}[!h]
\setlength{\unitlength}{1.0cm}
{\includegraphics[width=\columnwidth]{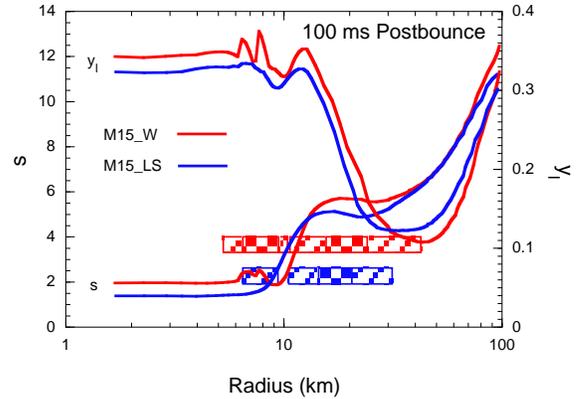}}
\caption{\label{fig:w_ls_15_100}
Entropy and lepton fraction profiles of models M15\_W and M15\_LS (as described in Table \ref{tab:S}) 100 ms after bounce. Also shown are the Ledoux unstable regions (checkerboard hatching) and the neutron finger unstable regions (diagonal pattern) as given by the Livermore group's criteria.}
\end{figure}

\begin{figure}[!h]
\setlength{\unitlength}{1.0cm}
{\includegraphics[width=\columnwidth]{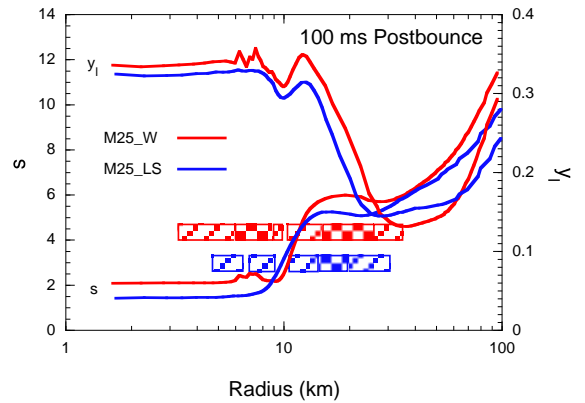}}
\caption{\label{fig:w_ls_25_100}
Same as Figure \ref{fig:w_ls_15_100} but for models M25\_W and M25\_LS.}
\end{figure}

\begin{figure}[!h]
\setlength{\unitlength}{1.0cm}
{\includegraphics[width=\columnwidth]{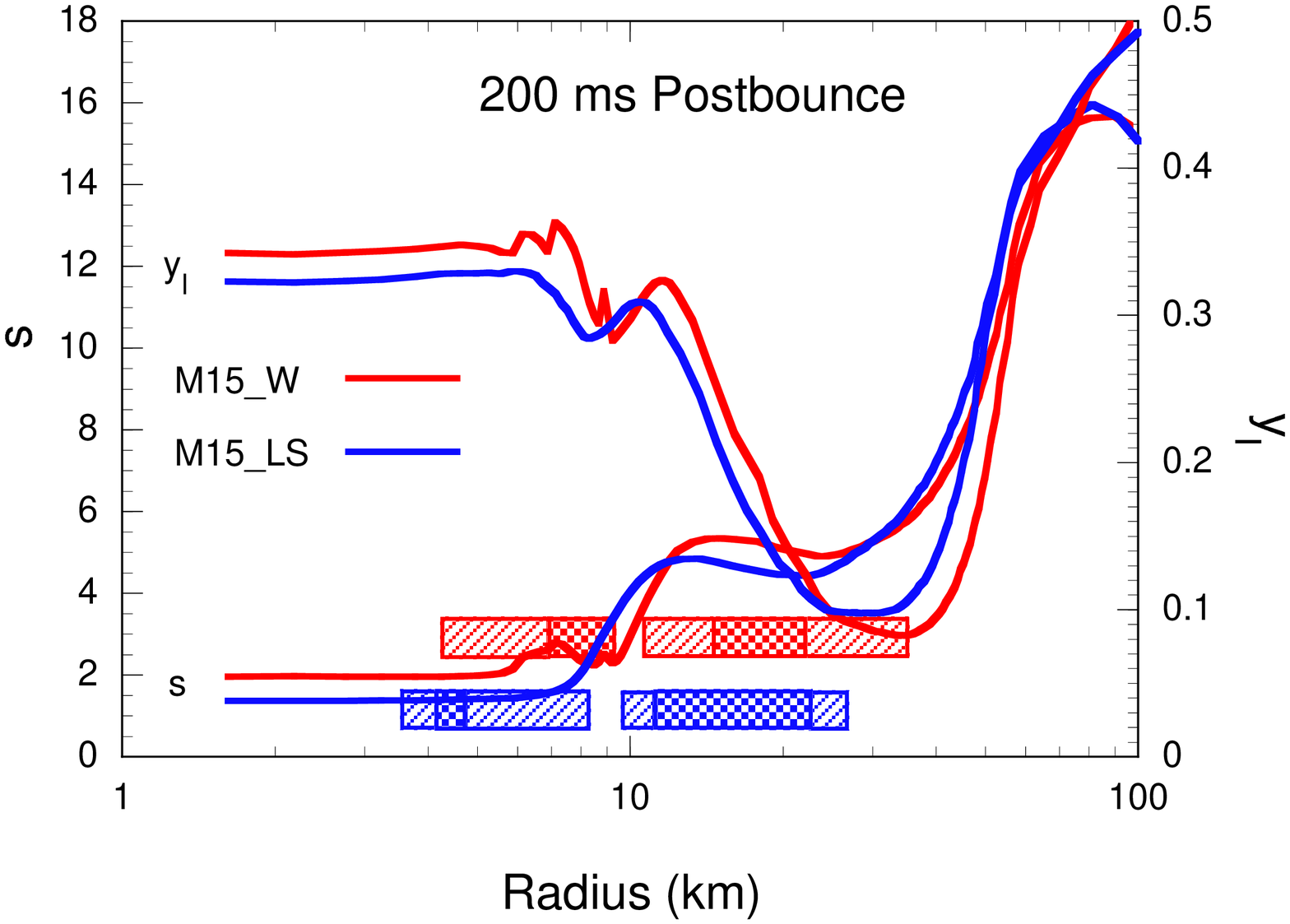}}
\caption{\label{fig:w_ls_15_200}
Same as Figure \ref{fig:w_ls_15_100} but at 200 ms after bounce.}
\end{figure}

\begin{figure}[!h]
\setlength{\unitlength}{1.0cm}
{\includegraphics[width=\columnwidth]{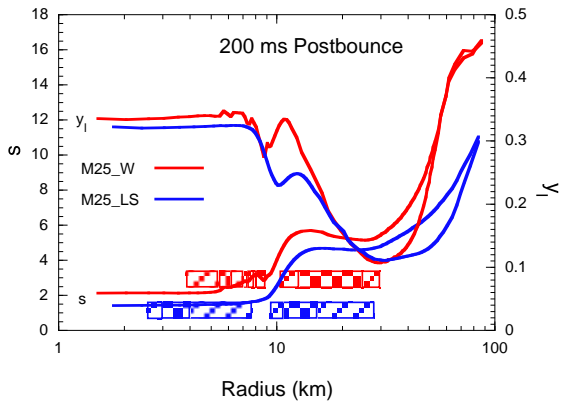}}
\caption{\label{fig:w_ls_25_200}
Same as Figure \ref{fig:w_ls_25_100} but at 200 ms after bounce.}
\end{figure}

Before analyzing the models listed in Table \ref{tab:S} for stability using the method described in the preceding Sections, let us first compare the core profiles produced by simulations using the Livermore equation of state with those produced by the Lattimer-Swesty equation of state, and ascertain the Ledoux unstable regions and the regions unstable to neutron fingers by the criteria given by the Livermore group \citep{wilsonm88}. Recall that according to the Livermore group's criteria a fluid is neutron finger unstable if a fluid element displaced outwards at constant composition but in temperature and pressure equilibrium with the background ends up after the displacement with a density less than that of the background.  The profiles in entropy and lepton fraction of models M15\_W and M15\_LS are shown at 100 ms and 200 ms after bounce in Figures \ref{fig:w_ls_15_100} and \ref{fig:w_ls_15_200}, respectively. Likewise, the profiles of models M25\_W and M25\_LS are shown at 100 ms and 200 ms after bounce in Figures \ref{fig:w_ls_25_100} and \ref{fig:w_ls_25_200}, respectively.

Figures \ref{fig:w_ls_15_100} - \ref{fig:w_ls_25_200} show that the core collapse simulations using the Livermore equation of state produce profiles in entropy and lepton fraction at 100 and 200 ms similar in shape to the simulations using the Lattimer-Swesty equation of state. The main differences are the slightly higher overall values of the entropy and lepton fraction given by the Livermore equation of state. These differences arise mainly during infall because of the smaller free proton fraction given by the Livermore equation of state in comparison with the Lattimer-Swesty equation of state for a given thermodynamic state. A smaller free proton fraction during infall leads to a reduced rate of electron capture, and this results in a higher lepton fraction at trapping. The electron captures that occur after trapping as the \nue's equilibrate with the matter produces a greater increase of entropy, as the smaller free proton fraction and reduced electron capture rate given by the Livermore equation of state results in a greater disequilibrium between matter and \nue's at trapping.

All models exhibit a region of Ledoux instability, indicated by the checkerboard patterns in the horizontal stripes, in the vicinity of 20 km from the core center at 100 ms, and models M15\_W and M25\_W exhibit an additional region of Ledoux instability in the vicinity of 8 km from the core center. At 200 ms the inward advection of matter has moved these Ledoux unstable regions slightly inward, and additional (very mild) Ledoux unstable regions have developed in models M15\_LS and M25\_LS at small radii. In most cases these Ledoux unstable regions are driven by negative entropy gradients. The location and even existence of these Ledoux unstable regions must not be taken very seriously, however, as convection, not included in these simulations, will rapidly smooth the entropy and lepton fraction gradients driving the convection.

Also shown in these figures by the slanted line pattern in the horizontal stripes are the regions that are neutron finger unstable by the Livermore group's criteria. The neutron finger instability considerably extends the regions of fluid instability to either side of the Ledoux unstable regions. In particular, the region of fluid instability extends nearly up to the neutrinospheres, which are located near the outer parts of the troughs in the lepton fraction profiles. This illustrates the claim by the Livermore group that fluid motions resulting from the neutron finger instability will advect lepton rich matter to the neutrinosphere from regions below, and thereby enhance the \nue\ luminosity.

Let us now consider the analysis of the models listed in Table \ref{tab:S} for stability using the methodology developed in the preceding Sections. To perform this analysis, each model was considered in turn and a double loop over the model was performed. The outer loop was over the spherical mass shells between the core center and the neutrino sphere (which comprised between 55 and 70 mass shells, depending on the model). For each mass shell the thermodynamic state and the gravitational acceleration were noted, and an inner loop was performed over the fluid element size. For this inner loop the radius of the fluid element was set initially to the \nue\ flux-averaged mean free path for the mass shell under consideration and then increased successively by a factor of $\sqrt{10}$ as the loop was executed until the radius of the fluid element exceeded one-half the distance of the mass shell from the center of the core. When the latter occurred, the loop was completed by setting the fluid element equal to one-half the distance of the mass shell from the center of the core. Thus, fluid element radii were considered spanning the range $\langle \lambda_{\nue} \rangle \le R_{\rm blob} \le \frac{1}{2} R$, where $\langle \lambda_{\nue} \rangle$ is the flux-averaged \nue\ mean free path, $R_{\rm blob}$ is the radius of the fluid element, and $R$ is the distance from the core center to the mass shell in question. Executing this inner loop for each mass shell typically involved 6 to 12 fluid element radii. For each mass shell and fluid element radius two liinearly independent equilibration simulations were performed as described in Section \ref{sec:Response}. The results of this pair of equilibration simulations were the four response functions, $\Sigma_{s}, \Sigma_{ Y_{\ell} }$, $\Upsilon_{s}$, and $\Upsilon_{ Y_{\ell} }$. Having computed these response functions, the gradients of $\ln s$ and $\ln Y_{{\ell}}$ at the mass shell in question were noted, and the roots of equation (\ref{eq:GC1}) were obtained.

From these roots the stability of the fluid element of the given radius and at the given mass shell was ascertained as follows. If all the roots were real and negative, or if a real root was negative and the real part of a complex conjugate pair of roots was also negative, then the fluid was deemed stable for the given fluid element radius. If one or more of the real roots was positive, or if a real root and/or the real part of a complex conjugate pair of roots was positive, then the fluid was deemed unstable with a growth rate given by the largest of the positive elements, i.e., the largest of the real roots or the largest of the real root and the real part of a complex conjugate pair, as the case may be. If the largest positive element was the real part of a complex conjugate pair, and lepton transport dominated thermal transport (i.e., $|\Upsilon_{ Y_{\ell} }| > |\Sigma_{s}|$. which was always the case), the fluid was deemed unstable to lepto-entropy semiconvection for the given fluid element radius. If the largest positive element was a real root, the fluid would otherwise be stable in the absence of thermal or lepton transport (i.e., Ledoux stable), and lepton transport dominated thermal transport (again this was always the case), the fluid was deemed unstable to lepto-entropy fingers for the given fluid element radius. If the largest positive element was a real root, and the fluid would otherwise be unstable in the absence of thermal or lepton transport (i.e., Ledoux unstable), the fluid was deemed unstable to convection for the given fluid element radius.

\begin{figure}[!h]
\setlength{\unitlength}{1.0cm}
{\includegraphics[width=\columnwidth]{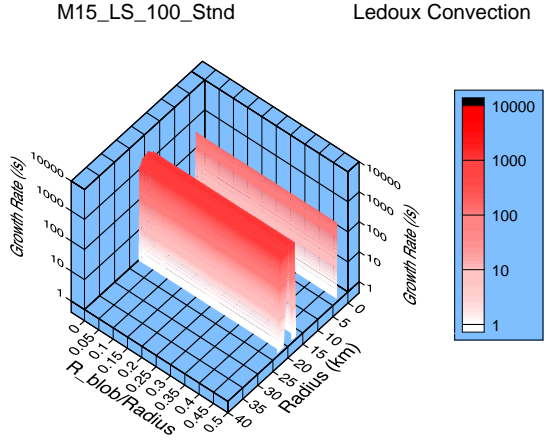}}
\caption{\label{fig:M15_LS_100_L}
\small{Model M15\_LS\_100\_Stnd: Growth rate as a function of core radius and fluid element size for Ledoux convection.}}
\end{figure}

\begin{figure}[!h]
\setlength{\unitlength}{1.0cm}
{\includegraphics[width=\columnwidth]{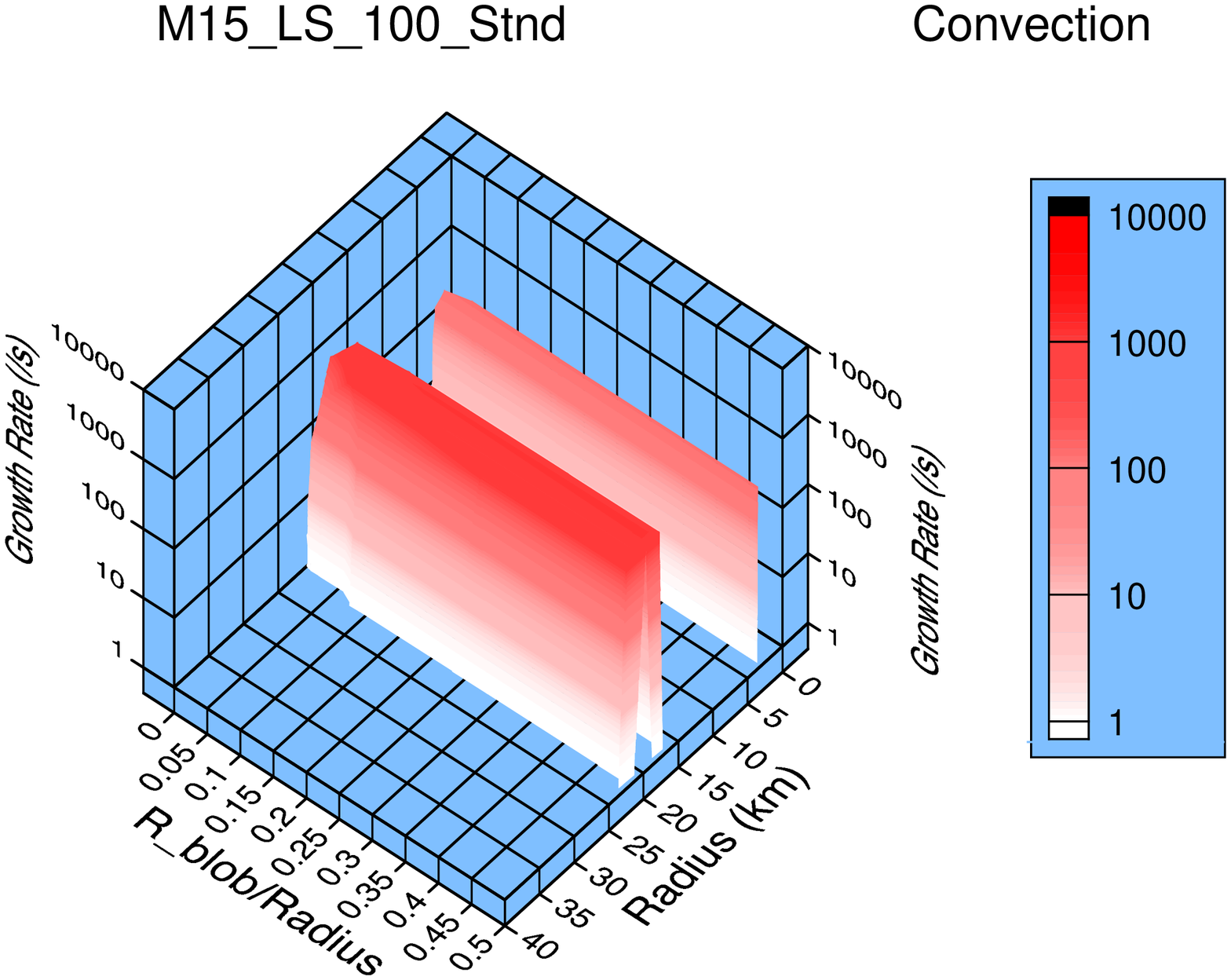}}
\caption{\label{fig:M15_LS_100_CV}
\small{Model M15\_LS\_100\_Stnd: Growth rate as a function of core radius and fluid element size for Convection.}}
\end{figure}

\begin{figure}[!h]
\setlength{\unitlength}{1.0cm}
{\includegraphics[width=\columnwidth]{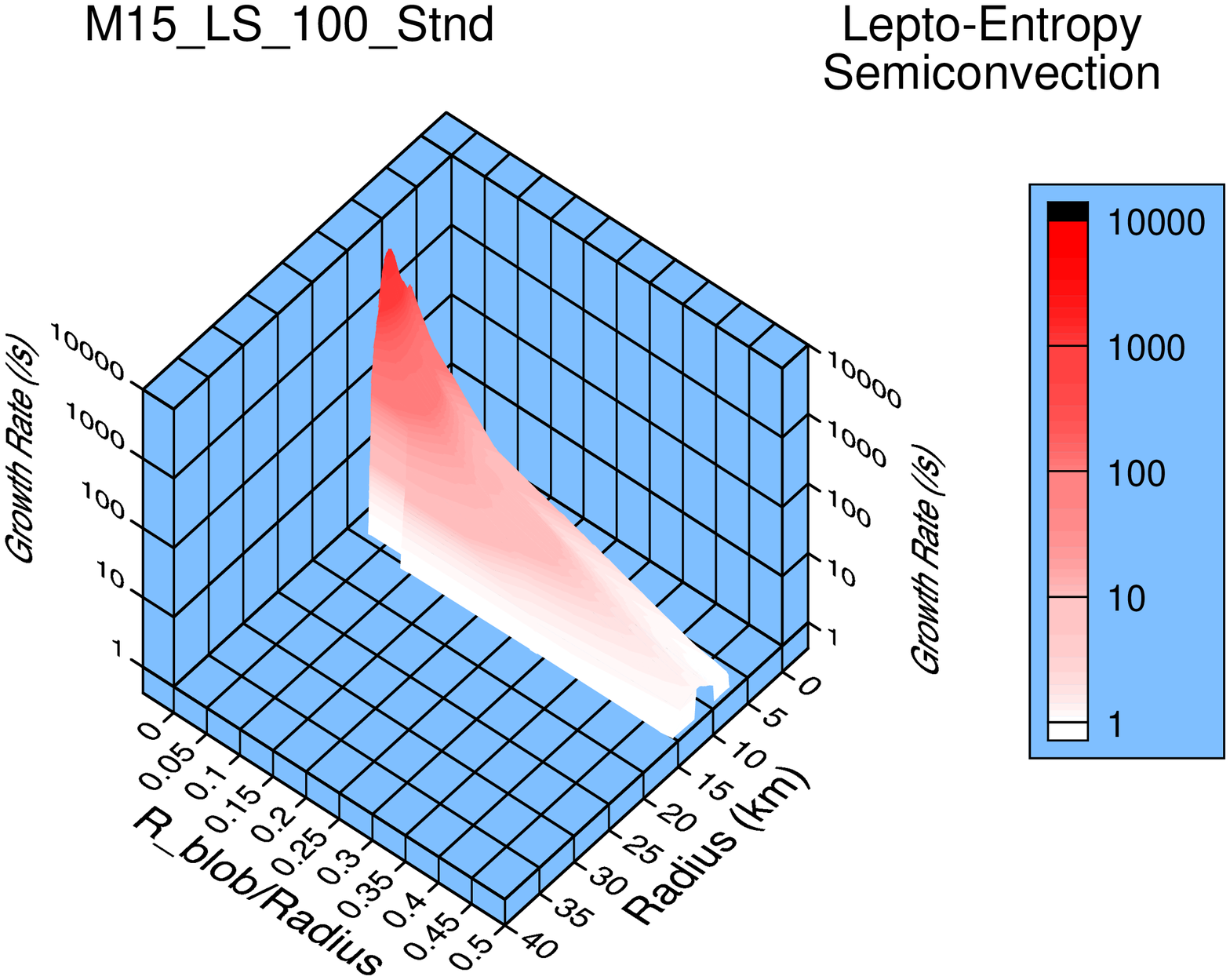}}
\caption{\label{fig:M15_LS_100_SC}
\small{Model M15\_LS\_100\_Stnd: Growth rate as a function of core radius and fluid element size for semiconvection.}}
\end{figure}

\begin{figure}[!h]
\setlength{\unitlength}{1.0cm}
{\includegraphics[width=\columnwidth]{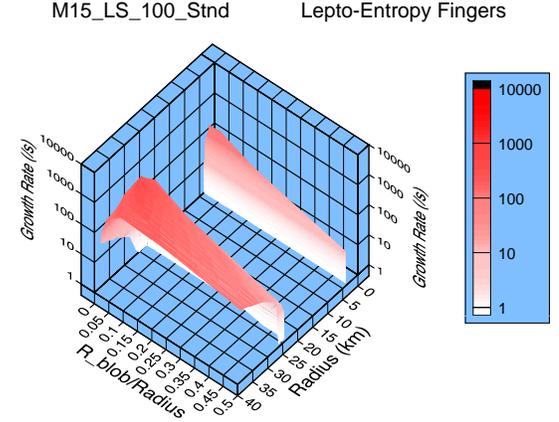}}
\caption{\label{fig:M15_LS_100_NF}
\small{Model M15\_LS\_100\_Stnd: Growth rate as a function of core radius and fluid element size for lepto-entropy fingers.}}
\end{figure}



Figures \ref{fig:M15_LS_100_L} - \ref{fig:M15_LS_100_NF} show the results of our analysis of the stability of model M15\_LS\_100\_Stnd. Figure (\ref{fig:M15_LS_100_L}) gives the growth rate as a function of core radius and fluid element size for Ledoux convection, i.e., the growth rates of the fluid instability that would aeise in the absence of thermal and lepton transport. Except for an  extremely narrow region near the core center, only the region between $R \sim 17$ km and $R \sim 22$ km is Ledoux unstable, where $R$ is the core radius, as is also evident from Figure (\ref{fig:w_ls_15_100}).

Figure \ref{fig:M15_LS_100_CV} gives the growth rates for the convectively unstable region, i.e., the region that is unstable in the presence of thermal and lepton transport that is also Ledoux  unstable. The unstable convectively unstable region is essentially the  same as  the Ledoux unstable region, except that  thermal and lepton transport slightly reduces the growth rates. This is because the convectively unstable region is buoyancy driven, and the tendency of thermal and lepton transport to equilibrate the fluid element with the background reduces the buoyancy that drives this convection.

Most interesting are Figures \ref{fig:M15_LS_100_SC} and \ref{fig:M15_LS_100_NF} which show regions that are unstable solely by virtue of thermal and lepton transport. Figure \ref{fig:M15_LS_100_SC} shows that the lepto-entropy semiconvectively unstable regions are located toward the inner part of the core, i.e., between $R \sim 8$ km and $R \sim 16$ km. Referring to the discussion in Section \ref{sec:LEFS} and Figure \ref{fig:CRF0_3_d14_semi}, instability to semiconvection is  anticipated in the cold and relatively lepton rich inner core when the entropy gradient is positive, as it is for $8 \simgreater R \simgreater 16$ according to Figure \ref{fig:w_ls_15_100}. The lepto-entropy semiconvection will be dominated by small scales, as the growth rate for this instability increases with decreasing fluid element size, as is evident from Figure \ref{fig:M15_LS_100_SC}. Figure \ref{fig:M15_LS_100_NF} shows the region unstable to lepto-entropy fingers. Significantly, it extends from the Ledoux unstable region to the vicinity of the neutrinosphere, and could be a very important source of lepton transport. The growth rates peak for fluid element radii $\sim 1/20 \times R$ near the neutrinosphere increasing in comparison with $R$ to  $\sim 1/4 \times R$ deeper in the core. 

\begin{figure}[!h]
\setlength{\unitlength}{1.0cm}
{\includegraphics[width=\columnwidth]{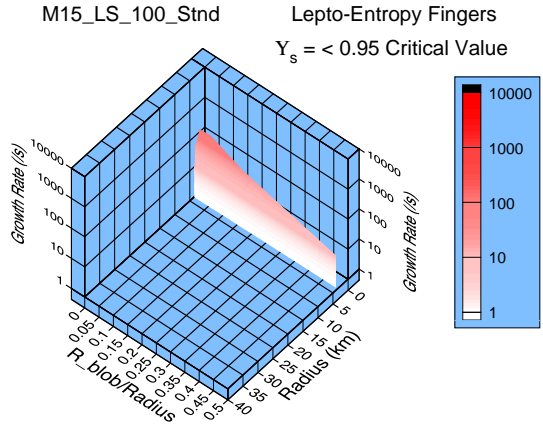}}
\caption{\label{fig:M15_LS_100_NF_UslCV}
Model M15\_LS\_100\_Stnd: Growth rate as a function of core radius and fluid element size for lepto-entropy fingers when $\Upsilon_{s}$ is limited to 95 percent of its critical value, as given by the inequality in equation (\ref{eq:GC10}).}
\end{figure}


As a test of the applicability of inequality (\ref{eq:GC10}) in predicting the lepto-entropy finger instability in the collapsed cores of supernova progenitors,  Figure $\ref{fig:M15_LS_100_NF_UslCV}$ shows that when the cross response function $\Upsilon_{s}$ is restricted to be no larger than 95 percent of the critical value, as given by the inequality in equation (\ref{eq:GC10}), the lepto-entropy finger instability disappears except for an insignificant region near the core center. Figure $\ref{fig:M15_LS_100_NF_UslCV}$ provides a global view of the significance of inequality (\ref{eq:GC10}) for a particular core collapse model. Figures \ref{fig: st_pl_new1} and \ref{fig:st_pl_.95ys} provide an expanded view of the critical role of inequality (\ref{eq:GC10}) at a given thermodynamic state. These latter two figures show the $\frac{d \ln \bar{Y}_{\ell}}{dz} - \frac{d \ln \bar{s}}{dz}$ plane for the thermodynamic state ($\rho = 3.4 \times 10^{12}$ g cm$^{-3}$, $T = 10.3$ MeV, $Y_{\rm e}$ = 0.1165), which pertains to one of the zones of Model M15\_LS\_100\_Stnd. The ratio of fluid element size to distance to core center (i.e., R$_{\rm blob}$/Radius) was chosen to be 0.047, for which the lepto-entropy growth rate was maxmimum. Shown is the convective regions (red), lepto-entropy semiconvective region (green), lepto-entropy finger region (purple) and stable region (blue). Also shown are iso-growth rate contours. Figure \ref{fig:st_pl_.95ys}

\begin{figure}[!h]
\setlength{\unitlength}{1.0cm}
{\includegraphics[width=\columnwidth]{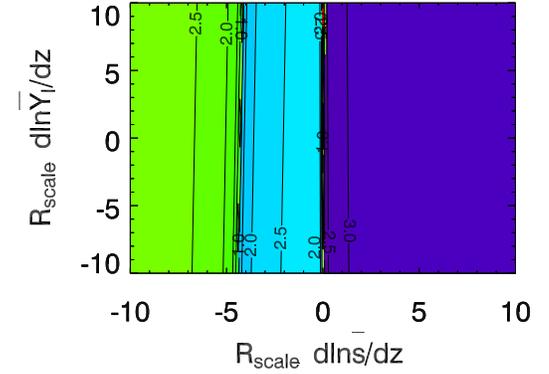}}
\caption{\label{fig:st_pl_new1}
The $\frac{d \ln \bar{Y}_{\ell}}{dz} - \frac{d \ln \bar{s}}{dz}$ plane for the thermodynamic state ($\rho = 3.4 \times 10^{12}$ g cm$^{-3}$, $T = 10.3$ MeV, $Y_{\rm e}$ = 0.1165), of one of the zones of Model M15\_LS\_100\_Stnd, for which inequality (\ref{eq:GC10}) is satisfied, and R$_{\rm blob}$/Radius = 0.047. The color coding is red: convective, blue: stable, green: lepto-entropy semiconvective, and purple: lepto-entropy fingers.}
\end{figure}

\begin{figure}[!h]
\setlength{\unitlength}{1.0cm}
{\includegraphics[width=\columnwidth]{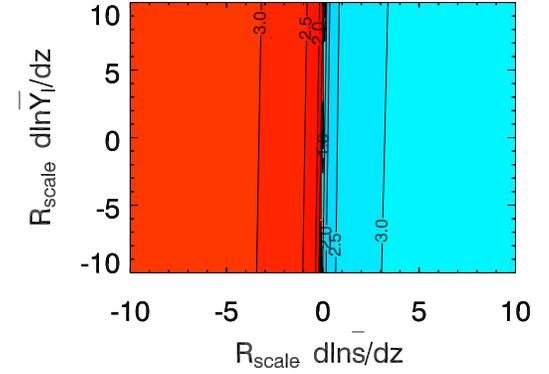}}
\caption{\label{fig:st_pl_.95ys}
The same as Figure \ref{fig: st_pl_new1} except that $\Upsilon_{s}$ was restricted to be 95 percent of the value needed to satisfy inequality (\ref{eq:GC10}).}
\end{figure}


\begin{figure}[!h]
\setlength{\unitlength}{1.0cm}
{\includegraphics[width=\columnwidth]{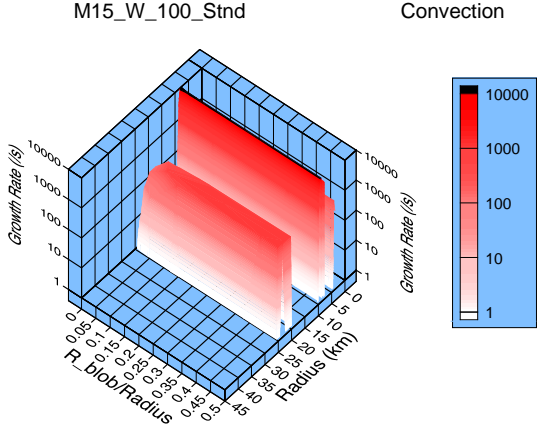}}
\caption{\label{fig:M15_W_100_CV}
Model M15\_W\_100\_Stnd: Growth rate as a function of core radius and fluid element size for Convection.}
\end{figure}


Figure \ref{fig:M15_W_100_CV} shows the regions of convective instability for model M15\_W\_100\_Stnd, a model evolved with the Wilson EOS. There is a convectively unstable region deep in the core ($7 \simless R \simless 9$ km), and another very narrow region near the core center. The convectively unstable region at $7 \simless R \simless 9$ km, not seen in model M15\_LS\_100\_Stnd, corresponds to the negative gradients in both $s$ and $Y_{\ell}$, which can be seen in Figure (\ref{fig:w_ls_15_100}), and which causes this region to be Ledoux unstable. Figures {\ref{fig:M15_W_100_SC} and {\ref{fig:M15_W_100_NF} show the regions of the lepto-entropy semiconvective and lepto-entropy finger instabilities, respectively. These regions are located in the core very similarly to the locations of the corresponding regions in model M15\_LS\_100\_Stnd, except that the lepto-entropy growth rate is down by a factor of 2 or 3.

\begin{figure}[!h]
\setlength{\unitlength}{1.0cm}
{\includegraphics[width=\columnwidth]{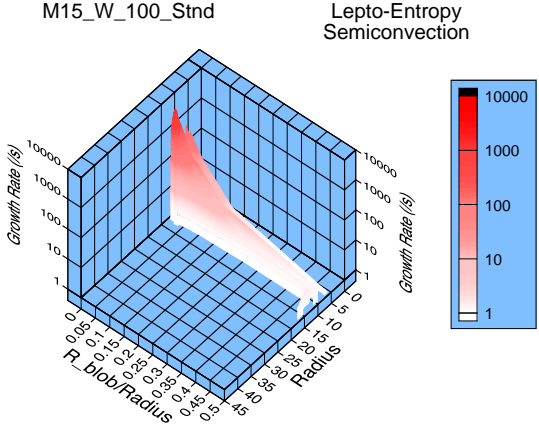}}
\caption{\label{fig:M15_W_100_SC}
Model M15\_W\_100\_Stnd: Growth rate as a function of core radius and fluid element size for semiconvection.}
\end{figure}

\begin{figure}[!h]
\setlength{\unitlength}{1.0cm}
{\includegraphics[width=\columnwidth]{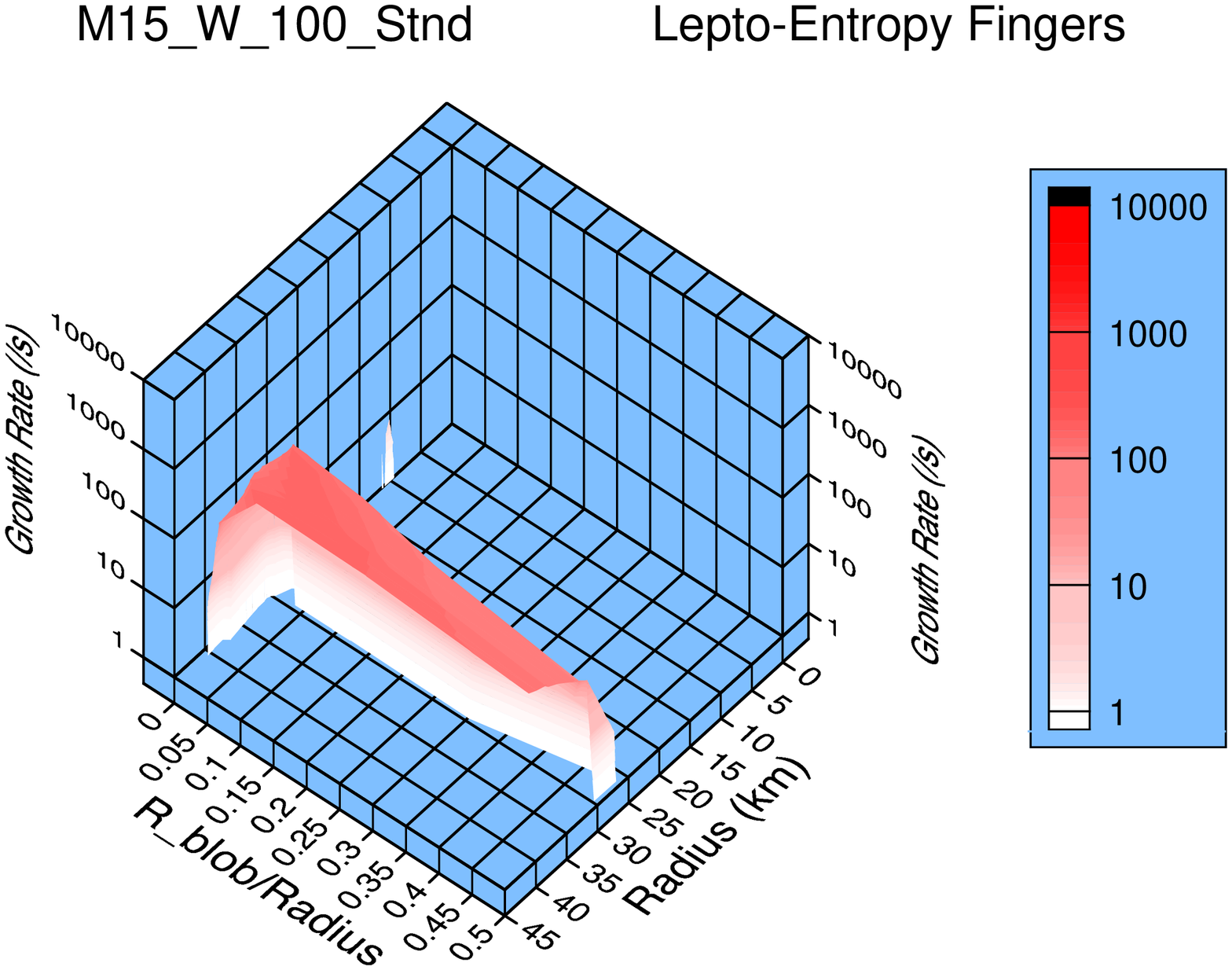}}
\caption{\label{fig:M15_W_100_NF}
Model M15\_W\_100\_Stnd: Growth rate as a function of core radius and fluid element size for lepto-entropy fingers.}
\end{figure}

\begin{figure}[!h]
\setlength{\unitlength}{1.0cm}
{\includegraphics[width=\columnwidth]{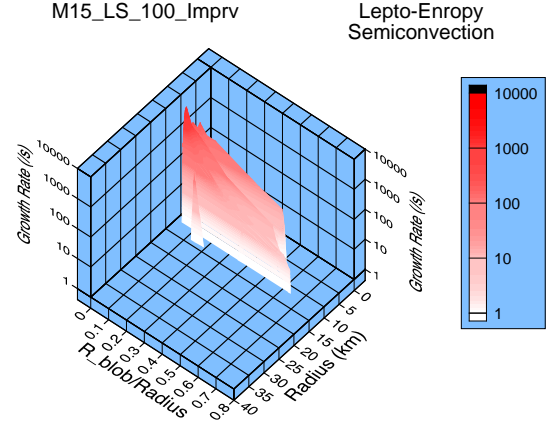}}
\caption{\label{fig:M15_LS_100_Imprv_SC}
Model M15\_LS\_100\_Imprv: Growth rate as a function of core radius and fluid element size for semiconvection computed with ``improved'' neutrino rates.}
\end{figure}

\begin{figure}[!h]
\setlength{\unitlength}{1.0cm}
{\includegraphics[width=\columnwidth]{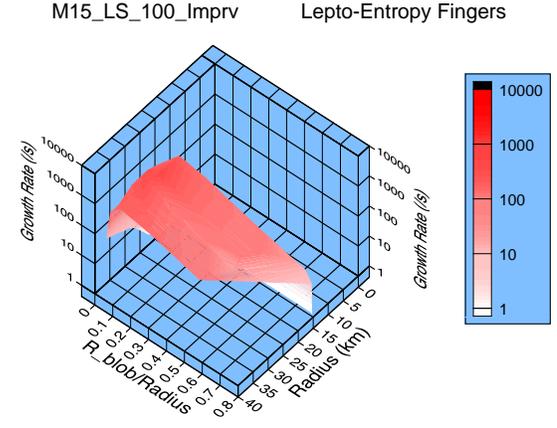}}
\caption{\label{fig:M15_LS_100_Imprv_NF}
Model M15\_LS\_100\_Imprv: Growth rate as a function of core radius and fluid element size for lepto-entropy fingers computed with ``improved'' neutrino rates..}
\end{figure}



Figures \ref{fig:M15_LS_100_Imprv_SC} and \ref{fig:M15_LS_100_Imprv_NF} show the regions of lepto-entropy semiconvective and lepto-entropy finger instabilities, respectively, for model M15\_LS\_100\_Imprv, a model whose regions and stability and instability were analyzed on the basis of ``improved'' neutrino microphysics. Lepto-entropy semiconvective is dominated at small scales, as in the case of the standard neutrino microphysics (Figure {\ref{fig:M15_LS_100_SC}), but extends over a smaller region of $R$, viz., $8 \simgreater R \simgreater 12$ km. Lepto-entropy fingers extends over a larger range of $R$, from $R = 15$ km up to the vicinity of the neutrinosphere, instead of from 23 km in the case of the standard neutrino microphysics. The lepto-entropy finger growth rates are about the same in both analyses, being about several hundred $s^{-1}$ throughout most of the unstable region.

\begin{figure}[!h]
\setlength{\unitlength}{1.0cm}
{\includegraphics[width=\columnwidth]{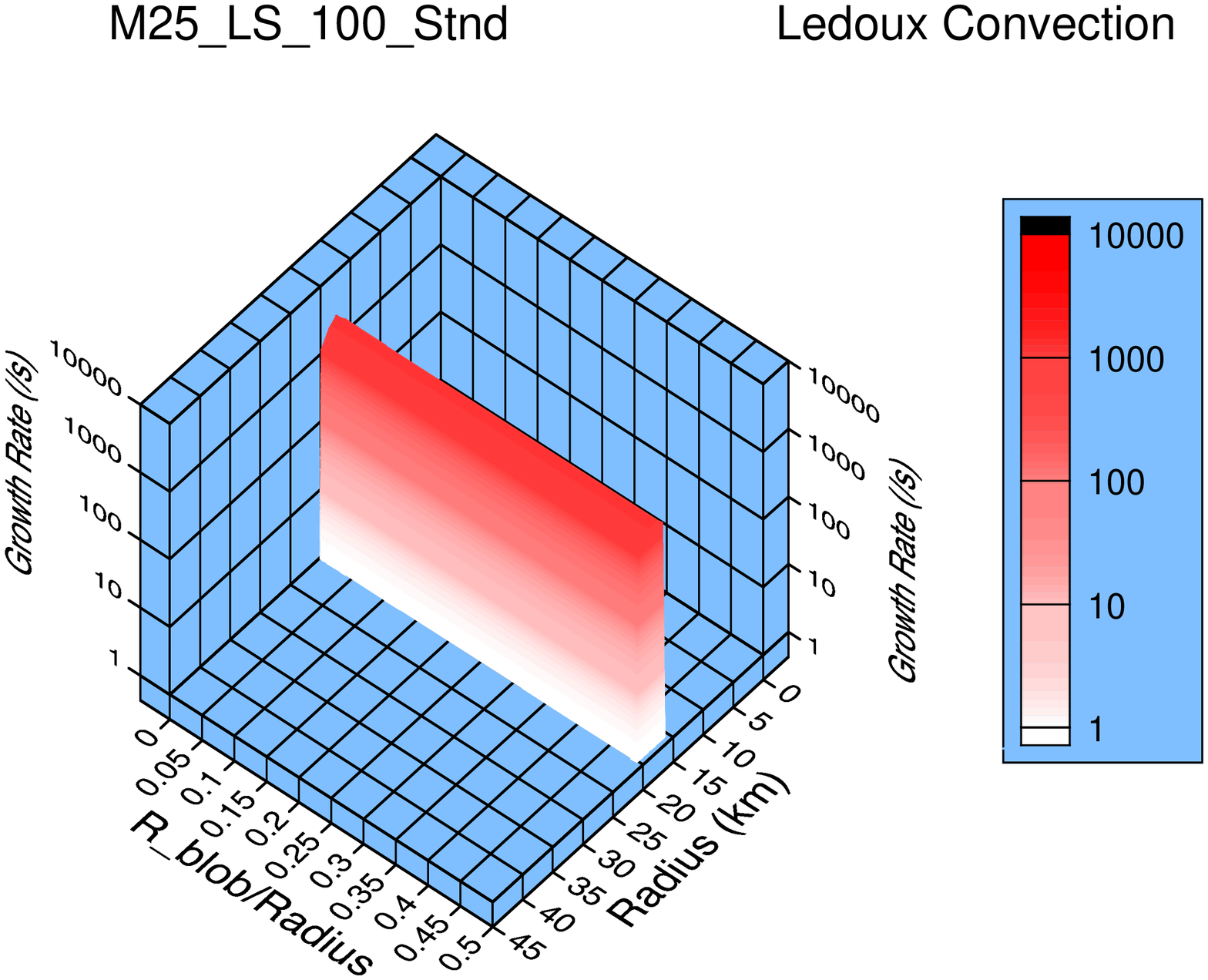}}
\caption{\label{fig:M25_LS_100_L}
Model M25\_LS\_100\_Stnd: Growth rate as a function of core radius and fluid element size for Ledoux convection.}
\end{figure}

\begin{figure}[!h]
\setlength{\unitlength}{1.0cm}
{\includegraphics[width=\columnwidth]{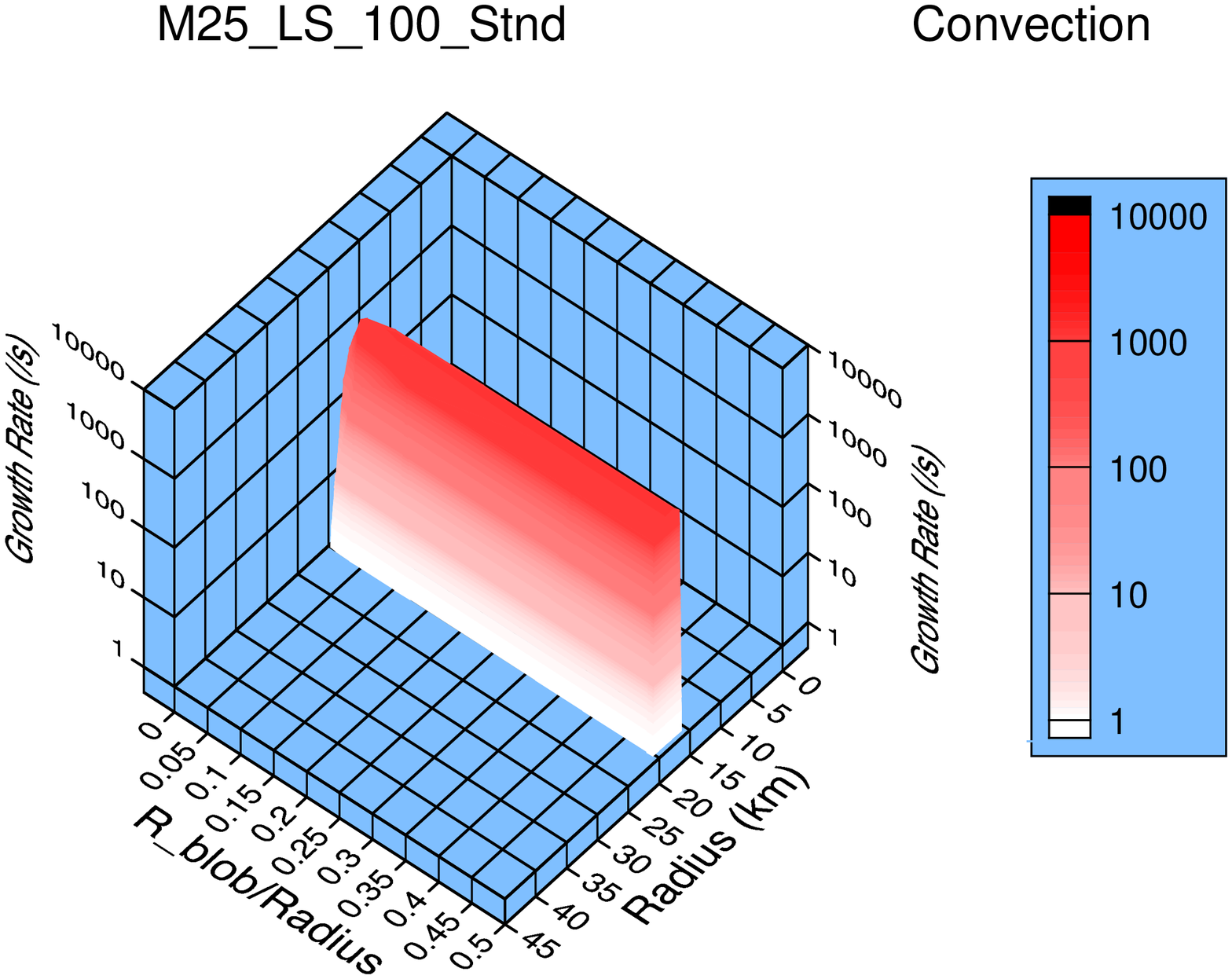}}
\caption{\label{fig:M25_LS_100_CV}
Model M25\_LS\_100\_Stnd: Growth rate as a function of core radius and fluid element size for Convection.}
\end{figure}

\begin{figure}[!h]
\setlength{\unitlength}{1.0cm}
{\includegraphics[width=\columnwidth]{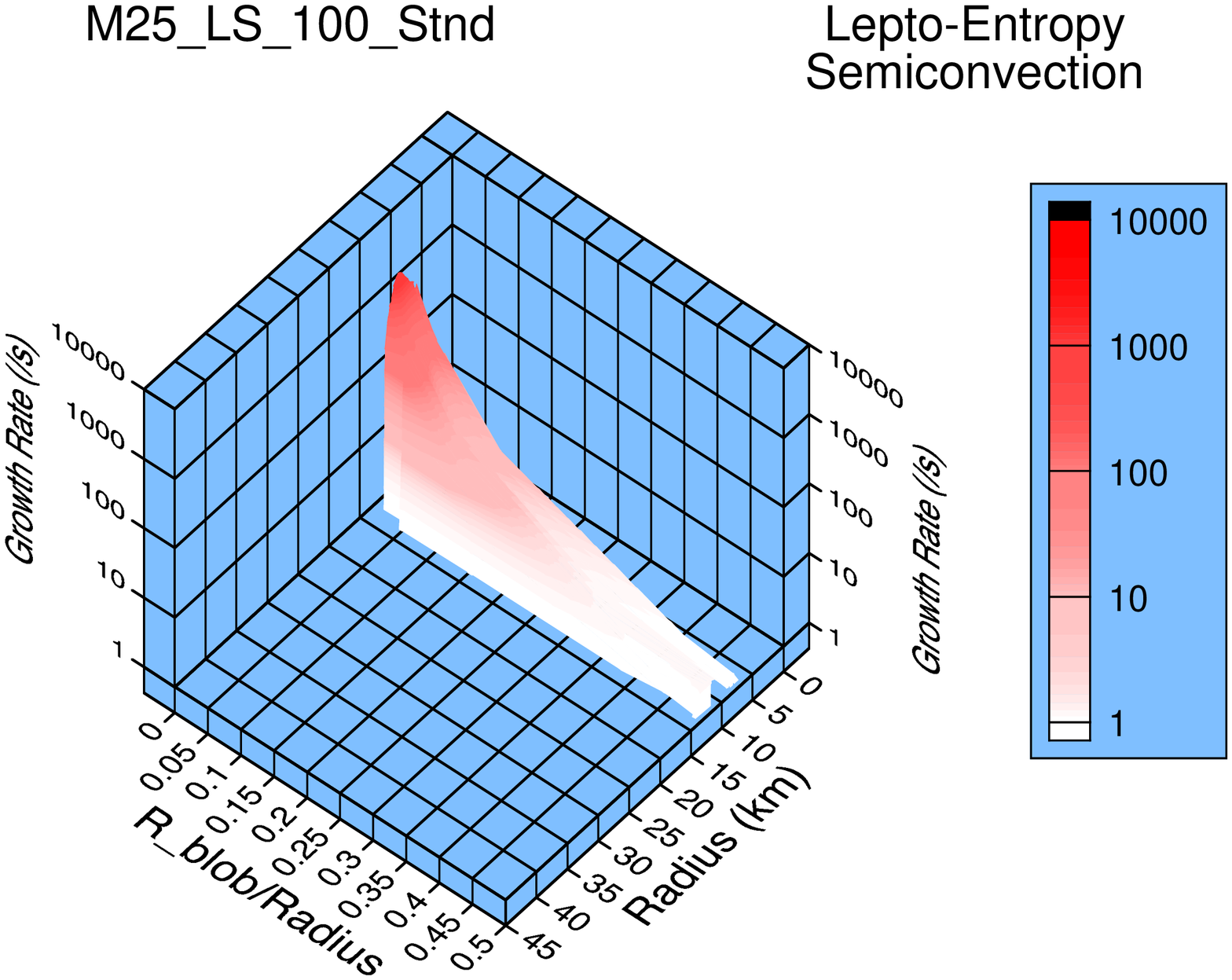}}
\caption{\label{fig:M25_LS_100_SC}
Model M25\_LS\_100\_Stnd: Growth rate as a function of core radius and fluid element size for semiconvection.}
\end{figure}

\begin{figure}[!h]
\setlength{\unitlength}{1.0cm}
{\includegraphics[width=\columnwidth]{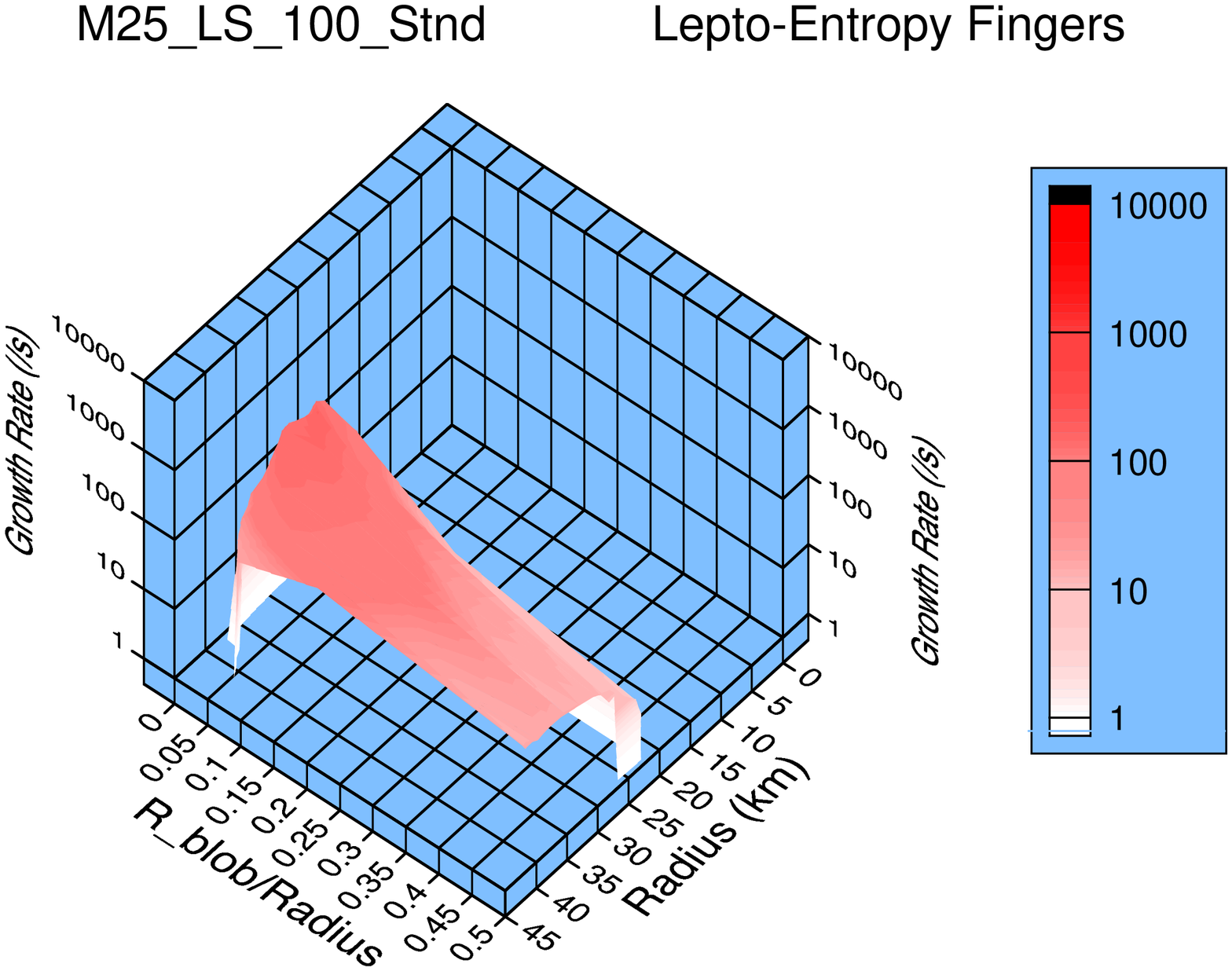}}
\caption{\label{fig:M25_LS_100_NF}
Model M25\_LS\_100\_Stnd: Growth rate as a function of core radius and fluid element size for lepto-entropy fingers.}
\end{figure}

Figures \ref{fig:M25_LS_100_L} - \ref{fig:M25_LS_100_NF} show the various instability regions for model M25\_LS\_100\_Stnd and Figures \ref{fig:M25_W_100_CV} - \ref{fig:M25_W_100_NF} do the same for M25\_W\_100\_Stnd (we omit the ``Ledoux Convection''  graph here because it is very similar to the ``Convection'' graph). Figures \ref{fig:M25_LS_100_Imprv_NF} and \ref{fig:M25_LS_100_Imprv_NF} show the regions of lepto-entropy semiconvective and lepto-entropy finger instabilities, respectively, for model M25\_LS\_100\_Imprv (we omit the graphs of Ledoux Convection and Convection because they are very similar to the results for model M25\_LS\_100\_Stnd. The graphs for models M25\_LS\_100\_Stnd, M25\_W\_100\_Stnd, and M25\_LS\_100\_Imprv\_NF are similar, respectively, to the corresponding graphs for models M15\_LS\_100\_Stnd, M15\_W\_100\_Stnd, and M15\_LS\_100\_Imprv\_NF. The various modes of instability are qualitatively in the same regions of the core, and the growth rates of these instabilities are similar.

We omit showing graphs of the various instability regions for the models at 200 ms as it is not clear how meaningful they are at this time. The fluid motions resulting from the instabilities would have modified the background gradients in $s$ and $Y_{\ell}$, thus modifying the instabilities and their growth rates. Indeed, they may already have done so by 100 ms. All that can be said without detailed multi-dimensional radiation-hydrodynamical simulations is that the lepto-entropy semiconvective and lepto-entropy finger instabilities are likely to appear for for some period of time after core bounce below the neutrinosphere of a proto-supernova.

\begin{figure}[!h]
\setlength{\unitlength}{1.0cm}
{\includegraphics[width=\columnwidth]{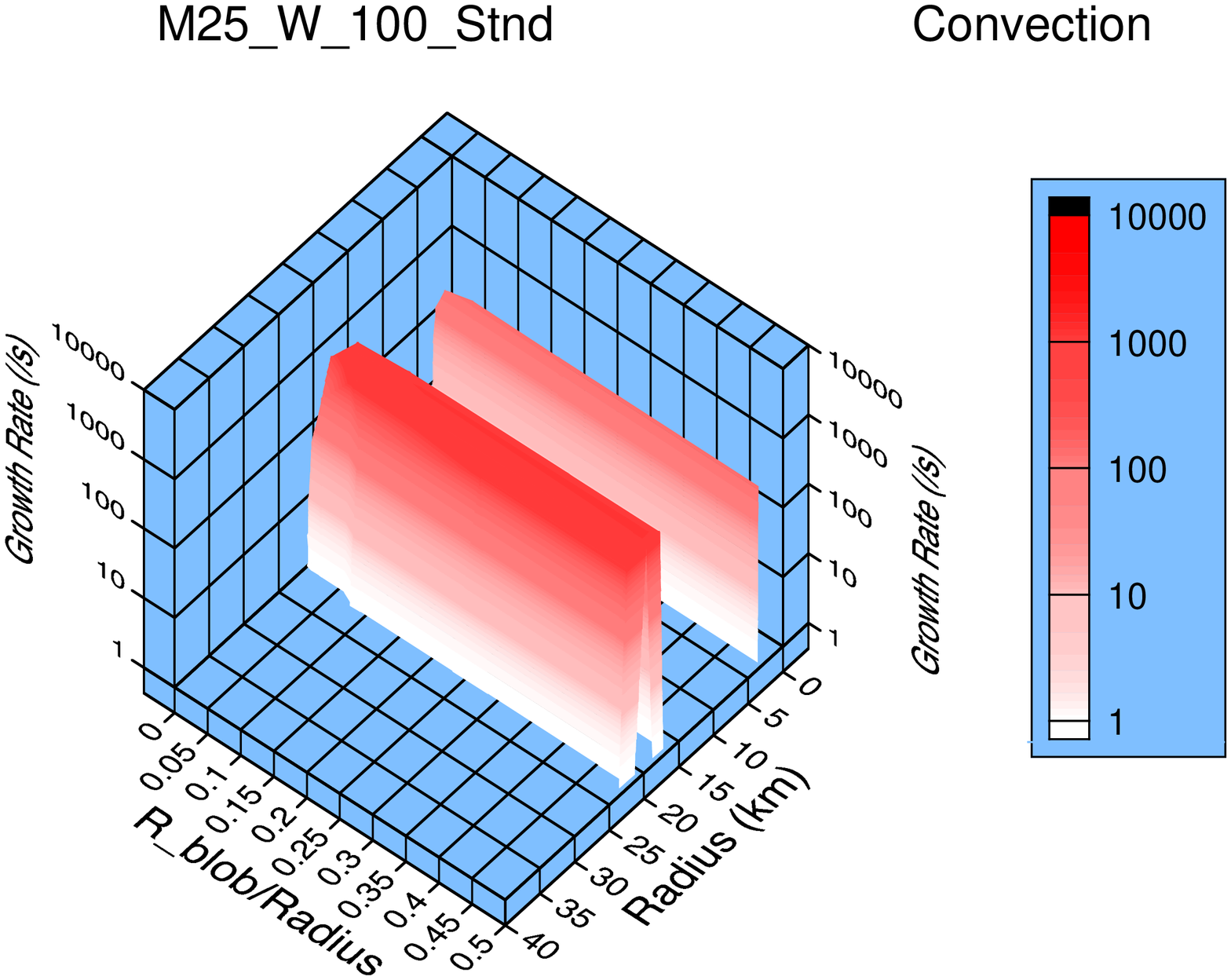}}
\caption{\label{fig:M25_W_100_CV}
Model M25\_W\_100\_Stnd: Growth rate as a function of core radius and fluid element size for Convection.}
\end{figure}

\begin{figure}[!h]
\setlength{\unitlength}{1.0cm}
{\includegraphics[width=\columnwidth]{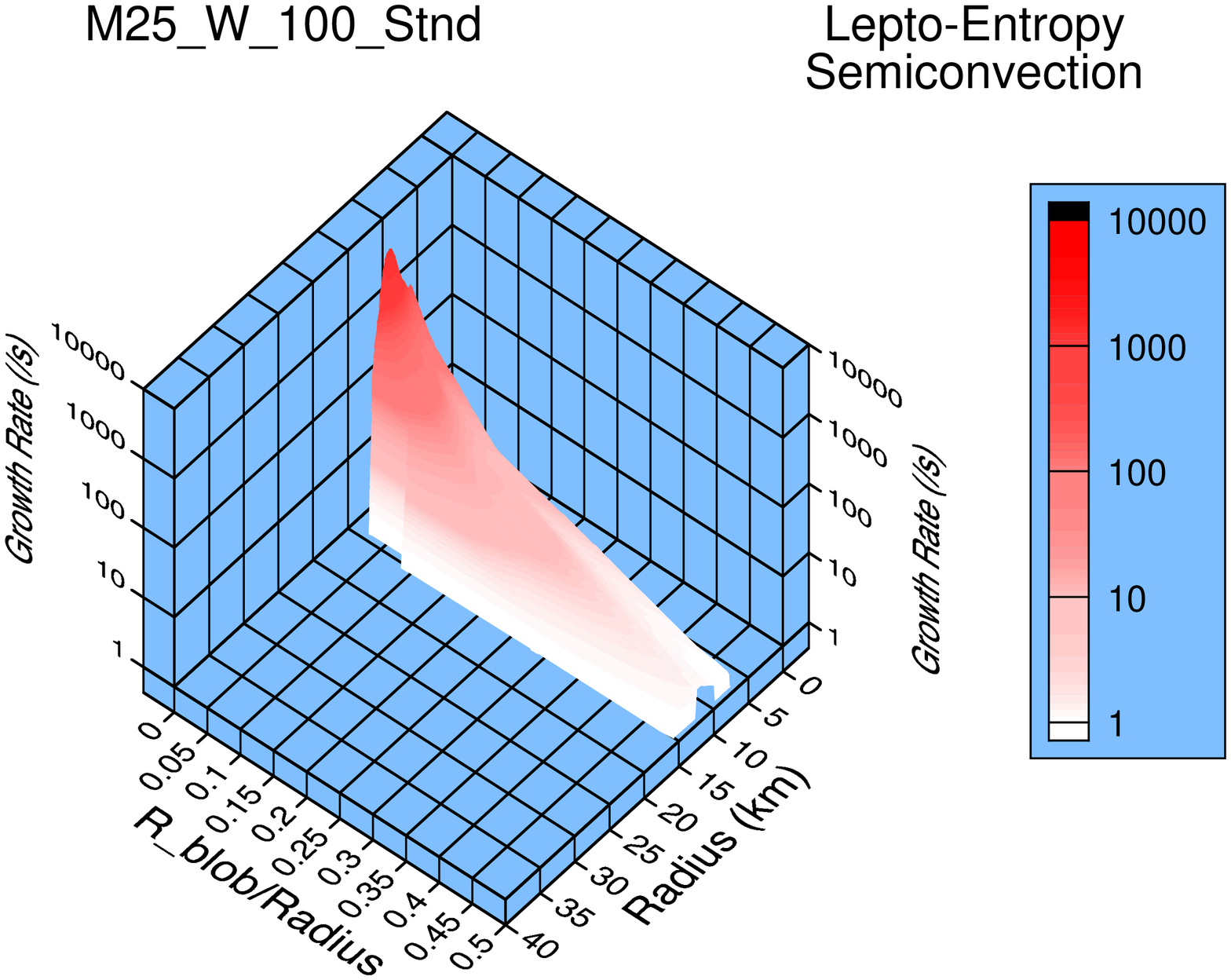}}
\caption{\label{fig:M25_W_100_SC}
Model M25\_W\_100\_Stnd: Growth rate as a function of core radius and fluid element size for semiconvection.}
\end{figure}

\begin{figure}[!h]
\setlength{\unitlength}{1.0cm}
{\includegraphics[width=\columnwidth]{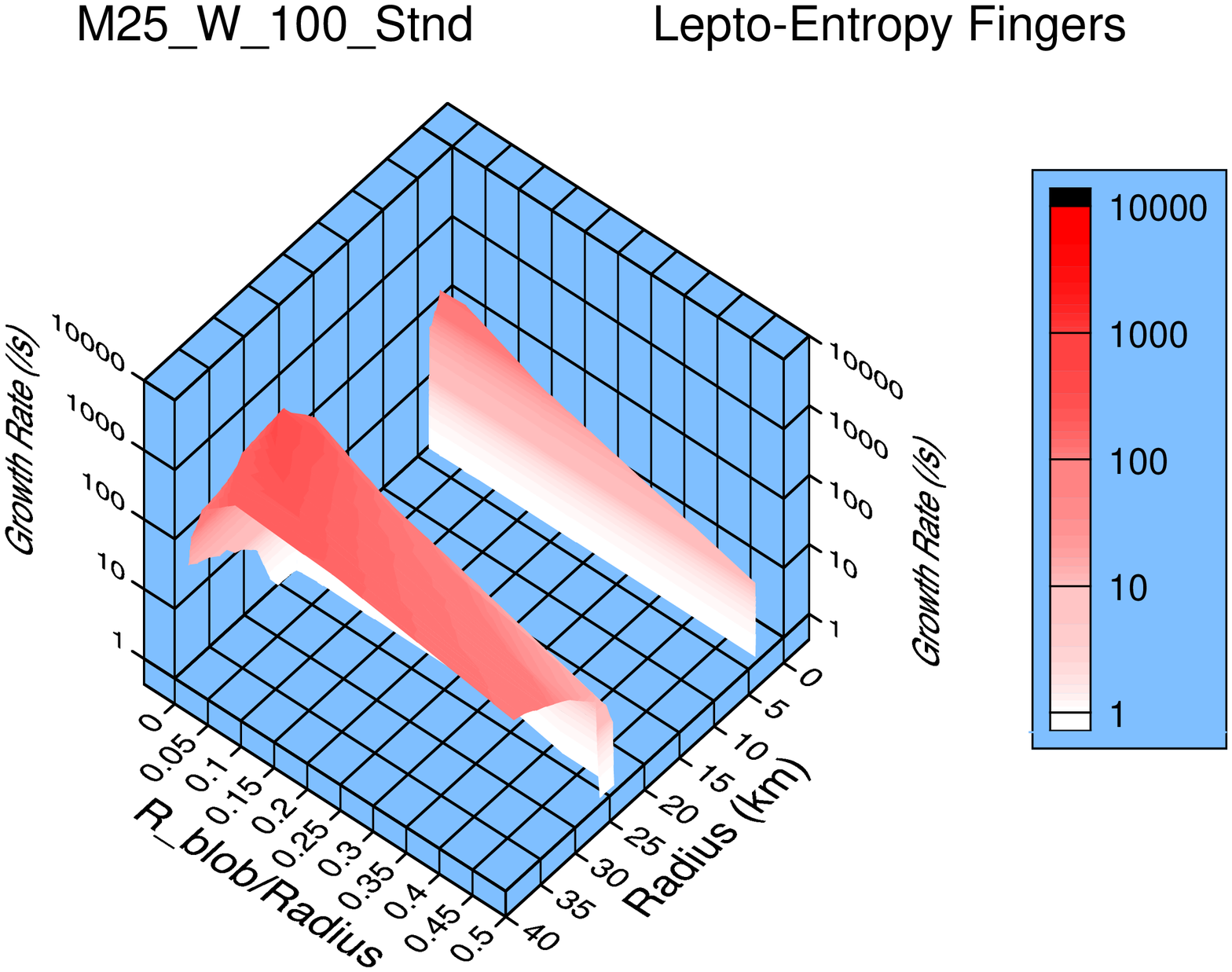}}
\caption{\label{fig:M25_W_100_NF}
Model M25\_W\_100\_Stnd: Growth rate as a function of core radius and fluid element size for lepto-entropy fingers.}
\end{figure}

\begin{figure}[!h]
\setlength{\unitlength}{1.0cm}
{\includegraphics[width=\columnwidth]{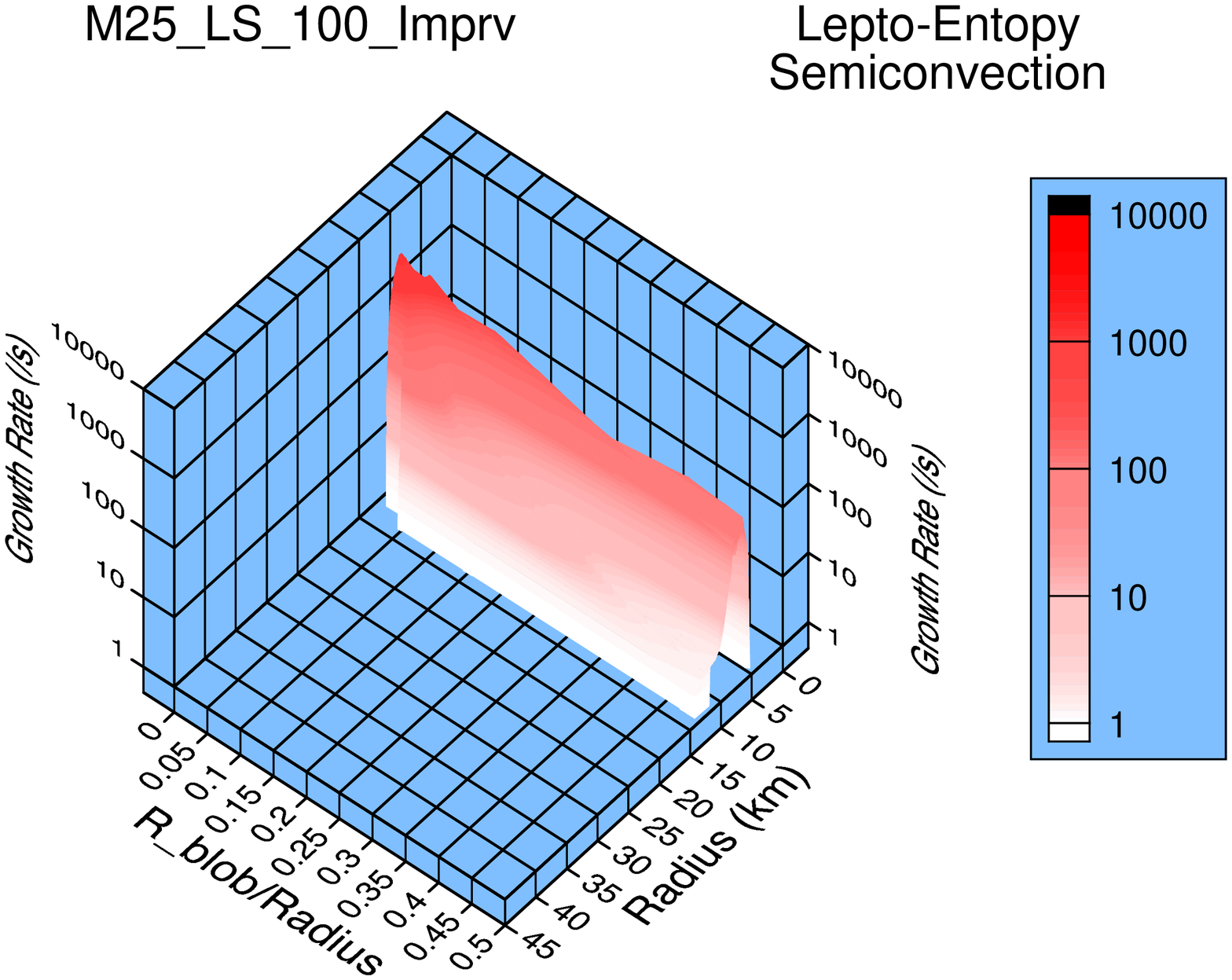}}
\caption{\label{fig:M25_LS_100_Imprv_SC}
Model M25\_LS\_100\_Imprv: Growth rate as a function of core radius and fluid element size for lepto-entropy fingers.}
\end{figure}

\begin{figure}[!h]
\setlength{\unitlength}{1.0cm}
{\includegraphics[width=\columnwidth]{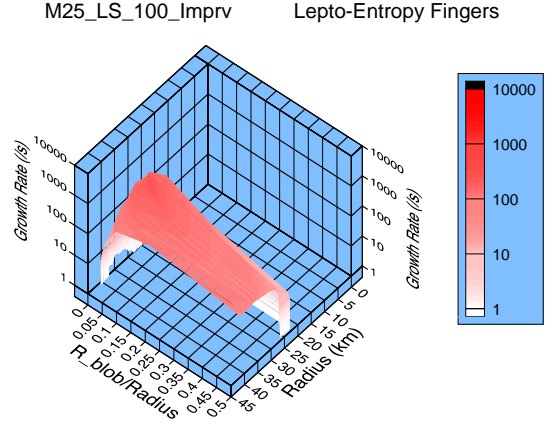}}
\caption{\label{fig:M25_LS_100_Imprv_NF}
Model M25\_LS\_100\_Imprv: Growth rate as a function of core radius and fluid element size for Convection.}
\end{figure}

\section{Comparisons With Other Work}
\label{sec:Comp}

In this Section we will compare our the results of stability analyses with the few other analyses of doubly diffusive instabilities in the literature that are relevant to the collapsed cores of supernova progenitors, and we will examine results of extant multi-dimensional radiation hydrodynamical simulations of stellar core collapse to see if the lepto-entropy finger instability has been seen in any of these simulations

\subsection{Comparisons With Prior Stability Analyses}
\label{sec:Compstb}

A number of authors have recognized the potential role of convection below the neutrinosphere in enhancing the neutrino luminosity. Most of this work e.g., \citep{epstein79, bruennbl79, liviobc80, colgatep80, smarrwbb81, lattimerm81, burrowsf93, jankam93, jankam94, bruennm94, keiljm96}, was focussed on Ledoux convection or Raleigh-Taylor overturns. \citet{smarrwbb81} pointed out the possibility of doubly diffusive instabilities, in particular, neutron fingers, arising in the core after bounce. The Livermore group \citep{wilsonmww86, wilsonm88, wilsonm93} have emphasized the role of doubly diffusive instabilities below the neutrinosphere and have incorporated it in their simulations. In their treatment of these instabilities, they have assumed that the rate of thermal equilibration greatly exceeds that of lepton equilibration, using arguments reviewed in Section \ref{sec:DDInstabilities}, and on the basis of this found that material below the neutrinosphere in their core collapse simulations was unstable to neutron fingers in an extensive region. Modeling the resulting convection by a mixing length algorithm they found an enhancement of the \nue\ luminosity which helped power explosions in their simulations. 

\citet{bruennmd95} applied the single particle analysis of \citet{grossmanna93} to stellar material with neutrino mediated thermal and lepton transport. On the basis of some numerical equilibration experiments they questioned the assumption that thermal equilibration rate exceeded the lepton equilibration rate, and concluded that the neutron finger instability would probably not be present in the collapsed cores of supernova progenitors. They also rediscovered the the observation of \citet{lattimerm81} that the derivative $\left( \pderiv{ \ln \rho }{ \ln Y_{\ell} } \right)_{p,s}$ changes from negative  to positive when the value of $Y_{\ell}$ falls below a critical value, and that the core was likely stable when that occurred.

\citet{bruennd96} also applied the single particle analysis of \citet{grossmanna93}, and extended the analysis by introducing the four response functions and the methodology for determining their values. They also pointed out that the rate of lepton transport substantially exceeded the rate of thermal transport, and that the appearance of neutron fingers in the collapsed core of a supernova progenitor was very unlikely. Like \citet{bruennmd95}, however, they applied their analysis to only a very limited region of the thermodynamic state space of a core and concluded that the material would either be semiconvective or stable, depending on the value of $Y_{\ell}$.

\citet{mirallespu00} have examined the stability of the material in an evolving protoneutron star. By employing an equilibrium diffusion approximation for neutrino transport they were able to derive a dispersion equation (a cubic equation analogous to equation (\ref{eq:GC1})) involving both the growth or decay rate and the horizontal wave number or length scale of the perturbation. They were thus able to solve for the length scale of a perturbation giving the minimum thermal and lepton gradients needed for instability or, equivalently, the length scale of a perturbation giving the maximum growth rate for given thermal and lepton gradients. In our work it was necessary to consider different length scales explicitly. However, the use of an equilibrium diffusion approximation for neutrino transport may cause some of the subtleties of the transport to be missed and result in different relative values for the response functions.

The \citet{mirallespu00} analysis was applied to the interior of a neutron star during its long term evolution. Their results are therefore probably not directly comparable to the results of our analysis here which pertain to the collapsed cores of supernova progenitors before the excess entropy of the shocked material has had a chance to be radiated away. Nevertheless we note that very early in the evolution of the protoneutron star studied by \citet{mirallespu00} they find a convectively unstable region near the surface surrounded by diffusively unstable region which they attribute to neutron fingers. They do not find semiconvection anywhere. This appears to be not altogether dissimilar to our results which show a convectively unstable region bounded above by lepto-entropy fingers and below by semiconvection. The appearance of semiconvection below the convectively unstable region in our models and not in the model examined by \citet{mirallespu00} might be due to the fact that the lepton fraction in the deep interior of our models is relatively large (so that $\left( \pderiv{ \ln \rho }{ \ln Y_{\ell} } \right)_{p,s} < 0$ whereas the opposite might have been true in their model. According to the discussion at the end of Section \ref{sec:GC}, this could explain the different results.

The \citet{mirallespu00} work also included viscosity, the magnitude of which is uncertain and not included in this work, They found that viscosity, as given by \citet{vandenhornv81}, did not greatly affect the diffusively unstable regions, but found that the convectively unstable regions to be more extended with the inclusion of viscosity..

The role of magnetic fields in the development of proto-neutron
star instabilities was investigated in \citet{mirallespu02}. Magnetic
fields have a dual role: stabilization and destabilization. The
essential results presented in \citet{mirallespu02} indicate that magnetic
field strengths in excess of $10^{13}$ G may do both: stabilize
regions unstable to convection, limiting the spatial extent of
such regions, and destabilize regions to new modes of instability.
However, their fundamental conclusion is that only very strong
magnetic fields ($> 10^{16}$ G) would significantly alter the
stability properties of proto-neutron stars. These fields might
be generated by dynamo action from convective activity present
in the core prior to the escalation of field strengths. Once the
field strength increases, convective activity is shut off. Thus,
magnetic fields may affect both the spatial extent and the temporal
duration of convective activity in the proto-neutron star.

\subsection{Comparisons With Multi-Dimensional Core Collapse Simulations}
\label{sec:Compccs}

Most of the extant  multi-dimensional radiation-hydrodynamical simulations of stellar core collapse have employed gray neutrino transport, and it is not clear how much of the dynamics of the lepto-entropy finger instability these simulations are capable of capturing. With this caveat in mind, we will proceed to  examine the results of these simulations that have been described in the literature.

The simulations of \citet{herantbhfc94} used a smooth Newtonian two-dimensional SPH (smooth particle hydrodynamics) code coupled to gray neutrino transport. The letter transports number of energy of each neutrino species in a two-part scheme consisting of flux-limited diffusion scheme for optically thick regions and a light bulb approximation for optically thin regions. For their core collapse simulation of a 25 \msol Weaver and Woosley \citet{weaverw93} an instability below the neutrinosphere developed over a time scale of 15 ms after bounce with maximum turnover velocities varying over time between 1000 and 4000 km s$^{-1}$. This unstable region extended 15 km below the neutrinosphere which had a radius of 40 km at the time, with $\sim$ 5 convective cells in a 90 degree sector. \citet{herantbhfc94} attributed this instability to the neutron finger mechanism described by the Livermore group. The time for the growth of the instability, the scale of the turbulence, and the region over which it was observed is consistent with the results displayed in Figure \ref{fig:M15_LS_100_NF}. We suggest that they were seeing lepto-entropy fingers.

The  simulations of \citet{fryer98, fryer99}, and \citet{fryerw02} used basically the  smooth Newtonian dimensional SPH code coupled to gray neutrino transport of the \citet{herantbhfc94} work, but with improvements in the SPH code (e.g., technical improvements and the addition of GR effects), the equation of state, and the neutrino transport (the latter still being gray, however). Additionally. the simulations performed perfromed by \citet{fryerw02} were three-dimensional. Unfortunately most of the discussion in these works was concerned with the entropy driven convection occurring above the neutrinosphere, rather than convection below. However, Figure 2 of \citet{fryerw02}, which is a two-dimensional slice of the core of a 15 \msol progenitor 75 ms after bounce, does show evidence of convective flows below the neutrinosphere, although it is difficult to infer details.

Both the simulations of \citet{millerwm93} and \citet{jankam96} cut out the innermost part of the collapsed stellar core and imposed boundary conditions, so fluid motions below the neutrinosphere were not computed. In the simulations of  \citet{mezzacappacbbgsu98a}, a PPM hydrodynamics scheme was coupled to the zero and first moments of neutrino fields previously computed in spherical symmetry with a multigroup, flux-limited transport. While these simulations were able to capture some convective instabilities, the imposed neutrino background resulted in the inability of simulations to compute the fluid element-to-background neutrino transport that would arise in response to differences between the thermodynamic state of the fluid element and the background. Consequently, doubly diffusive instabilities could not be captured by these simulations.

Finally, the simulations of \citet{burasrjk02} used Newtonian PPM hydrodynamic with some general relativistic corrections coupled to radial-ray transport via a multigroup, tangent ray scheme with closure provided by the solution of the Boltzmann equaiton on an angularly averaged stellar background. While details are sparse, the authors describe a convection that sets in below the neutrinosphere by 40 ms after bounce, is persistent up to 0.26 s after bounce when the simulations were terminated, and which slowly extended deeper into the core. The authors attributed this convection to  Ledoux convection, but it may have been lepto-entropy fingers.

\section{Conclusions}
\label{sec:Cncl}

The fluid below the neutrinosphere of the collapsed core of a supernova progenitor can be driven unstable by thermal and lepton diffusion as well as by a gravitationally unstable stratification of the fluid. We have developed the methodology introduced by \citet{bruennd96} for analyzing fluid instabilities in the presence of neutrino mediated thermal and lepton transfer. In this analysis four ``response functions'' are introduced to model the thermal and lepton equilibration of a fluid element perturbed in entropy and lepton fraction from that of the background. These response functions depend on the thermodynamic state of the background and the size of the fluid element, and are obtained by detailed neutrino transport simulations. Given values of the response functions and the ambient gravitational field at the position of the fluid element, a cubic equation can be derived whose roots characterize the response of the fluid element to a perturbation in either its motion or thermodynamic state. From a determination of these roots, the stability or mode of instability of the fluid can be ascertained as a function of the gradients in entropy and lepton fraction. This methodology is applied both to resolve a long standing controversy regarding the presence of the neutron finger instability below the neutrinosphere, and to perform an extensive analysis of the fluid instabilities that do occur below the neutrinosphere for a number of representative post bounce supernova progenitors.

The Livermore group \citep{wilsonmww86, wilsonm88, wilsonm93} have argued that thermal equilibration proceeds much more rapidly than lepton equilibration in the fluid below the neutrinosphere. As a consequence,  given the gradients in entropy and lepton fraction that arise in their supernova models subsequent to the initial collapse of the core and bounce they found that the neutron finger instability will arise in an extensive region below the neutrinosphere. Modeling by a mixing length algorithm the fluid motions that are expected to result from this instability, they found an enhancement of the \nue\ luminosity that helped to produce successful explosions in their sophisticated spherically symmetric supernova simulations. Their work is very important as it has emphasized the potential role of fluid instabilities below the neutrinosphere in the supernova mechanism.

We have performed core collapse simulations using the Livermore \citep{wilson03} equation of state. Applying their criteria for the neutron finger instability, we do find an extensive region below the neutrinosphere, on either side of a Ledoux unstable region, that is neutron finger unstable. A similar exercise for core collapse simulations computed with the \citet{lattimers91} equation of state shows a similar region of neutron finger instability, although not quite as extensive. Applying our methodology to the region below the neutrinosphere of these models we always find that lepton equilibration proceeds much more rapidly than thermal equilibration, and therefore that the neutron finger instability never occurs.

On the other hand, we have also found that the cross response functions, viz., the tendency for a perturbation in entropy of the fluid element relative to the background to drive a net lepton flow between them, and the tendency of a lepton fraction perturbation to drive a net energy flow, have substantial positive magnitudes for all thermodynamic states, and that the ratio of the second to the first of these response functions is large. These cross response functions cause a difference in lepton fraction between the fluid element and the background to develop primarily in response to an entropy difference. This is in contradistinction to the case without diffusion (the Ledoux case) in which a lepton fraction difference develops from a displacement of the fluid element through a gradient in the background lepton fraction.

As a result of this, several new doubly diffusive instabilities appear which we have referred to as lepton-entropy fingers and lepton-entropy semiconvection. These instabilities are expected to present in significant regions below the neutrinospheres of post collapsed stellar cores, may play an important role in the supernova mechanism. We have analyzed these instabilities in detail.

For each of a number of post bounce core collapse supernova models we have performed an extensive survey of fluid  instabilities on a two-dimensional grid of fluid element size and radial location in the core. Our results show a common pattern of unstable regions in these core collapse models. Outside of a Ledoux unstable region is a region where the lepton fraction is small, due to prior shock dissociation and deleptonization, and the entropy gradient is positive, due to the powering up of the shock. The fluid is leptom-entropy finger unstable in this region, with maximum growth rates for fluid element scales approximately 1/20 the distance of the fluid element from the core center. This unstable region can extend up almost to the neutrinosphere, and may therefore be important for the same reason adduced by the Livermore group for their neutron fingers. Below the Ledoux unstable region is region of unshocked material of relatively high lepton fraction. This region is unstable to lepto-entropy semiconvection with growth rates favoring small scales. Reviewing the multidimensional core collapse simulations that have been reported in the literature we conclude that the lepto-entropy finger instability may have already been seen.

\section{Acknowledgments}
\label{sec:Ack}

S.W.B. was partially funded by a grant from the DOE Office of Science Scientific Discovery through Advanced Computing Program.

A.M. is supported at the Oak Ridge National Laboratory, which is managed by UT-Battelle, LLC for the DOE under contract DE-AC05-00OR22725. He is also supported in part by a DOE ONP Scientific Discovery through Advanced Computing Program grant.

\clearpage

\end{document}